\documentclass[aps,prb,reprint,twocolumn,groupedaddress,longbibliography]{revtex4-2}

\usepackage{amsfonts}
\usepackage{amsmath}
\usepackage{amssymb}

\usepackage[T1]{fontenc}
\usepackage{newtxtext}
\usepackage{newtxmath}

\usepackage{graphicx}
\usepackage{bm}
\usepackage{braket}
\usepackage{xcolor}
\usepackage{hyperref}
\hypersetup{colorlinks=true, linkcolor=blue, citecolor=blue, urlcolor=blue}

\begin{document}

\title{Anyon Coherent States in Fractional Chern Insulators}

\author{Pedro Fittipaldi de Castro}
\author{Haylen Gerhard}
\author{Wladimir A. Benalcazar}
\email{benalcazar@emory.edu}

\affiliation{Department of Physics, Emory University, Atlanta, GA 30322, USA}

\date{\today}

\begin{abstract}
In fractional quantum Hall (FQH) systems, coherent anyonic excitations are the natural entities for describing anyon dynamics, their interactions, and braiding, yet no general expression for their wave functions is known beyond Laughlin's quasiholes. We introduce a systematic, operator-based construction of maximally localized wave functions of anyons of either charge in FQH states and fractional Chern insulators, including those in non-ideal, non-vortexable bands. We find that these wave functions describe coherent quasiparticles: zero modes of local annihilation operators that saturate projected-density uncertainty bounds. These coherent states owe their existence to a nontrivial topology of the anyon band, with the winding of their Bloch coefficients encoding an anyon band’s Chern number. As such, their guiding centers exist anywhere within the lattice, and their mutual overlaps decay as Gaussians determined by the effective magnetic length of the anyon band.
\end{abstract}

\maketitle

Anyons are the defining excitations of topologically ordered
phases~\cite{leinaas1977theory,wilczek1982quantum,wen1990topological,moore1989classical}.
Their fractional charge and statistics, long studied in fractional quantum Hall (FQH)
liquids~\cite{TsuiStormerGossard82,laughlin1983,Haldane1983,halperin1984,arovas1984fractional,saminadayar1997observation,de1998direct,bartolomei2020fractional,nakamura2020direct},
now also arise at zero magnetic field in fractional Chern insulators
(FCIs)~\cite{neupert2011fractional,sheng2011fractional,tang2011fractional,regnault2011fractional}
realized in moir\'e MoTe$_2$ and rhombohedral
graphene~\cite{cai2023signatures,zeng2023thermodynamic,park2023observation,xu2023observation,Lu2024FQAH},
and local probes have begun to image individual
anyons~\cite{deng2026realspaceimagingquantumhall,park2026localspectroscopyanyonsbound}.
These advances make a basic question pressing: what does a single anyon look like in a
crystal? The answer determines how anyons move, interact, and
braid~\cite{kitaev2003fault,nayak2008nonabelian}, and sets the intrinsic size and shape against which images of anyons can be compared.

In the continuum, the canonical answer is Laughlin's quasihole, obtained by threading a
vortex through the FQH ground state~\cite{laughlin1983}. Within the lowest Landau level
(LLL), it is a coherent state of the quasihole guiding-center
coordinates~\cite{kjonsberg1997anyon,shn5-vwgt,gzdq-wkxj}: a minimum-uncertainty wave packet
that can be centered anywhere. For the opposite charge, however, the problem has remained
open since the 1980s. Inserting an antivortex excites higher Landau levels, and projecting
back onto the LLL~\cite{laughlin1983,jain1989,jain2007composite,Hansson2009} yields
quasielectrons that are neither coherent nor maximally
localized~\cite{PhysRevB.108.L201106}. 

In FCIs, which possess only discrete translation symmetry, the situation is worse. Laughlin-type
quasiholes, which have been used to describe dispersive anyon
bands~\cite{shn5-vwgt,iyer2026dispersionanyonblochbands}, can be constructed only in
\emph{ideal} Chern bands, whose quantum geometry mimics that of the
LLL~\cite{PhysRevB.90.165139,jackson2015geometric,ledwith2023vortexability}, but the
bands of experimental FCIs are not ideal. Numerical studies have localized anyons with
pinning
potentials~\cite{macaluso2020charge,ravciunas2018creating,Liu2025FCIquasiparticle,palm2026interferometric,wang2026anyon,cfm5-wgz1,adlong2026anyonsimpuritiesfractionalcharge}
or by binding them to other excitations~\cite{Lu_2026,lu2026anyonboundstatesmote2}, approaches that bias the anyon profile and thus are not representative of stand-alone states that carry the intrinsic characteristics of the anyon band.

A more basic question is what form the maximally localized states of an anyon band can take in a crystal at all. For electrons, the answer is set by band topology. Topologically trivial bands admit
exponentially localized Wannier states pinned to the lattice. A nonzero Chern number $C$
forbids them~\cite{thouless1984wannier,brouder2007exponential}; localized states instead form
an overcomplete family of coherent states that can be centered anywhere. In FCIs, the low-energy
single-anyon states also form a band, labeled by crystal momentum, and whether its maximally
localized states are lattice-pinned Wannier states or coherent states, and what sets
their size, has remained unknown. 

Here, we introduce a systematic, operator-based construction of maximally localized anyons of positive or negative charge in FQH and FCI systems and show they are coherent states: (i) they are zero modes of local annihilation operators and saturate an uncertainty bound
of the projected density operators, and (ii) together they resolve the projector onto the
single-anyon band. To obtain them, we extend the Spatial Localizer, recently developed for insulators~\cite{SpatLoc} and excitons~\cite{gerhard2026spatiallocalizerconstituentresolvedexciton}, to systems of many interacting electrons. In the $1/3$ FQH state, it recovers Laughlin's quasihole and, on equal footing, produces a quasielectron coherent state. In a kagome FCI~\cite{PhysRevLett.106.236803} whose Chern band is neither ideal nor vortexable~\cite{ledwith2023vortexability}, the constructed anyons still form a family of coherent states with guiding centers $\bm r$ anywhere in the lattice, rather than Wannier states pinned to fixed positions in the unit cell. 

Without the ladder algebra that guarantees a continuum of coherent states in FQH liquids~\cite{kjonsberg1997anyon}, we argue that anyon coherent states in FCIs are instead tied to the emergent topology of the anyon band, in analogy with how, for single-particle insulators with Chern number $C$, the Atiyah-Singer index theorem~\cite{nakahara2018geometry,PhysRevB.110.245112} guarantees $|C|$ orthogonal coherent states at each position $\bm r$ within the unit cell. In the kagome FCI under periodic boundary conditions at $\nu=1/3$, we show that the anyon band has $C_\text{anyon}=+3$ ($-3$) for quasiholes (quasielectrons), matching the three orthogonal coherent states we find at every guiding center~$\bm r$, and that the winding of their Bloch coefficients is equal to $C_{\mathrm{anyon}}$. Furthermore, while the shapes of the states' density profiles vary widely with~$\bm r$, their spread is nearly constant and matches, without free parameters, the effective magnetic length $\ell=(2\pi|C_\text{anyon}|/A_\text{BZ})^{1/2}$ predicted from the anyon Chern number, with $A_\text{BZ}$ the Brillouin-zone area.

\emph{Anyon coherent states in general fractional quantum Hall systems}\label{sec:LocalizerSetUp}---
\begin{figure*}
    \centering
    \includegraphics[width=\linewidth]{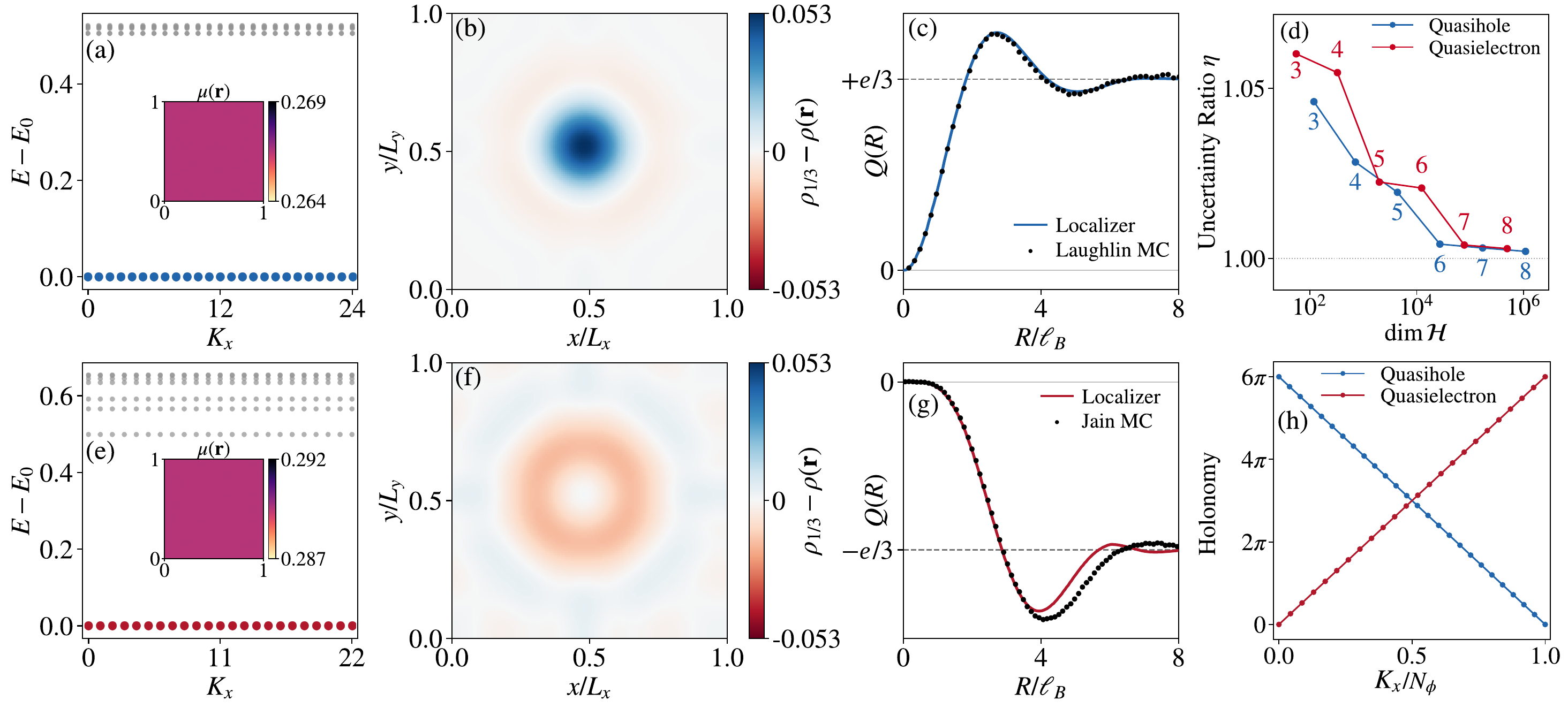}
    \caption{\textbf{Anyon coherent states in the FQHE.}
(a) Low-energy spectrum of the $V_1$ pseudopotential at $N_\phi=3N_e+1=25$, with the anyon band in blue.
Inset: localizer indicator function \eqref{LIF}.
(b) Electron-density deviation from the uniform
$\nu=1/3$ background for the quasihole localizer state
$\ket{\psi(\bm 0)}$.
(c) Integrated charge $Q(R)$ of $\ket{\psi(\bm 0)}$ (blue),
compared with Monte Carlo results for a Laughlin quasihole.
(d) Uncertainty ratio $\eta_{\bm r}$ [Eq.~\eqref{UncertaintyMain}]
at $\bm r=\bm 0$ versus Hilbert-space dimension
$\binom{N_\phi}{N_e}$ for quasiholes (blue) and
quasielectrons (red); labels give $N_e$, and the dotted
line marks $\eta=1$.
(e) Low-energy spectrum of the Coulomb interaction at $N_\phi=3N_e-1=23$.
(f,g) Same as (b,c) for the quasielectron, compared
with a Jain quasielectron.
(h) Holonomy winding of the quasihole (blue) and
quasielectron (red) bands versus magnetic momentum $K$.
}
    \label{fig:4}
\end{figure*}
We focus on continuum FQH states in the Jain sequence at fillings $\nu=p/q$~\cite{jain1989,jain2007composite} and on FCIs at the same
fillings whose elementary excitations carry charge $\pm e/q$, where
$p,q\in\mathbb{Z}^{+}$ are coprime. Taking $e>0$ to be the magnitude of the electronic charge, the positive (negative) sign corresponds to quasiholes (quasielectrons). On a finite torus, single-quasihole (single-quasielectron) states belong to particle sectors satisfying $pN_{\phi}-qN_e=+1$ ($-1$),
where $N_e$ is the number of electrons and $N_{\phi}$ denotes the
number of magnetic flux quanta in the continuum system or the
number of unit cells in the FCI.

\begin{figure*}
    \centering
    \includegraphics[width=\linewidth]{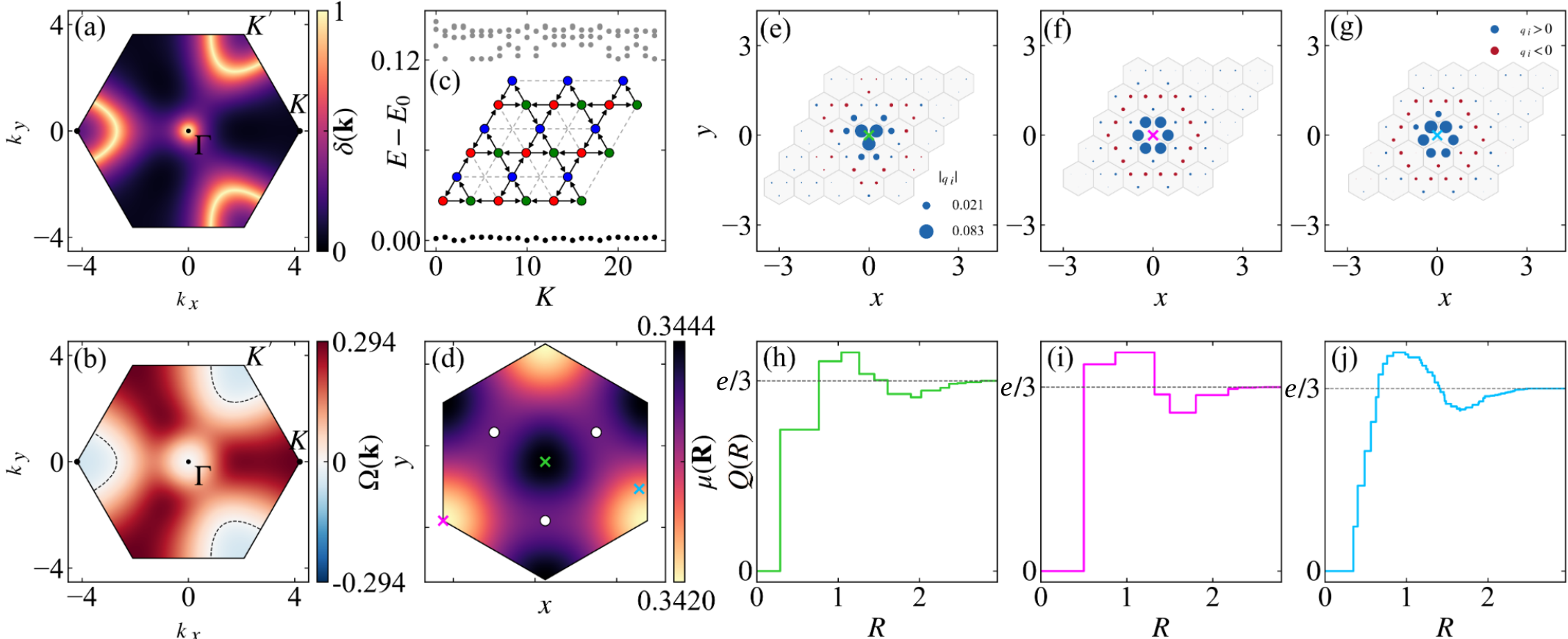}
    \caption{\textbf{Quasihole coherent states in the kagome FCI.} (a) Deviation $\delta(\bm{k})=(\mathrm{tr}\,g(\bm{k})-|\Omega(\bm{k})|)/(\mathrm{tr}\,g(\bm{k})+|\Omega(\bm{k})|)$ of the $C=1$ kagome band from the ideal trace condition. (b) Berry curvature $\Omega(\bm{k})$. (c) Spectrum of the interacting kagome Hamiltonian at $N_{\phi}=3N_{e}+1=25$, with the quasihole band in black. Inset: kagome lattice; arrows indicate the $+\mathrm{i}$ direction of the NN hoppings. (d) LIF \eqref{LIF} on the Wigner-Seitz cell; white circles mark sites and colored crosses extraction points.(e--g) Charge density, relative to filling $\nu=1/3$, of Spatial Localizer quasiholes centered at different points of the unit cell.  (h--j) Integrated charge $Q(R)$ around the cores of the quasiholes in (e--g).}
    \label{fig:5}
\end{figure*}

For the systems considered here, the sectors
$pN_{\phi}-qN_e=\pm1$ exhibit an isolated low-energy
\emph{anyon band}, separated from higher-energy states by a gap,
with one anyon Bloch state $\ket{\Psi(\bm{K})}$ per momentum
sector $\bm{K}$~\footnote{Throughout, $\bm{K}$ denotes magnetic
momentum in the continuum FQH system and crystal momentum in
the FCI.}.
Representative spectra are shown in Figs.~\ref{fig:4}(a),(e)
for the continuum FQH system and Fig.~\ref{fig:5}(c) for the FCI. Hence, the orthogonal projector onto the anyon band is given by
\begin{equation}
    P = VV^{\dagger}, \quad V = \sum_{\bm{K}}\ket{\Psi(\bm{K})}\bra{\bm{K}}, \label{Projectors}
\end{equation}
where the vectors $\ket{\bm{K}}$ form an orthonormal basis of the space $\mathbb{C}^{N_{\phi}}$ with the dimensionality of the anyon band.

The building blocks for obtaining anyon coherent states are the periodic density operators projected onto the anyon band and centered at point $\bm{r}$,
\begin{align}
    \tilde{\rho}_{j}(\bm{r}) &= \frac{e^{-\mathrm{i}\bm{q}_{j}\cdot\bm{r}}}{N_e}V^{\dagger}\left(\sum_{\alpha,\beta}[U_{j}]_{\alpha\beta}c^{\dagger}_{\alpha}c_{\beta}\right)V \label{projecteddensityMain}
\end{align}
Here $\alpha,\beta$ label single-electron states, including
spatial and internal degrees of freedom. The matrix $U_j$
represents the single-particle unitary periodic position
operator associated with the spatial coordinate $X_{j}$ and the wavevector $\bm q_j$. The expectation value of $\tilde{\rho}_{j}$ therefore encodes the center-of-mass (or charge) coordinate of a many-body state along $\bm{q}_{j}$. The factor $e^{-\mathrm{i}\bm q_j\cdot\bm r}$ shifts the origin of the coordinate system to $\bm r$. Explicit expressions for $\tilde{\rho}_{j}(\bm{r})$ in the FQH system and the kagome FCI are given in the Supplemental Material (SM)~\cite{SM}.

In general, projected density operators associated with different directions need not commute and therefore need not admit a common eigenbasis. Moreover, as in single-particle band theory, a topological obstruction to a smooth, periodic Bloch gauge prevents the construction of a complete orthonormal Wannier basis that is exponentially localized in all directions~\cite{thouless1984wannier, brouder2007exponential}. This obstruction does not preclude localized states forming a nonorthogonal, overcomplete set.

To determine the set of maximally localized states within the anyon band
without assuming anything about its topology, we extend
the Spatial Localizer of Ref.~\cite{SpatLoc} to many-body systems. We split each periodic density operator \eqref{projecteddensityMain} into Hermitian and anti-Hermitian components, $C_{j}(\bm{r}) = [\tilde{\rho}_{j}(\bm{r})+\tilde{\rho}^{\dagger}_{j}(\bm{r})]/2 - \mathrm{I}_{N_{\phi}}$ and $S_{j}(\bm{r}) = [\tilde{\rho}_{j}(\bm{r})-\tilde{\rho}^{\dagger}_{j}(\bm{r})]/(2\mathrm{i})$. Then, we define the $d$-dimensional PBC Spatial Localizer as
\begin{equation}
    L(\bm{r}) \equiv \sum_{j=1}^{d'} [C_{j}(\bm{r}) \otimes \Gamma_{2j-1}+S_{j}(\bm{r}) \otimes \Gamma_{2j}], \label{Localizer}
\end{equation}
where $\Gamma_{j}$ are the generators of the Clifford algebra $\mathrm{Cl}_{2d',0}(\mathbb{R})$, satisfying $\{\Gamma_{i},\Gamma_{j}\}=2\delta_{ij}$~\footnote{Concretely, in our calculations we adopt the Weyl-Brauer representation of the Clifford elements~\cite{brauer1935spinors}}. The embedding dimensionality $d'$ is equal (greater) than the number $d$ of spatial dimensions for isotropic localization in square (non-square) lattices.

The smallest eigenvalue magnitude of $L(\bm r)$,
\begin{equation}
    \mu(\bm r)\equiv\min|\mathrm{spec}\,L(\bm r)|, \label{LIF}
\end{equation}
defines the \emph{localizer indicator function} (LIF) over the torus. In atomic
insulators, the LIF has sharp minima that coincide with
Wannier centers, whereas in Chern insulators it is nearly
flat, vanishes in the thermodynamic limit, and is associated with a continuous family of localized
states~\cite{SpatLoc}. This behavior mirrors continuum
integer quantum Hall systems, where localized coherent
states can be centered at arbitrary positions~$\bm r$.

The eigenvectors $\ket{\mu_\kappa(\bm{r})}$ of $L(\bm r)$ with eigenvalue magnitude $\mu(\bm r)$, where the sector index $\kappa$ labels a possible degeneracy, live in the tensor product of the physical and Clifford embedding spaces and can be Schmidt-decomposed as
\begin{equation}
    \ket{\mu_\kappa(\bm r)}
    = \sum_i s_{\kappa i}(\bm r)
    \ket{p_{\kappa i}(\bm r)}\otimes\ket{\xi_{\kappa i}(\bm r)},
    \label{LocalizerState}
\end{equation}
where $s_{\kappa i}(\bm r)\geq s_{\kappa,i+1}(\bm r)\geq0$,
$\sum_i s_{\kappa i}^2(\bm r)=1$, and $\ket{p_{\kappa i}(\bm r)}$
and $\ket{\xi_{\kappa i}(\bm r)}$ are the physical and embedding
Schmidt vectors, respectively. We construct the anyon \emph{localizer states}, $\ket{\psi_\kappa(\bm{r})}$, by orthonormalizing the leading physical Schmidt vectors $\ket{p_{\kappa 1}(\bm r)}$ (SM Sec.~\ref{Appendix:Localizer}).

In the $1/3$ FQH and FCI systems studied here, we obtain three
nearly degenerate lowest eigenvalues,
$\lambda_0(\bm r),\lambda_1(\bm r),\lambda_2(\bm r)
\approx\mu(\bm r)$, yielding three orthonormal localizer
states centered around each position $\bm{r}$ in the torus, which is consistent with the $m$-fold degeneracy of $1/m$ Laughlin quasiholes generated by the $m$ topological sectors of the parent FQH ground state on the torus~\cite{PhysRevB.93.245156} (SM Sec.~\ref{Appendix:degeneracyLaughlin}). Their charge distributions are nearly identical
(SM Sec.~\ref{Appendix:Localizer}), reflecting the expected local
indistinguishability of topological sectors in the
thermodynamic limit~\cite{nielsen2013local,PhysRevB.91.045126}. We therefore
show the charge distribution of one representative state $\ket{\psi_{\kappa=0}(\bm{r})}$ at each selected position $\bm{r}$ in the main figures and drop the $\kappa$ subscript from here on, unless explicitly stated otherwise.

In SM Sec.~\ref{Appendix:Localizer}, we show that, in two dimensions and in the thermodynamic limit, $\ket{\psi(\bm r)}$ saturates the bound
\begin{equation}
    \eta_{\bm{r}} \equiv \frac{\Delta\tilde X_{\bm{r}}^2+\Delta\tilde Y_{\bm{r}}^2}{|\langle[\tilde X_{\bm{r}},\tilde Y_{\bm{r}}]\rangle|}
    \geq 1,
    \label{UncertaintyMain}
\end{equation}
which follows from the Heisenberg-Robertson (HR) relation $\Delta\tilde X_{\bm r}\Delta\tilde Y_{\bm r}\ge|\langle[\tilde X_{\bm r},\tilde Y_{\bm r}]\rangle|/2$~\cite{robertson1929uncertainty}. Equality in Eq.~\eqref{UncertaintyMain} requires both saturation of the HR relation and equal variances $\Delta\tilde X_{\bm r}^2=\Delta\tilde Y_{\bm r}^2$, and thus singles out isotropic (unsqueezed) minimum-uncertainty states.
Here, $\tilde X_{\bm{r}}$ ($\tilde Y_{\bm{r}}$) is the centered projected position operator under open boundary conditions, obtained from Eq.~\eqref{projecteddensityMain} by replacing $e^{-\mathrm{i}\bm q_j\cdot\bm r}U_j$ with $X-r_x$ ($Y-r_y$) [Eq.~(\ref{OBCdensities})].

We determine $\ket{\psi(\bm{r})}$ for the $1/3$ FQH phase on a torus with repulsive interactions between electrons in the lowest Landau level (LLL) described by Haldane's $V_1$ pseudopotential~\cite{Haldane1983,PhysRevB.32.6924,Trugman1985,Papic2015} 
(Coulomb potential~\cite{yoshioka1984ground,PhysRevLett.50.1219}) in the single-quasihole (quasielectron) sectors (SM Sec.~\ref{Numerics}). For $N_e=8$, these sectors contain $N_\phi=25$ ($N_{\phi}=23$) flux quanta. Their spectra, labeled by the one-dimensional magnetic momentum index $K_x$ in the Landau gauge, exhibit isolated anyon bands [Figs.~\ref{fig:4}(a),(e)], from which the Spatial localizer \eqref{Localizer} produces states $\ket{\psi(\bm{r})}$ with localized electronic charge deficits (quasihole) and excesses (quasielectron) [Figs.~\ref{fig:4}(b),(f)] relative to $1/3$ filling. The flat LIF [insets of Figs.~\ref{fig:4}(a),(e)] shows no preferred extraction position, which implies a continuous family of localized anyon states centered anywhere on the torus.

Integrating the charge densities of Figs.~\ref{fig:4}(b),(f) over disks of increasing radii yields the profiles of Figs.~\ref{fig:4}(c),(g), which reach their peaks before approaching the quasihole and quasielectron fractional charges $+e/3$ and $-e/3$, respectively. Monte Carlo sampling of the Laughlin quasihole~\cite{laughlin1983} and Jain quasielectron~\cite{jain1989,PhysRevB.41.7653,PhysRevB.45.1223,doi:10.1142/S0217979297001301,jain2007composite} on the infinite plane with $N_e=45$ gives a Laughlin profile that agrees closely with our quasihole, whereas the Jain profile overshoots at a larger radius and converges to $-e/3$ more slowly, showing that our quasielectron is more compact than the Jain trial state.

In the thermodynamic limit, the projected-density commutator is proportional to the identity, $[\tilde X_{\bm r},\tilde Y_{\bm r}]=\pm\mathrm{i}\,3\ell_B^2$, where $\ell_B$ is the magnetic length. The bound~\eqref{UncertaintyMain} then takes the same value, $\Delta\tilde X_{\bm r}^2+\Delta\tilde Y_{\bm r}^2\ge3\ell_B^2$, for every state in the anyon band, so any state that saturates it is maximally localized within that subspace. Moreover, the lowering operator $Z_{\bm r}=(\tilde X_{\bm r}+\mathrm{i}\epsilon\tilde Y_{\bm r})/(\sqrt6\,\ell_B)$, with $\epsilon=\operatorname{sgn}(-\mathrm{i}\langle[\tilde X_{\bm r},\tilde Y_{\bm r}]\rangle)$, obeys the canonical relation $[Z_{\bm r},Z_{\bm r}^\dagger]=1$, and $\|Z_{\bm r}\ket{\psi(\bm r)}\|^2=(\eta_{\bm r}-1)/2$. Saturating the bound is therefore equivalent to being a coherent state, i.e., a zero mode of $Z_{\bm r}$.

Our finite-size results approach this limit. As the system grows, $\eta_{\bm r}$ of
the Spatial Localizer anyons decreases toward 1 [Fig.~\ref{fig:4}(d)], and for
$N_e=8$ we find $\langle[Z_{\bm r},Z_{\bm r}^\dagger]\rangle=1$ up to corrections of
order $10^{-4}$ and $\|Z_{\bm r}\ket{\psi(\bm r)}\|^2\sim10^{-3}$, decreasing
monotonically with system size. The Spatial Localizer
anyons thus converge to maximally localized coherent states of $\tilde X_{\bm r}$ and
$\tilde Y_{\bm r}$. Exact coherence is reached only in the thermodynamic limit, where
the LIF vanishes; the LIF of the systems studied here decreases monotonically with
system size (SM Sec.~\ref{Numerics}).

We now extend this coherent-state description to an FCI of spin-polarized electrons on a kagome lattice with repulsive nearest-neighbor (NN) interactions, complex NN hoppings, and real next-nearest-neighbor hoppings~\cite{PhysRevLett.106.236803,Lu_2026}. The model is illustrated in the inset of Fig.~\ref{fig:5}(c), and its Hamiltonian is given in SM Sec.~\ref{Numerics}. For the parameters considered here, the lowest single-particle band is nearly flat and has Chern number $C=1$. 

The geometry of this Chern band departs substantially from the ideal condition $\mathrm{tr}\,g(\bm{k})=|\Omega(\bm{k})|$ relating the quantum metric $g(\bm{k})$ and Berry curvature $\Omega(\bm{k})$ [Fig.~\ref{fig:5}(a)], most strongly near $\Gamma$, where the Berry curvature nearly vanishes, and around $K'$, where it changes sign [Fig.~\ref{fig:5}(b)]. The band is also non-vortexable: no discrete vortex function of the form of Ref.~\cite{PhysRevB.110.245112} preserves the lowest-band subspace (SM Sec.~\ref{sec:vortexability}). Hence, the FCI ground state and anyons cannot be analytically built by band-preserving vortex attachment~\cite{ledwith2023vortexability}.

The Spatial Localizer, however, requires neither of these two conditions: it requires only that the Hamiltonian has a spectrally isolated anyon subspace represented in any basis, such as the energy eigenstates obtained by exact diagonalization [Fig~\ref{fig:5}(c)]. For this kagome lattice, we construct \eqref{Localizer} from three projected-density operators associated with the three lattice directions of the triangular Bravais lattice. We focus on the single-quasihole (quasielectron) sector, which accommodates $N_e=8$ ($N_e=7$) electrons on a torus of $5\times5$ ($5 \times 4$) unit cells. To investigate finite-size dependence, we repeat the analysis for a smaller $4 \times 4$ lattice with $N_{e}=5$ in the single quasihole sector.

In the main text and figures we present results for quasiholes in the $5 \times 5$ lattice; the corresponding results for quasielectrons are presented in SM Sec.~\ref{Numerics}. The resulting LIF is nearly flat [Fig.~\ref{fig:5}(d)], allowing us to extract a continuous family of three linearly independent localized anyons $\ket{\psi(\bm{r})}$ centered at any point $\bm{r}$ on the torus. Unlike in the LLL, the lattice lacks continuous translational and rotational symmetries. Consequently, the shape of the charge distribution depends on the position of an anyon within the unit cell [Figs.~\ref{fig:5}(e)--(g)]. Despite this variation, the corresponding integrated charges behave similarly [Figs.~\ref{fig:5}(h)--(j)], resulting in a nearly uniform variance of the projected charge density $\Delta\tilde{X}_{\bm{r}}^2+\Delta\tilde{Y}_{\bm{r}}^2=(0.413 \pm 0.005)  \ a^2$, where $a$ is the magnitude of the primitive lattice vectors. Such variance yields an effective magnetic length $\ell = \sqrt{\Delta\tilde{X}_{\bm{r}}^2+\Delta\tilde{Y}_{\bm{r}}^2} \approx 0.643 \ a$. 

\begin{figure*}
    \centering
    \includegraphics[width=\linewidth]{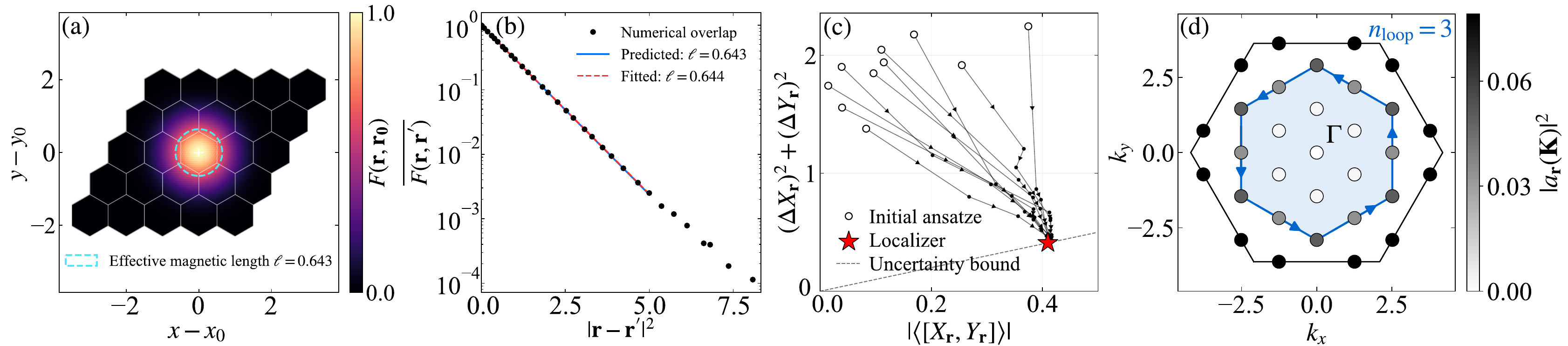}
    \caption{\textbf{Overcompleteness, maximal localization, and topology of FCI anyon coherent states.} (a) Subspace fidelity $F(\bm{r},\bm{r}_0)$ between coherent states at fixed $\bm{r}_0$ and varying $\bm{r}$. (b) $F$ averaged over pairs at fixed separation, versus separation squared. (c) Density variance under iterative minimization of the wavefunction spread functional from random initial states, converging to that of the Spatial Localizer quasihole. (d) $|a_{\bm r}(\bm K)|^2$ of the Bloch coefficients~\eqref{CoherentBloch} for the coherent state of Fig.~\ref{fig:5}(f); the light blue region encloses vorticity $n_{\mathrm{loop}}=3$. System size: $5 \times 5$ lattice.}
    \label{fig:Beautiful}
\end{figure*}

In the absence of continuous translational symmetry and uniform Berry curvature, $[\tilde X_{\bm r},\tilde Y_{\bm r}]$ is not proportional to the identity as in the continuum FQH case; the commutator expectation value depends on the state. The natural generalization of $Z_{\bm r}$ is then normalized by this expectation value, $Z_{\bm r}=(\tilde X_{\bm r}+\mathrm{i}\epsilon\tilde Y_{\bm r})/\sqrt{2|\langle[\tilde X_{\bm r},\tilde Y_{\bm r}]\rangle|}$, evaluated in $\ket{\psi(\bm r)}$. This choice preserves the exact relation $\|Z_{\bm r}\ket{\psi(\bm r)}\|^2=(\eta_{\bm r}-1)/2$, so zero modes of $Z_{\bm r}$ again coincide with saturation of Eq.~\eqref{UncertaintyMain}. We obtain $\|Z_{\bm r}\ket{\psi(\bm r)}\|^2\sim10^{-3}$ for the $4\times4$ and $5 \times 4$ lattices and $\sim10^{-4}$ for the $5\times5$ lattice.

Beyond saturating the uncertainty bound, coherent states form an overcomplete family that resolves the identity. In SM Sec.~\ref{sec:klauder}, we verify that the localizer states $\ket{\psi_{\kappa}(\bm{r})}$, $\kappa\in\{0,1,2\}$, centered at all positions on the torus span the $N_{\phi}$-dimensional single-anyon band and resolve the projector \eqref{Projectors},
\begin{equation}
    \frac{1}{3A_{\mathrm c}}\int d^2r\,\sum_{\kappa=0}^{2}\ket{\psi_\kappa(\bm{r})}\bra{\psi_\kappa(\bm{r})}
    = P + O(10^{-2}), \label{GeneralizedCoherent}
\end{equation}
where $A_{\mathrm c}$ is the unit-cell area. The maximum deviation of diagonal elements decreases from $4\%$ on the $4\times4$ lattice to $2.7\%$ on the $5 \times 4$ and $0.7\%$ on the $5\times5$ lattice. Together with their normalization and continuity in $\bm r$, Eq.~\eqref{GeneralizedCoherent} makes $\ket{\psi_{\kappa}(\bm{r})}$ coherent states in the sense of Klauder~\cite{klauder1963continuous,klauder1985coherent}. Figures~\ref{fig:Beautiful}(a,b) show that their overlaps, measured by the subspace fidelity $F(\bm r,\bm r')=\frac13\sum_{\kappa,\kappa'}|\langle\psi_\kappa(\bm r)|\psi_{\kappa'}(\bm r')\rangle|^2$, are isotropic and decay as $F\propto\exp[-|\bm{r}-\bm{r}'|^2/(2\ell_{\mathrm{fit}}^2)]$, with $\ell_{\mathrm{fit}}=0.644\,a$, matching the spread $\ell$ of each individual state, as for
Landau-level coherent states.

Because $\langle[\tilde X_{\bm r},\tilde Y_{\bm r}]\rangle$ is state-dependent, saturating~\eqref{UncertaintyMain} does not by itself imply minimal spread (i.e., maximal localization). We therefore test maximal localization independently by minimizing a periodic density-spread functional whose initial ansatze are states randomly picked within the anyon band (SM Sec.~\ref{sec:spread}); all iterative processes converge to the variance of the Spatial Localizer anyons [Fig.~\ref{fig:Beautiful}(c)].

The Spatial Localizer thus reveals that the maximally
localized anyon states form a continuous, overcomplete
family of coherent states, rather than an orthonormal
Wannier basis with fixed centers in each unit cell.
At each position, there are three linearly independent
states that are approximate zero modes of a local
annihilation operator. This multiplicity suggests a
topological interpretation: in single-particle Chern bands, the Atiyah-Singer index theorem~\cite{nakahara2018geometry} relates the number of projected-position zero modes to the Chern number~\cite{PhysRevB.110.245112}. Our findings are therefore consistent with interpreting the low-energy many-body subspace identified in Fig.~\ref{fig:5}(c) as an emergent quasiparticle band with $|C_{\mathrm{anyon}}|=3$.

\emph{Anyon band topology}\label{sec:ApplicatiotoFCI}---For many-body states of indistinguishable particles, parallel transport between neighboring points of the BZ is given by the momentum-space matrix elements $t_j(\bm K)=
\bra{\bm K}\tilde{\rho}_{\bm q_j}
\ket{\bm K+\bm q_j}$ of the projected density operator~\eqref{projecteddensityMain}, which encodes the state's center-of-mass coordinate~\cite{davenport2026compositequantumgeometrysemiclassical,hwang2026exciton,paiva2024shiftpolarizationexcitonsquantum,hwang2026stablewavefunctionzerosindicate,gerhard2026spatiallocalizerconstituentresolvedexciton}. Therefore, we define the Berry flux through an elementary plaquette of the BZ as
\begin{equation}
    \theta_{\bm K}=\mathrm{Arg}\left[t_1(\bm K)t_2(\bm K+\bm q_1) t_1^*(\bm K+\bm q_2)t_2^*(\bm K)\right] \label{PlaquettePhases}
\end{equation}
For continuum FQH systems with a uniform magnetic field, such elementary Berry flux is equal to $\pm 2\pi q/N_{\phi}$ for anyons of charge $\pm e/q$ (SM Sec.~\ref{AppendixBerryConnection}). Figure ~\ref{fig:4}(h) shows the opposite accumulated flux over magnetic unit cells $K_x$ for FQH quasiholes and quasielectrons. The net Berry flux $\Phi = -2\pi C_{\mathrm{anyon}}$ over the magnetic BZ is quantized and defines the anyon Chern number $C_{\mathrm{anyon}}=\pm q$. For the kagome FCI, we evaluate the plaquette phases \eqref{PlaquettePhases} numerically. At the sectors $N_{\phi}-3N_e=\pm 1$, their sums yield $C_{\mathrm{anyon}}=\pm 3$, in agreement with the analytical FQH result. In all cases studied here, $C_{\mathrm{anyon}}$ reflects the $q$-fold degeneracies of the parent FQH or FCI ground state and, consequently, of the anyon coherent states $\ket{\psi(\bm{r})}$.

The anyon Chern number also sets the size of the coherent states. The commutator $[\tilde X_{\bm r},\tilde Y_{\bm r}]$ approximates the average Berry curvature of the anyon band, $2\pi|C_{\mathrm{anyon}}|/A_{\mathrm{BZ}}$, where $A_{\mathrm{BZ}}$ is the BZ area. Saturating the uncertainty bound~\eqref{UncertaintyMain} therefore predicts an effective magnetic length $\ell=\sqrt{2\pi|C_{\mathrm{anyon}}|/A_{\mathrm{BZ}}}$. For the kagome lattice with $|C_{\mathrm{anyon}}|=3$, this gives $\ell=0.643\,a$, in agreement with the anyon spread and overlap length.

Finally, the topology of the anyon band manifests in the coefficients $a_{\bm{r}}(\bm{K})$ of the coherent states
\begin{equation}
    \ket{\psi(\bm r)}
    = \sum_{\bm K}a_{\bm r}(\bm K)\ket{\Psi(\bm K)}. \label{CoherentBloch}
\end{equation}
expanded in the anyon Bloch basis. Chern bands force the momentum coefficients of smooth, localized wavepackets to vanish at certain points, and the total vorticity of those coefficients around those points equals the Chern number~\cite{li2024constraints}. Figure~\ref{fig:Beautiful}(d) shows $|a_{\bm r}(\bm K)|^2$ for the coherent quasihole in Fig.~\ref{fig:5}(f). Although our $5\times5$ momentum grid is too coarse to resolve the number and positions of all individual zeros, we can distinguish a highly suppressed region highlighted in blue in Fig.~\ref{fig:Beautiful}(d) around the $\Gamma$ point, with $|a_{\bm{r}}(\Gamma)|$ vanishing to machine precision.

A gauge-invariant circulation of the phases of $a_{\bm r}(\bm K)$, corrected by the Berry flux~\eqref{PlaquettePhases}, counts the vorticity $n_R$ of the coefficients inside a region $R$ of the BZ~\cite{fukui2005chern} (SM Sec.~\ref{sec:vorticity}). Around the light blue region in Fig.~\ref{fig:Beautiful}(d), we obtain $n_{\mathrm{loop}}=3=C_{\mathrm{anyon}}$.

\emph{Discussion}\label{sec:discussion}---We have introduced a gauge-invariant, ansatz-free construction of maximally localized fractional excitations in FQH and FCI systems and have found, in all cases, that they are coherent states. Their coherent-state nature reveals the nontrivial topology of the quasiparticle band, whose Chern number has a real-space signature encoded in the vorticity of the momentum-space coefficients of anyon coherent states. In the systems studied here, the magnitude of $C_{\mathrm{anyon}}$ equals the parent ground-state degeneracy rather than the Chern number of the underlying single-particle band.

Because the construction does not require spatial symmetries, it provides a route to investigating anyon trapping at lattice defects and its possible relation to weak topological indices~\cite{PhysRevLett.132.116602,PhysRevB.89.224503,ran2009one,PhysRevB.82.115120,PhysRevLett.105.030403}. Extending the method to spatially separated non-Abelian anyons could also facilitate braiding calculations: within an isolated low-energy manifold, the Spatial Localizer \eqref{Localizer} produces localized states without repeatedly diagonalizing the many-body Hamiltonian with moving pinning potentials. Finally, localized multi-anyon configurations at controlled separations could provide a basis for extracting effective anyon-anyon interactions and investigating the emergence of bound states, \emph{i.e.} molecular anyons~\cite{scl5-8pv6}, in FCIs.

\emph{Note added:} While this manuscript was in preparation, we became aware of three works~\cite{pu2026compositefermionsidealchern,xiao2026anyonchernbandsopposite,sarkar2026latticeinducedsplittingconnectivitytopology} that define anyon parallel transport and the Chern number in a similar manner to ours. Our results agree where they overlap.

\emph{Acknowledgments}---This work was supported by startup funds from Emory University.

\clearpage
\onecolumngrid
\setcounter{equation}{0}
\setcounter{figure}{0}
\setcounter{table}{0}
\setcounter{section}{0}
\setcounter{subsection}{0}
\setcounter{footnote}{0}
\renewcommand{\thesection}{S\Roman{section}}
\renewcommand{\theequation}{S\arabic{equation}}
\renewcommand{\thefigure}{S\arabic{figure}}
\renewcommand{\thetable}{S\arabic{table}}
\renewcommand{\theHsection}{S\arabic{section}}
\renewcommand{\theHequation}{S\arabic{equation}}
\renewcommand{\theHfigure}{S\arabic{figure}}
\renewcommand{\theHtable}{S\arabic{table}}
\makeatletter
\def\NAT@bibsetnum#1{\setlength{\topsep}{\z@}\NATx@bibsetnum{\ref{LastBibItemSM}}}
\def\endthebibliography{\auto@bib@innerbib\edef\@currentlabel{\arabic{NAT@ctr}}\label{LastBibItemSM}\endNAT@thebibliography\aftergroup\auto@bib@empty}
\makeatother

\begin{center}
{\large\bfseries Supplemental Material for ``Anyon Coherent States in Fractional Chern Insulators''\par}
\vspace{1.5ex}
Pedro Fittipaldi de Castro, Haylen Gerhard, and Wladimir A. Benalcazar\par
\vspace{0.5ex}
{\small\itshape Department of Physics, Emory University, Atlanta, GA 30322, USA\par}
\end{center}
\vspace{2ex}

Equations, figures, and references without an ``S'' prefix refer to the main text.

\section{Periodic density operators}\label{app:density}

\subsection{Continuum FQH system}

For the continuum FQH system, we use a Landau-gauge orbital
basis labeled by $j=0,\ldots,N_\phi-1$. The two periodic
guiding-center density operators restricted to the anyon band are
\begin{align}
    \tilde{\rho}_x(\bm r)
    &= \pm\frac{e^{-\mathrm{i}\bm q_1\cdot\bm r}}{N_e}
    V^\dagger\left(
    \sum_{j=0}^{N_\phi-1}e^{\mathrm{i}2\pi j/N_\phi}
    c_j^\dagger c_j\right)V, \\
    \tilde{\rho}_y(\bm r)
    &= \pm\frac{e^{-\mathrm{i}\bm q_2\cdot\bm r}}{N_e}
    V^\dagger\left(
    \sum_{j=0}^{N_\phi-1}c_{j+1}^\dagger c_j\right)V, \label{FQHEdensityoperators}
\end{align}
where orbital indices are understood modulo $N_\phi$ and the $+$ ($-$) signs generate electron (hole) density operators. In the scaled coordinates used here,
$\bm q_x\cdot\bm r=2\pi x/N_\phi$ and
$\bm q_y\cdot\bm r=2\pi y/N_\phi$.
These are guiding-center density harmonics, with the
Gaussian lowest-Landau-level form factor omitted.

\subsection{FCI}

We assume our system is defined on a $d$-dimensional lattice with $N_{j}$ cells along the primitive vectors $\bm{a}_{j}$, where $j \in \{1,\dots,d\}$, and a total of $N_{\phi}=N_{1} \times \dots \times N_{d}$ unit cells. Let $\bm{b}_{j}$ be the reciprocal lattice vectors, such that $\bm{a}_{j}\cdot\bm{b}_{j'}=2\pi\delta_{jj'}$, and let $\bm{q}_{j}=\bm{b}_{j}/N_{j}$ be the vectors connecting neighboring points in the Brillouin zone. Under open boundary conditions, the natural object for localizing states along the $j$ direction is the projected density operator
\begin{align}
    \tilde{X}_{j} &= V^{\dagger}\sum_{\bm{R},s} \left[\frac{\bm{b}_{j}}{2\pi}\cdot(\bm{R}+\bm{r}_s)\right]n_{\bm{R},s}V,\\
    V &= \sum_{n=0}^{N_{\phi}-1}\ket{\Psi_n}\bra{n},
\end{align}
where $n_{\bm{R},s}$ is the particle number operator at unit cell $\bm{R}$ and orbital position $\bm{r}_{s}$ and $\ket{n} \in \mathbb{C}^{N_{\phi}}$ with $\langle m| n\rangle=\delta_{mn}$. Notice that $\tilde{X}_{j}$ acts not on the many-body Hilbert space, but on the auxiliary space $\mathbb{C}^{N_{\phi}}$ with the dimensionality of the low-energy manifold. The expectation value $\langle \tilde{X}_{j} \rangle$ on a given state represents the center-of-charge coordinate of the many-electron configuration along $\bm{a}_{j}$, while $\Delta \tilde{X}_{j}^2 = \langle \tilde{X}_{j}^2 \rangle - \langle \tilde{X}_{j} \rangle^2$ is the spatial variance of the charge distribution. Under periodic boundary conditions (PBC), we embed each coordinate onto the unit circle through a periodic density operator
\begin{align}
    \tilde{\rho}_{j} &= \frac{e^{-\mathrm{i}\bm{q}_{j}\cdot\bm{r}}}{N_e}V^{\dagger}\left(\sum_{\bm{R},s}e^{\mathrm{i}\bm{q}_{j}\cdot(\bm{R}+\bm{r}_s)}n_{\bm{R},s}\right)V, \label{projecteddensity}
\end{align}
where $n_{\bm{R},s}=c^{\dagger}_{\bm{R},s}c_{\bm{R},s}$ ($n_{\bm{R},s}=c^{\dagger}_{\bm{R},s}c_{\bm{R},s}$) for electrons (holes). For our kagome lattice, the primitive vectors are $\bm{a}_1=(1,0)$ and $\bm{a}_2=\frac{1}{2}(1,\sqrt{3})$. We define three operators of the form \eqref{projecteddensity} for $j \in \{1,2,3\}$ using wavevectors $\bm{q}_1 = \frac{\bm{b}_1}{N_1}$, $\bm{q}_2 = \frac{\bm{b_2}}{N_2}$ with $\bm{b}_1 = 2\pi (1,-\frac{1}{\sqrt{3}})$, $\bm{b}_2 = (0,\frac{4\pi}{\sqrt{3}})$, and $\bm{q}_3 = -\frac{\bm{b}_1+\bm{b}_2}{N_{1}}$ for a lattice with aspect ratio $N_{1}/N_{2}=1$.

At first, one might consider working with the unitary Resta position operator~\cite{SM:Resta1998} instead of $\tilde{\rho}_{j}$. After all, it successfully describes the electronic polarization and, more generally, the moments and cumulants of the charge distribution of non-degenerate many-body ground states~\cite{SM:SouzaWilkensMartin2000}. However, the Resta operator does not uniquely determine the center of charge of a many-electron state, as we now show.

\section{Unitary Resta versus density operators}

\begin{figure}
    \centering
    \includegraphics[width=0.6\linewidth]{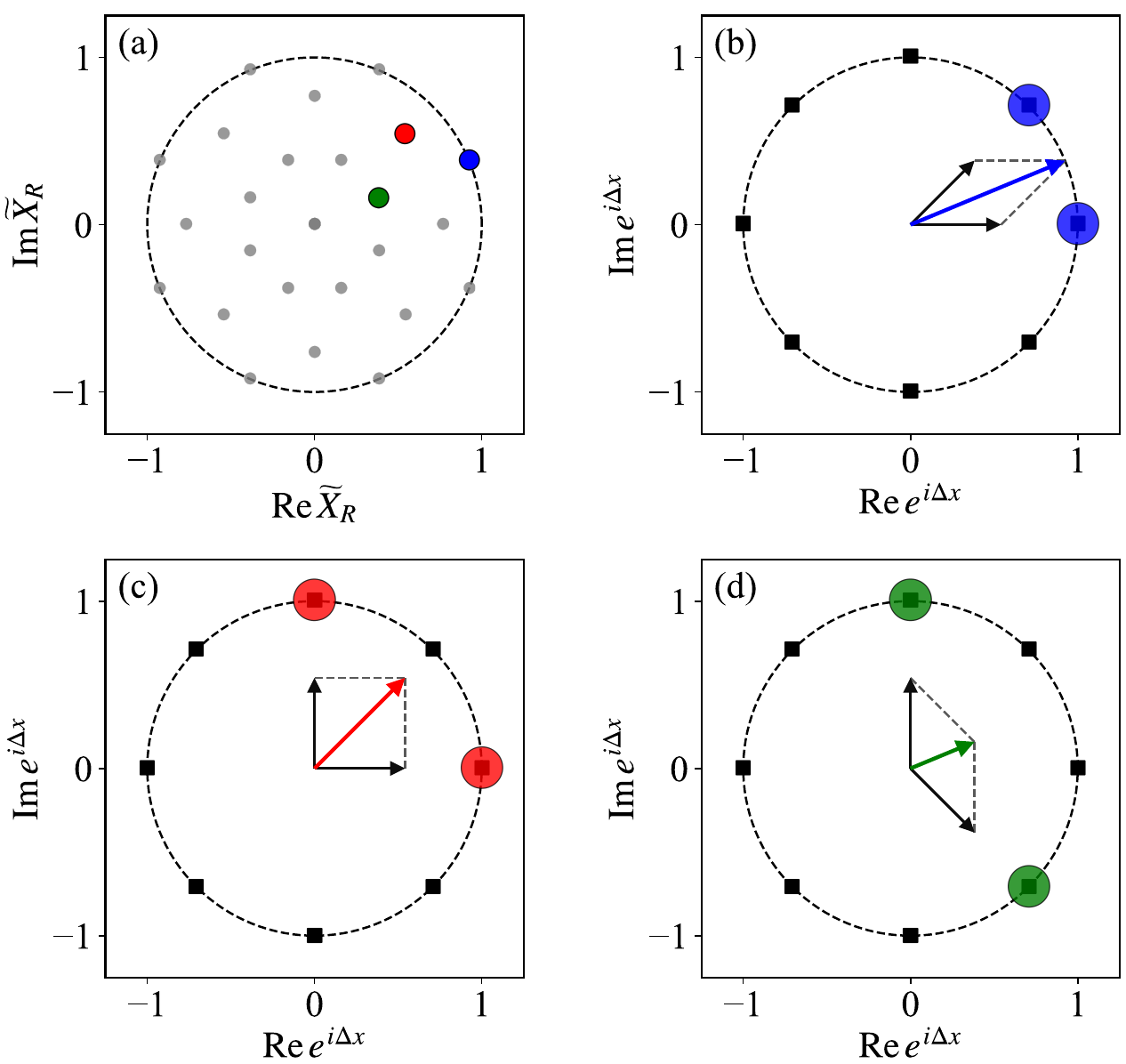}
    \caption{\textbf{Periodic density operator.} (a) Spectrum of the periodic density operator \eqref{AdditiveResta} acting on the two-particle sector of a one-dimensional fermionic system with periodic boundary conditions. The phase of each eigenvalue determines the center of mass $\langle \hat{X}\rangle$, while its modulus encodes the spatial spread of the corresponding eigenstate. (b)--(d) Charge distributions of the eigenstates associated with the blue, red, and green eigenvalues highlighted in panel (a), respectively. Black squares denote lattice sites. Black arrows represent the single-particle phasors, while the colored arrow denotes their vector sum. The direction of this sum specifies the center of mass, whereas its magnitude encodes the spatial spread of the corresponding configuration. Here, $L=8$ and $N_e=2$.}
    \label{fig:1}
\end{figure}

For simplicity, we consider one-dimensional systems with $N_1=L$ sites $x$, such that $q_{1}=\frac{2\pi}{L}$. Before any projection onto the subspace of interest, Resta's unitary periodic position operator reads
\begin{equation}
    \hat{U}_{\mathrm{Resta}}
    \equiv
    e^{\mathrm{i}\frac{2\pi}{L}\hat{X}}
    =
    \exp\left(
    \mathrm{i}\frac{2\pi}{L}
    \sum_{x=0}^{L-1}x\hat{n}_x
    \right),
    \label{Resta}
\end{equation}
where $\hat{n}_x$ is the particle-number operator on site $x$. Fock states
$\ket{x_1,x_2,\ldots,x_{L}}$ form an eigenbasis of~\eqref{Resta} with eigenvalues $\lambda_{\mathrm{Resta}}=e^{\mathrm{i}\frac{2\pi}{L}P_{\mathrm{el}}}$, where $P_\mathrm{el} = \sum_{j=1}^{N_e}x_j n_{x_j} \ (\mathrm{mod} \ L)$ is the state's electronic polarization. In general, there are many configurations $\{x_{j}\}$ leading to the same polarization. As a result, the spectrum of \eqref{Resta} is degenerate, and states within an eigenspace share a common $P_{\mathrm{el}}$ but have very different localization properties. For example, at $N_{e}=2$, the vectors $\ket{\dots 001100 \dots}$, $\ket{ \dots 010010 \dots }$, $\ket{\dots 100001\dots}$, and so on all belong to the same eigenspace. Therefore, the spectrum of the unitary Resta operator \eqref{Resta} does not resolve the spatial compactness of the charge distribution of states within a many-body manifold and is therefore not sufficient for identifying maximally localized electronic configurations.

More importantly, the operator \eqref{Resta} is pathological when used to calculate the center of mass of a many-body state. If the total polarization is only defined modulo $L$ as $P_\mathrm{el} = \frac{L}{2\pi}\mathrm{Arg}\langle \hat{U}_{\mathrm{Resta}}\rangle +mL$ where $m \in \mathbb{Z}$, a system containing $N_{e}$ particles admits $N_{e}$ inequivalent center-of-mass coordinates,
\begin{equation}
    x_{\mathrm{cm}}^{(m)} = \frac{L}{2\pi N_{e}}\mathrm{Arg}\langle \hat{U}_{\mathrm{Resta}}\rangle + \frac{mL}{N_{e}}, \quad m \in \{0,1,\dots,N_{e}-1\}.
\end{equation}
Thus, in addition to being highly degenerate, the eigenvalues of the many-body Resta operator determine the center of mass only modulo $L/N_e$.

Both limitations can be circumvented by replacing the unitary operator~\eqref{Resta} with the periodic density operator,
\begin{equation}
    \rho \equiv \frac{1}{N_e}\sum_{x=0}^{L-1}e^{\mathrm{i}\frac{2 \pi}{L} x}\hat{n}_{x}, \label{AdditiveResta}
\end{equation}
The above expression is simply the Fourier component of $n_x$ with momentum $k=\frac{2\pi}{L}$. The operators~\eqref{Resta} and~\eqref{AdditiveResta} share the Fock basis $\ket{x_1,x_2,\ldots,x_L}$ as an eigenbasis. Their eigenvalues, however, combine phasors $e^{\mathrm{i}\frac{2 \pi}{L} x_j}$ in fundamentally different ways:
\begin{equation}
    \lambda_{\mathrm{Resta}}
    =
    \prod_{j=1}^{N_e}
    e^{\mathrm{i}\frac{2 \pi}{L} x_j}
    =
    e^{\mathrm{i}\frac{2 \pi}{L}\sum_j x_j},
    \qquad
    \lambda_\rho
    =
    \frac{1}{N_e}\sum_{j=1}^{N_e}
    e^{\mathrm{i}\frac{2 \pi}{L} x_j}.
\end{equation}
While $\lambda_{\mathrm{Resta}}$ is necessarily unimodular and encodes nothing beyond the total polarization, $\lambda_\rho$ is generally nonunimodular, as shown in Fig.~\ref{fig:1}(a). Crucially, as illustrated in Figs.~\ref{fig:1}(b)--(d), the phase of $\lambda_\rho$ determines the center of mass of the corresponding Fock configuration, while its magnitude encodes the spread of its charge distribution.

For generic many-body states, the center of mass is given by
\begin{equation}
    \langle X \rangle = \frac{L}{2\pi}\mathrm{Arg} \ \langle\rho\rangle \ \ \  (\mathrm{mod} \ L),
\end{equation}
where $X$ is the OBC position operator. The variance of the charge distribution converges to
\begin{equation}
    \Delta X^2 = \langle X^2\rangle - \langle X \rangle^2 =
    -
    \left(\frac{L}{2\pi}\right)^2
    \ln
    \left|
    \langle\rho\rangle
    \right|^2 
\end{equation}
in the thermodynamic limit $\frac{2 \pi}{L} \to 0$, as derived by Ref.~\cite{SM:PhysRevLett.82.370}. In $d$ dimensions, taking the limit $|\bm{q}_{j}| \to 0$, and projecting $\rho_{j}$ onto the subspace of interest as in Eq.~~\ref{projecteddensity}, the center-of-mass coordinate and its variance become
\begin{align}
    \langle{\tilde{X}_{j}}\rangle &= \frac{L_{j}}{2\pi}\mathrm{Arg}\langle\tilde{\rho}_{j}\rangle, \quad \\
    \Delta \tilde{X}_{j}^2 &= -\left(\frac{L_{j}}{2\pi}\right)^2\ln|\langle\tilde{\rho}_{j}\rangle|^2, \label{Variance}
\end{align}
where $L_{j}=|\bm{a}_{j}|N_{j}$ is the length of the system along the $j$ direction. According to \eqref{Variance}, the eigenstates of $\tilde{\rho}_{j}$ with eigenvalues closest to being unimodular are maximally localized along the direction $j$. They form an orthonormal basis of the low-energy manifold and can be interpreted as many-body hybrid Wannier functions (localized $j$ but extended in all other directions). In general, projected density operators $\tilde{\rho}_{j}$ and $\tilde{\rho}_{j'}$ along different directions $j$ and $j'$ need not commute, so they do not necessarily share common eigenstates. As in single-particle band theory, when a many-body band has a topological obstruction to a smooth, periodic Bloch gauge, no orthonormal Wannier basis can be exponentially localized in all directions~\cite{SM:thouless1984wannier,SM:brouder2007exponential}.

\section{Degeneracy of Laughlin quasiholes on the torus}\label{Appendix:degeneracyLaughlin}

\subsection{Degeneracy of the Laughlin ground state}

Let $z=x+\mathrm{i}y$, with $x\equiv x + L_x$ and $y\equiv y+ L_y$, be the periodic coordinates on a rectangular torus with total flux quanta $L_xL_y=2\pi N_{\phi}\ell_{B}^2$. Let $\tau =\mathrm{i}\frac{L_y}{L_x}$ encode the aspect ratio. Ref.~\cite{SM:PhysRevB.93.245156} shows that the Laughlin ground states at filling $N_{e}/N_{\phi}=1/m$ on a torus are
\begin{equation}
    \Psi^{\mathrm{GS}}_{\kappa}(\{z_{i}\}) = \exp\left(\sum_{i=1}^{N_e}\frac{z_i^2-|z_i|^2}{4\ell_B^2}\right) F_\kappa(\mathcal{Z})\prod_{i<j}^{N_e}\vartheta_{\frac{1}{2},\frac{1}{2}}(z_{i}-z_{j}|\tau)^m, \label{TorusGroundState}
\end{equation}
where
\begin{equation}
    \vartheta_{a,b}(z|\tau) = \sum_{n=-\infty}^{\infty} e^{\mathrm{i}\pi(n+a)^2\tau}e^{\mathrm{i}2\pi(n+a)(\frac{z}{L_x}+b)}
\end{equation}
is the Jacobi theta function~\cite{SM:mumford1983tata}, defined for $a,b \in \{0,\frac{1}{2}\}$, and satisfying the quasiperiodicity conditions
\begin{align}
    \vartheta_{a,b}(z+L_x|\tau) &= e^{\mathrm{i}2\pi a}\vartheta_{a,b}(z|\tau) \\
    \vartheta_{a,b}(z+\mathrm{i}L_y|\tau)&= e^{-\mathrm{i}\pi\tau}e^{-\mathrm{i}2\pi(\frac{z}{L_x}+b)}\vartheta_{a,b}(z|\tau)
\end{align}
In \eqref{TorusGroundState},
\begin{equation}
    \mathcal{Z} = \sum_{i=1}^{N_e}z_{i}
\end{equation}
is $N_e$ times the complex center-of-mass coordinate, while
\begin{equation}
    F_\kappa(\mathcal{Z}) = e^{\mathrm{i}K^{(\kappa)}\mathcal{Z}}\prod_{\nu=1}^{m}\vartheta_{\frac{1}{2},\frac{1}{2}}(\mathcal{Z}-\mathcal{Z}_{\nu}^{(\kappa)}|\tau),
\end{equation}
and $K^{(\kappa)}$ and $\mathcal{Z}_{\nu}^{(\kappa)}$ satisfy the periodic boundary conditions
\begin{align}
    (-1)^{N_{\phi}}\exp(\mathrm{i}K^{(\kappa)}L_x)&=1, \label{PBC1} \\
    (-1)^{N_{\phi}}\exp(-K^{(\kappa)}L_y)\exp\left(\frac{\mathrm{i}2\pi}{L_x}\sum_{\nu=1}^{m}\mathcal{Z}_{\nu}^{(\kappa)}\right) &= 1. \label{PBC2}
\end{align}

Equations \eqref{PBC1} and \eqref{PBC2} have $m$ linearly-independent solutions $\{K^{(\kappa)},\sum_{\nu=1}^{m}\mathcal{Z}_{\nu}^{(\kappa)}\}$~\cite{SM:PhysRevB.93.245156}. Choosing a basis $\{F_\kappa\}$, with $\kappa=0,\ldots,m-1$ therefore yields $m$ linearly-independent Laughlin ground states~\eqref{TorusGroundState}.

\subsection{Degeneracy of quasiholes at each position}

Ref.~\cite{SM:PhysRevB.93.245156} constructs $m$ linearly independent
quasihole states at the same position
$\bm{r}_0=(x_0,y_0)$, or equivalently
$\eta=x_0+\mathrm{i}y_0$, corresponding to the $m$
center-of-mass sectors of the Laughlin state. At
$N_\phi=mN_e+1$, these states take the form
\begin{equation}
    \Psi^{\mathrm{qh}}_\kappa(\eta;\{z_i\})
    =
    e^{-G}f_\kappa(\eta;\{z_i\})
    e^{\frac{\eta^2}{4m\ell_B^2}}
    \prod_{i=1}^{N_e}e^{\frac{z_i^2}{4\ell_B^2}}, \label{QuasiHoleS}
\end{equation}
up to normalization, where
\begin{equation}
    G =
    \sum_{i=1}^{N_e}\frac{|z_i|^2}{4\ell_B^2}
    +\frac{|\eta|^2}{4m\ell_B^2},
\end{equation}
\begin{equation}
    f_\kappa(z_0;\{z_i\})
    =
    F_\kappa^{\mathrm{qh}}(\eta;\mathcal{Z})
    \prod_{i=1}^{N_e}
    \vartheta_{\frac12,\frac12}(z_i-\eta|\tau)
    \prod_{i<j}
    \vartheta_{\frac12,\frac12}(z_i-z_j|\tau)^m,
\end{equation}
and
\begin{equation}
    F_\kappa^{\mathrm{qh}}(\eta;\mathcal{Z})
    =
    e^{\mathrm{i}K^{(\kappa)}(\mathcal{Z}+\eta/m)}
    \prod_{\nu=1}^{m}
    \vartheta_{\frac12,\frac12}
    \left(\mathcal{Z}+\frac{\eta}{m}-\mathcal{Z}_\nu^{(\kappa)}\Big|\tau\right),
\end{equation}
with $\kappa=0,\ldots,m-1$. The parameters $K^{(\kappa)}$ and
$\mathcal{Z}_\nu^{(\kappa)}$ satisfy the boundary-condition constraints~\eqref{PBC1} and \eqref{PBC2}
at the quasihole flux $N_\phi=mN_e+1$.

\section{Many-body Spatial Localizer}\label{Appendix:Localizer}

The general framework for finding local state representations of electrons in insulators in $d$ dimensions is the Spatial Localizer developed in Ref.~\cite{SM:SpatLoc}, which we now extend to many-body systems. Its building blocks are centered density operators,
\begin{align}
    \tilde{\rho}_{j}(\bm{r})=e^{-\mathrm{i}\bm{q}_{j}\cdot\bm{r}}\tilde{\rho}_{j}, \label{Recentered}
\end{align}
which shift the origin to arbitrary points $\bm{r}$ in the continuum. Assembling the Spatial Localizer amounts to embedding every spatial degree of freedom in our set of density operators into a Clifford algebra. Each $\tilde{\rho}_{j}(\bm{r})$ encodes a periodic coordinate whose natural domain is a circle and therefore has two degrees of freedom. We can decouple them by splitting $\tilde{\rho}_{j}(\bm{r})$ into real and imaginary parts,
\begin{align}
    C_{j}(\bm{r}) &= \frac{\tilde{\rho}_{j}(\bm{r})+\tilde{\rho}^{\dagger}_{j}(\bm{r})}{2} - \mathrm{I}_N, \nonumber \\
    S_{j}(\bm{r}) &= \frac{\tilde{\rho}_{j}(\bm{r})-\tilde{\rho}^{\dagger}_{j}(\bm{r})}{2\mathrm{i}}, \label{realimaginarycomponents}
\end{align}
where $\mathbb{I}_N=V^{\dagger}V$ is the identity in the auxiliary space $\mathbb{C}^{N}$. Then, the $d$-dimensional PBC Spatial Localizer is defined as
\begin{equation}
    L(\bm{r}) \equiv \sum_{j=1}^{d'} C_{j}(\bm{r}) \otimes \Gamma_{2j-1}+S_{j}(\bm{r}) \otimes \Gamma_{2j}, \label{LocalizerAppendix}
\end{equation}
where $\Gamma_{n}$, $n=1,\dots,2d'$, are the generators of the Clifford algebra $\mathrm{Cl}_{2d',0}(\mathbb{R})$, satisfying $\{\Gamma_{n},\Gamma_{n'}\}=2\delta_{nn'}$, with $d'\geq d$.

We define the \emph{localizer indicator function} (LIF) as
\begin{equation}
    \mu(\bm{r})
    \equiv \min_{\lambda(\bm{r})\in\operatorname{spec}L(\bm{r})}|\lambda(\bm{r})|.
\end{equation}
By chiral symmetry, the nonzero Spatial Localizer eigenvalues occur
in opposite-sign pairs. Thus, for $\mu(\bm{r})>0$, the LIF
coincides with the minimum positive eigenvalue. Let
$\{\ket{\mu_\kappa(\bm{r})}\}_{\kappa=0}^{g-1}$ be an orthonormal basis
of its $g$-fold degenerate eigenspace,
\begin{equation}
    L(\bm{r})\ket{\mu_\kappa(\bm{r})}
    = \lambda(\bm{r})\ket{\mu_\kappa(\bm{r})}.
\end{equation}
The Spatial Localizer acts on the tensor product of the
physical and Clifford embedding spaces. Each eigenvector
therefore admits a Schmidt decomposition,
\begin{equation}
    \ket{\mu_\kappa(\bm{r})}
    = \sum_i s_{\kappa i}
    \ket{p_{\kappa i}(\bm{r})}\otimes\ket{\xi_{\kappa i}(\bm{r})},
    \qquad
    \sum_i s_{\kappa i}^2=1,
    \label{LocalizerStateAppendix}
\end{equation}
where $\ket{p_{\kappa i}}$ and $\ket{\xi_{\kappa i}}$ are the physical
and embedding Schmidt vectors, respectively.

\begin{figure}
    \centering
    \includegraphics[width=0.65\linewidth]{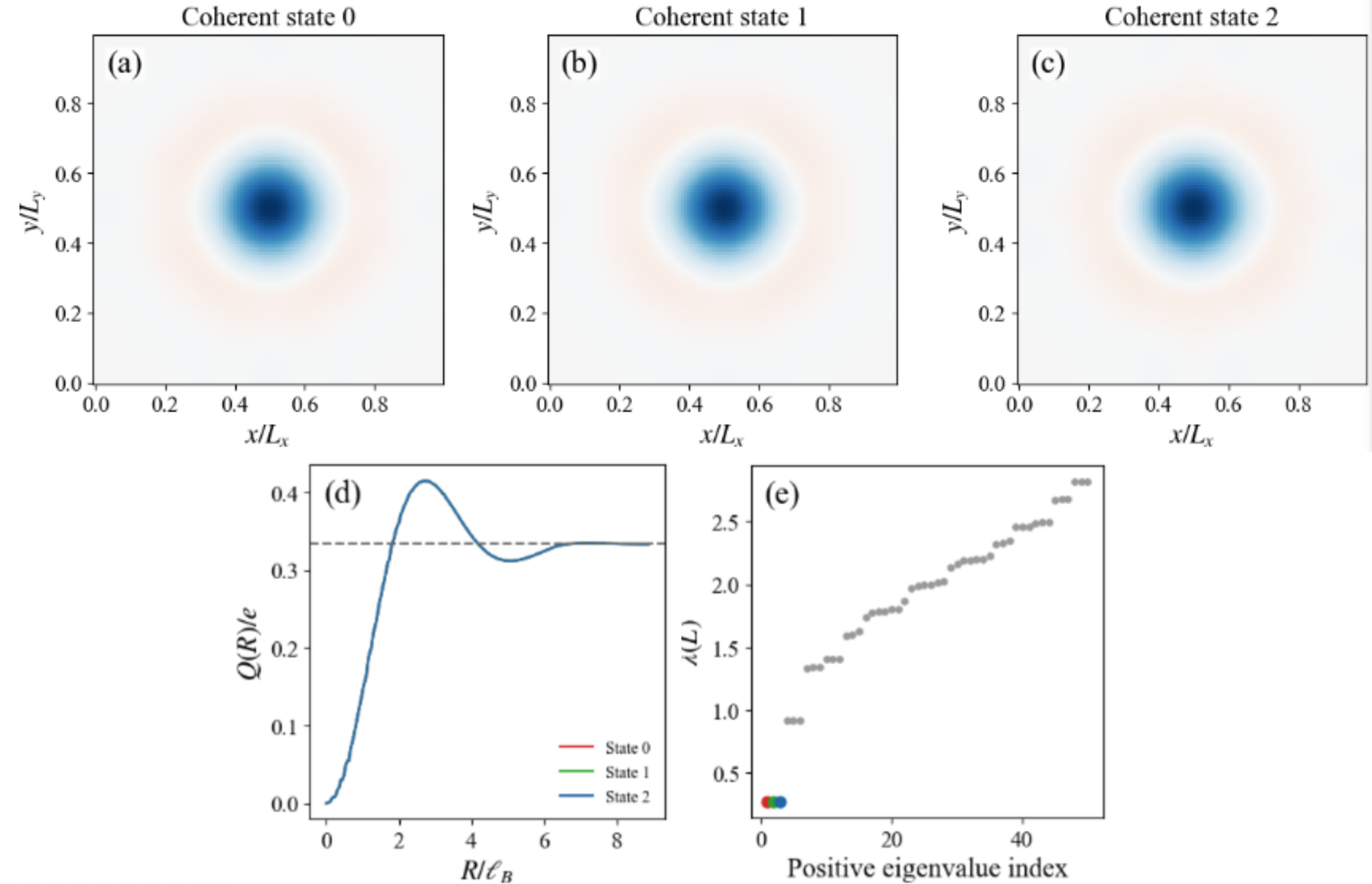}
    \caption{
\textbf{Three localizer states in the FQH quasihole manifold.}
(a--c) Charge profiles of the three mutually orthonormal
states extracted from the lowest positive localizer modes,
centered at the same position on the torus.
Blue (red) denotes positive (negative) charge relative
to the uniform background.
(d) Integrated excess charge within radius $R$ for each
state; the three curves overlap and approach $Q=e/3$
(dashed line).
(e) Positive localizer eigenvalues, with the three lowest
modes highlighted in the colors used in (d).
}
    \label{FHQqhcomparison}
\end{figure}

\begin{figure}
    \centering
    \includegraphics[width=0.65\linewidth]{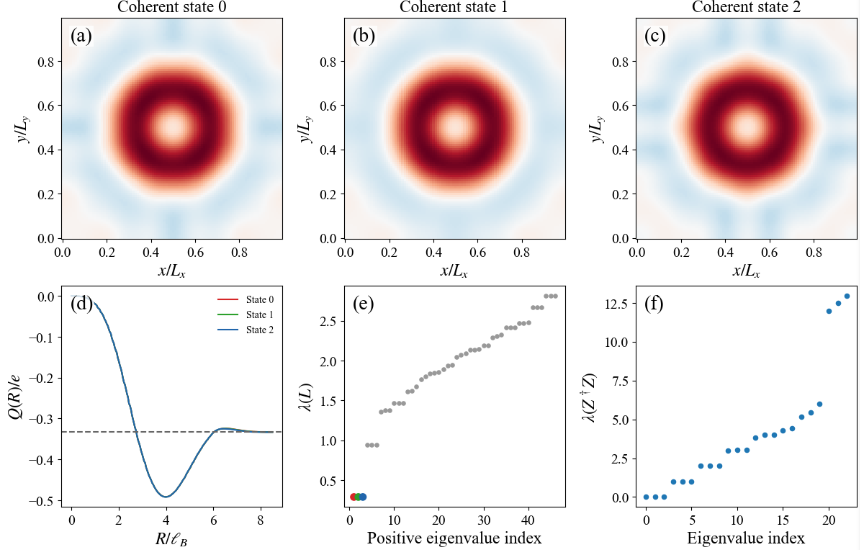}
    \caption{
\textbf{Three localizer states in the FQH quasielectron manifold.}
(a--c) Charge profiles of three mutually orthonormal
localizer states centered at the same position on the torus.
Blue (red) denotes positive (negative) charge relative
to the uniform background.
(d) Integrated excess charge within radius $R$ for each
state; the three curves overlap and approach $Q=-e/3$
(dashed line).
(e) Positive localizer eigenvalues, with the three lowest
modes highlighted in the colors used in (d).
}
    \label{FHQqecomparison}
\end{figure}

\begin{figure}
    \centering
    \includegraphics[width=0.65\linewidth]{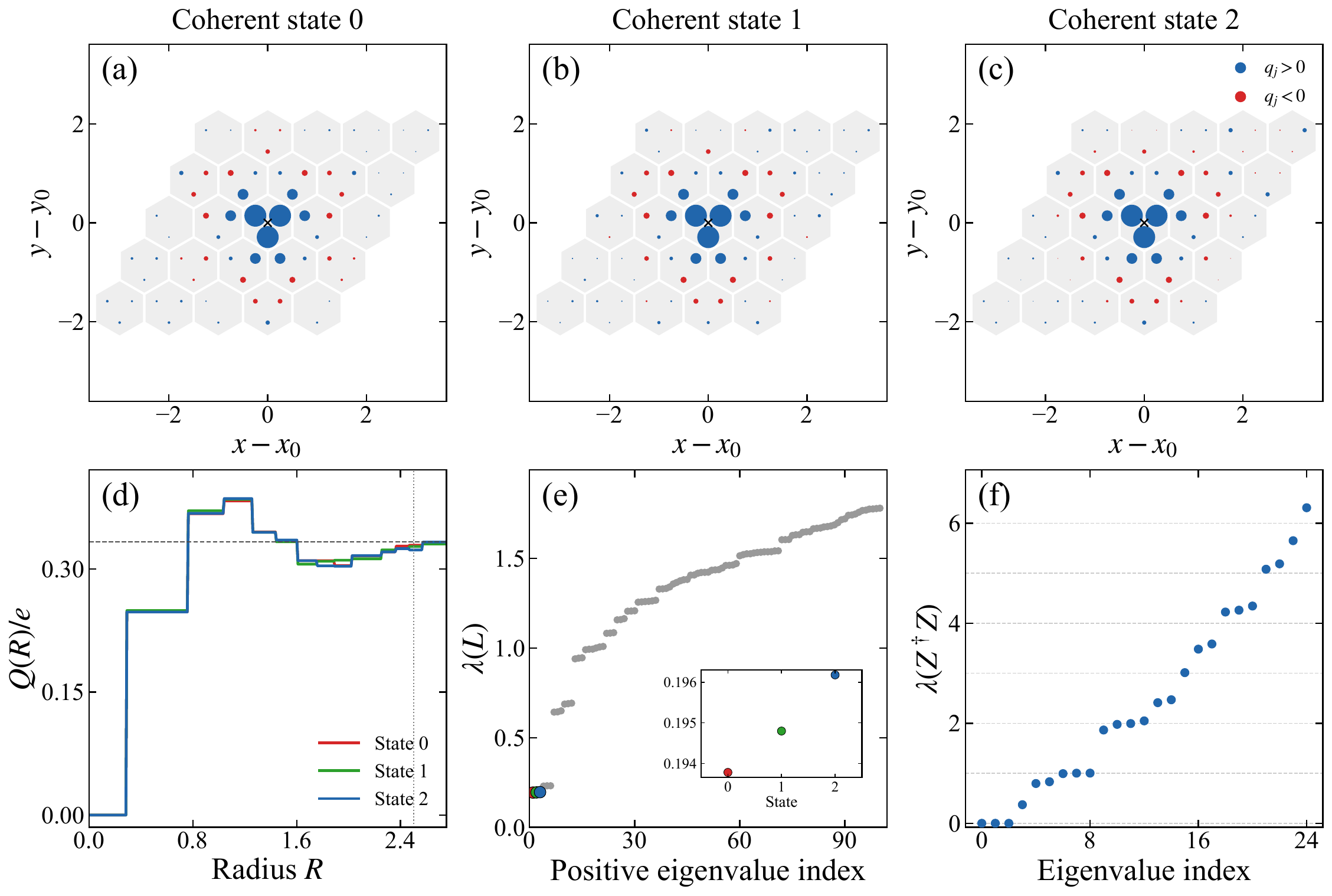}
    \caption{
\textbf{Three localizer states in the kagome FCI quasihole manifold.}
(a--c) Charge profiles of three mutually orthonormal
localizer states centered at the same position $\bm{r}_0$.
Light gray hexagons denote unit cells; disk areas are
proportional to the magnitude of the site-resolved charge,
with blue (red) indicating positive (negative) charge
relative to the uniform background.
(d) Integrated excess charge within radius $R$ for each
state, approaching $Q=e/3$ (horizontal dashed line).
(e) Positive localizer eigenvalues, with the three lowest
modes highlighted; the inset resolves their small splitting.
}\label{QuasiholeComparisons}
\end{figure}

We define the \emph{localizer states} as the leading
physical Schmidt vectors,
\begin{equation}
    \ket{\psi_\kappa(\bm{r})}\equiv\ket{p_{\kappa 1}(\bm{r})},
\end{equation}
and consider the subspace they span,
\begin{equation}
    \mathcal{S}(\bm{r})
    = \operatorname{span}
    \{\ket{\psi_\kappa(\bm{r})}\}_{\kappa=0}^{g-1}.
    \label{Localizerstate}
\end{equation}
For both the FQH and FCI systems considered here, we have an approximate degeneracy $g=3$, and we
numerically verify that the three states extracted
from the lowest positive localizer modes are mutually
orthonormal and share the same integrated charge profile within
each system [Figs.~\ref{FHQqhcomparison}, \ref{FHQqecomparison}, and \ref{QuasiholeComparisons}].

\begin{figure}
    \centering
    \includegraphics[width=0.65\linewidth]{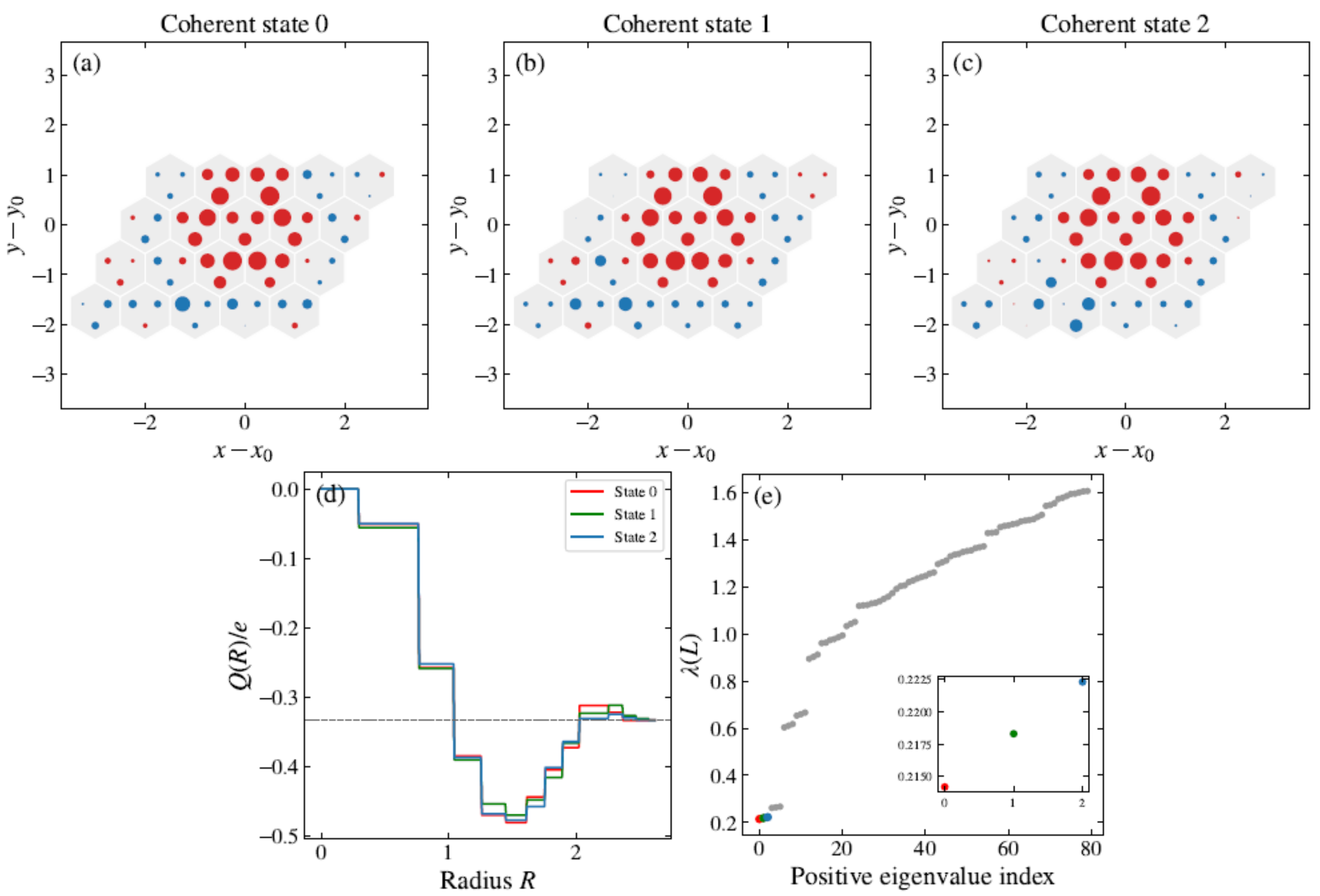}
    \caption{
\textbf{Three localizer states in the kagome FCI quasielectron manifold on a $4 \times 5$ lattice with $N_{e}=7$.}
(a--c) Charge profiles of three mutually orthonormal
localizer states centered at the same position $\bm{r}_0$.
Light gray hexagons denote unit cells; disk areas are
proportional to the magnitude of the site-resolved charge,
with blue (red) indicating positive (negative) charge
relative to the uniform background.
(d) Integrated excess charge within radius $R$ for each
state, approaching $Q=-e/3$ (horizontal dashed line).
(e) Positive localizer eigenvalues, with the three lowest
modes highlighted; the inset resolves their small splitting.
}\label{QuasielectronComparisons}
\end{figure}

\subsection{Bounds on the variances}

Under open boundary conditions, the $x$ component of the centered projected position operator is
\begin{equation}
    \tilde X_{\bm{r}}
    =\frac{1}{N_e}V^\dagger
    \Bigl(\sum_{\alpha,\beta}
    [X-r_x\mathbb I]_{\alpha\beta}
    c_\alpha^\dagger c_\beta\Bigr)V, \label{OBCdensities}
\end{equation}
where $[X-r_x\mathbb I]_{\alpha\beta} =\bra{\alpha}(X - r_x\mathbb I)\ket{\beta}$ are matrix elements of the centered single-particle position operator and $r_x=\hat{\bm{x}}\cdot\bm{r}$. The operator $\tilde{Y}_{\bm{r}}$ is defined in a similar manner.

From the definitions~\eqref{projecteddensityMain} and \eqref{OBCdensities}, and using \eqref{Variance}, we have
\begin{equation}
    \lim_{|\bm{q}_{j}|\to 0}\sum_{j}\frac{\Delta\tilde{\rho}_{j}(\bm{r})}{\bm{q}_{j}^2} = \sum_{j}\Delta \tilde{X}_{j}^2
\end{equation}
as $|\bm{q}_{j}| \to 0$. In that limit, reference~\cite{SM:SpatLoc} shows that the sums of the variances $\Delta \tilde{X}_{j}$ on a physical state $\ket{p_{\kappa 1}(\bm{r})}$ associated with a smallest-absolute eigenvalue $|\lambda_{\kappa}(\bm{r})|=\mu(\bm{r})$ satisfies the upper bound
\begin{align}
    \sum_{j}\Delta \tilde{X}_{j}^2 \leq \mu^{2}(\bm{r}) - \sum_{j}\left(\langle{\tilde{X}_{j}(0)\rangle - r_{j}}\right)^2 + \sum_{j<j'}|\langle[\tilde{X}_{j},\tilde{X}_{j'}]\rangle|+\sum_{j<j'}\sum_{i\neq 1}\frac{s_{\kappa i}}{s_{\kappa 1}}|\bra{p_{\kappa 1}(\bm{r})}[\tilde{X}_{j},\tilde{X}_{j'}]\ket{p_{\kappa i}(\bm{r})}|.
\end{align}
Since coherent states can be centered anywhere, we have $\langle{\tilde{X}_{j}(0)\rangle = r_{j}}$, so the second term vanishes. In the thermodynamic limit, the LIF $\mu(\bm{r})$ converges to zero, the leading Schmidt values $s_{\kappa 1}$ become $1$, and the subleading values become zero~\cite{SM:SpatLoc}. Therefore, in that limit we obtain
\begin{equation}
    \sum_{j}\Delta \tilde{X}_{j}^2 \leq \sum_{j<j'}|\langle[\tilde{X}_{j},\tilde{X}_{j'}]\rangle|.
\end{equation}
For two operators $\tilde{X}(\bm{r})$ and $\tilde{Y}(\bm{r})$, the above inequality becomes
\begin{equation}
    \Delta \tilde{X}(\bm{r}) + \Delta\tilde{Y}(\bm{r}) \leq [\tilde{X}(\bm{r}),\tilde{Y}(\bm{r})].
\end{equation}
saturates the usual Heisenberg-Robertson uncertainty relation~\cite{SM:robertson1929uncertainty}. This uncertainty saturation of the states $\ket{p_{\kappa 1}}$ is valid for all 2-dimensional systems, including those with non-square lattices that require more than two projected-density operators to build the Spatial Localizer.

In particular, for the kagome lattice, we have the three unit vectors
\begin{equation}
    \hat{\bm e}_1=\left(\frac{\sqrt{3}}{2},-\frac12\right),
    \qquad
    \hat{\bm e}_2=(0,1),
    \qquad
    \hat{\bm e}_3=\left(-\frac{\sqrt{3}}{2},-\frac12\right)
\end{equation}
satisfy
\begin{equation}
    \sum_{j=1}^{3}
    \hat{\bm e}_{j}\hat{\bm e}_{j}^{T}
    =\frac32\mathbb I_2.
\end{equation}
Defining the normalized OBC operators as
\begin{equation}
    \tilde X_{j,\bm r}
    =\sqrt{\frac23}\left(
        e_{jx}\tilde X_{\bm r}
        +e_{jy}\tilde Y_{\bm r}
    \right),
\end{equation}
their variances obey
\begin{align}
    \sum_{j=1}^{3}\Delta\tilde X_{j,\bm r}^{\,2}
    &=\frac23\sum_{j=1}^{3}
    \left[
        e_{jx}^{2}\Delta\tilde X_{\bm r}^{\,2}
        +e_{jy}^{2}\Delta\tilde Y_{\bm r}^{\,2}
        +e_{jx}e_{jy}
        \left\langle
            \{\delta\tilde X_{\bm r},
              \delta\tilde Y_{\bm r}\}
        \right\rangle
    \right] \nonumber\\
    &=\Delta\tilde X_{\bm r}^{\,2}
      +\Delta\tilde Y_{\bm r}^{\,2},
\end{align}
where $\delta A=A-\langle A\rangle$ and
$\Delta A^2=\langle(\delta A)^2\rangle$.
The mixed term vanishes because
$\sum_{j}e_{jx}e_{jy}=0$.
Without the normalization factor $\sqrt{2/3}$, the
sum of directional variances is instead
$\frac32(\Delta\tilde X_{\bm r}^{\,2}
+\Delta\tilde Y_{\bm r}^{\,2})$.

\section{Numerical results from finite-size exact diagonalization}\label{Numerics}

\subsection{Continuum FQH system}

We consider $N_{e}$ spin-polarized electrons restricted to the lowest Landau level (LLL) on a square torus of dimensions $L_x=L_y$ threaded by $N_{\phi}$ flux quanta. To benchmark the states obtained from the Spatial Localizer in the quasihole sector $N_{\phi}=3N_{e}+1$ against the Laughlin quasihole~\cite{SM:laughlin1983}, we assume the electrons interact through the Haldane $V_1$ pseudopotential~\cite{SM:Haldane1983}. We implement $V_1$ directly in second-quantized form in the single-particle Landau-gauge orbital basis $c^{\dagger}_{j}\ket{0}$, $j \in \{0,1,\dots,N_{\phi}-1\}$, following the constructions of Refs.~\cite{SM:PhysRevB.32.6924,SM:Trugman1985,SM:Papic2015},
\begin{equation}
    H_{V_1} = \sum_{n}\hat{b}_{n}^{\dagger}\hat{b}_{n}, \quad \hat{b}_{n} = \sum_{i+j=n}g_{ij}c^{\dagger}_{i}c^{\dagger}_j, \label{PseudoPotential}
\end{equation}
where $g_{ij}=x_{ij}e^{-(2\pi\ell_B x_{ij}/L_y)^2}$ and $x_{ij} = \frac{j-i}{2}$. Here, $\ell_B=\sqrt{\hbar/eB}$ is the magnetic length, $e$ is the electronic charge and $B$ is the magnitude of the external magnetic field in the $z$ direction. 

The best trial function for localized quasielectrons is the Jain state~\cite{SM:jain1989}, which is obtained not from the $V_1$ pseudopotential, but from the full Coulomb interaction projected onto the LLL,
\begin{align}
    H_{\mathrm{Coulomb}} &= \sum_{i \neq j}\sum_{t=0}^{N_{\phi}-1}W_{t}(i-j)c^{\dagger}_{i+t}c^{\dagger}_{j-t}c_{j}c_{i},
\end{align}
whose matrix elements in the Landau gauge read~\cite{SM:yoshioka1984ground,SM:PhysRevLett.50.1219}
\begin{align}
    W_{t}(d)=\frac{1}{2N_{\phi}}\sum_{s \in \mathbb{Z}}\sum_{m \in \mathbb{Z}}\frac{e^{-q^2/2}}{q}\cos\left(\frac{2\pi s(d+t)}{N_{\phi}}\right), \quad q^2 = \left(\frac{2\pi s}{L_x}\right)^2+\left(\frac{2\pi(t+mN_{\phi})}{L_y}\right)^2,\label{Coulomb}
\end{align}
where we leave the diverging $s=0$ and $t+mN_{\phi}=0$ terms out of the sum.

At one flux quantum above or below 1/3 filling of the Landau level ($N_{\phi}=3N_{e} \pm 1$), the spectra of \eqref{PseudoPotential} and \eqref{Coulomb} exhibit a gapped low-energy band consisting of $N_{\phi}$ degenerate Bloch anyons $\ket{\Psi(K)}$, one per magnetic momentum sector $K \in \{0,1,\dots,N_{\phi}-1\}$, as we can see in Figs.~\ref{fig:4} (a) and (e). We define periodic density operators along the guiding-center coordinates $x$ and $y$ as in Eq.~\eqref{FQHEdensityoperators} and feed them to the Spatial Localizer \eqref{Localizer} with $d=2$. Here, we show the values for quasiholes, which are qualitatively the same as those for quasielectrons. 

Upon diagonalizing $L(\bm{r})$ over a fine $\bm{r}$ mesh, we obtain LIFs that are uniform to machine precision, but decrease monotonically as we increase system size. At $N_{e}=8$ electrons, $\mu(\bm{r})=0.266$ ($\mu(\bm{r})=0.289$) for the quasihole (quasielectron) manifold, as shown in Figs.~\ref{fig:4} (a) and (e). A flat LIF indicates that the Spatial Localizer states \(\ket{\psi(\bm r)}\) have uniform localization properties across the torus. To quantify this, we locally construct OBC guiding-center operators from the unitary polar factors of the centered periodic densities~\eqref{FQHEdensityoperators}. We write \(\tilde{\rho}^{\mathrm{polar}}_{j}(\bm r)=U_{j,\bm r}^{\dagger}e^{\mathrm{i}\Theta_{j}(\bm r)}U_{j,\bm r}\), where \(\Theta_{j}(\bm r)\) is a diagonal matrix of phases in the interval $(-\pi,\pi]$, and define
\begin{equation}
    \tilde X_{j,\bm r}=(L_j/2\pi)U_{j,\bm r}^{\dagger}\Theta_{j}(\bm r)U_{j,\bm r}
\end{equation}
These Hermitian operators provide good approximations for the OBC densities~\eqref{OBCdensities} at \(\bm r\). Evaluated on \(\ket{\psi(\bm r)}\), they yield \(\Delta\tilde X_{\bm r}^{\,2}+\Delta\tilde Y_{\bm r}^{\,2}=3.005\,\ell_B^2\) throughout the torus, while \(\left|\langle[\tilde X_{\bm r},\tilde Y_{\bm r}]\rangle\right|=2.999\,\ell_B^2\), close to the exact Laughlin quasihole value \(3\ell_B^2\) at \(\nu=1/3\). Since
\begin{equation}
    \Delta\tilde X_{\bm r}^{\,2}+\Delta\tilde Y_{\bm r}^{\,2}\geq\left|\langle[\tilde X_{\bm r},\tilde Y_{\bm r}]\rangle\right|, \label{inequalityAppendix}
\end{equation}
these results show that the Spatial Localizer states nearly saturate the guiding-center uncertainty bound uniformly across the torus.

Crucially, we can locally define lowering and raising operators \(Z_{\bm r}=(\tilde X_{\bm r}+\mathrm{i}\tilde Y_{\bm r})/(\sqrt{6}\ell_B)\) and \(Z_{\bm r}^{\dagger}=(\tilde X_{\bm r}-\mathrm{i}\tilde Y_{\bm r})/(\sqrt{6}\ell_B)\), for which \(\langle[Z_{\bm r},Z_{\bm r}^{\dagger}]\rangle=0.999\). The near-saturation of the guiding-center uncertainty bound gives \(\|Z_{\bm r}\ket{\psi(\bm r)}\|^2=[3.005-2.999]/6\simeq10^{-3}\). As the system size increases, this norm decreases monotonically toward zero, indicating that \(\ket{\psi(\bm r)}\) approaches a zero mode of the locally centered lowering operator \(Z_{\bm r}\). In the thermodynamic limit, the Spatial Localizer states therefore approach guiding-center coherent states centered at \(\bm r\).

To assess whether our $N_e=8$ results are representative of the thermodynamic limit, we examine how closely the Spatial Localizer anyons approach saturation of \eqref{inequalityAppendix} as the system size increases. Thus, we define the quantity
\begin{equation}
    \eta_{\bm{r}} = \frac{\Delta X_{\bm{r}}^2+\Delta Y_{\bm{r}}^2}{|\langle [\tilde{X}_{\bm{r}},\tilde{Y}_{\bm{r}}] \rangle|} \geq 1, \label{etadefinition}
\end{equation}
should approach $1$ for all $\bm{r}$ as $N_{e},N_{\phi} \to \infty$ and $\nu \to 1/3$. Then, we note that \eqref{inequalityAppendix} is a particular case of the more general Heisenberg-Roberston-Schrodinger uncertainty relation
\begin{equation}
    \Delta \tilde{X}_{\bm{r}}^2 \Delta \tilde{Y}_{\bm{r}}^2 \geq \frac{1}{4}|\langle[\tilde{X}_{\bm{r}},\tilde{Y}_{\bm{r}}]\rangle|^2+\mathrm{Cov}(\tilde{X}_{\bm{r}},\tilde{Y}_{\bm{r}})^2, \label{HeisenbergRobertson}
\end{equation}
for when $\Delta \tilde{X}_{\bm{r}} = \Delta \tilde{Y}_{\bm{r}}$ and for when
\begin{equation}
    \mathrm{C_{ov}}(\tilde{X}_{\bm{r}},\tilde{Y}_{\bm{r}}) \equiv \frac{1}{2}\left\langle\{\tilde{X}_{\bm{r}}-\langle \tilde{X}_{\bm{r}} \rangle,\hat{Y}_{p}-\langle \tilde{Y}_{\bm{r}}\rangle\} \right\rangle \label{CovarianceExpression}
\end{equation}
vanishes.

\begin{figure*}
    \centering
    \includegraphics[width=0.8\linewidth]{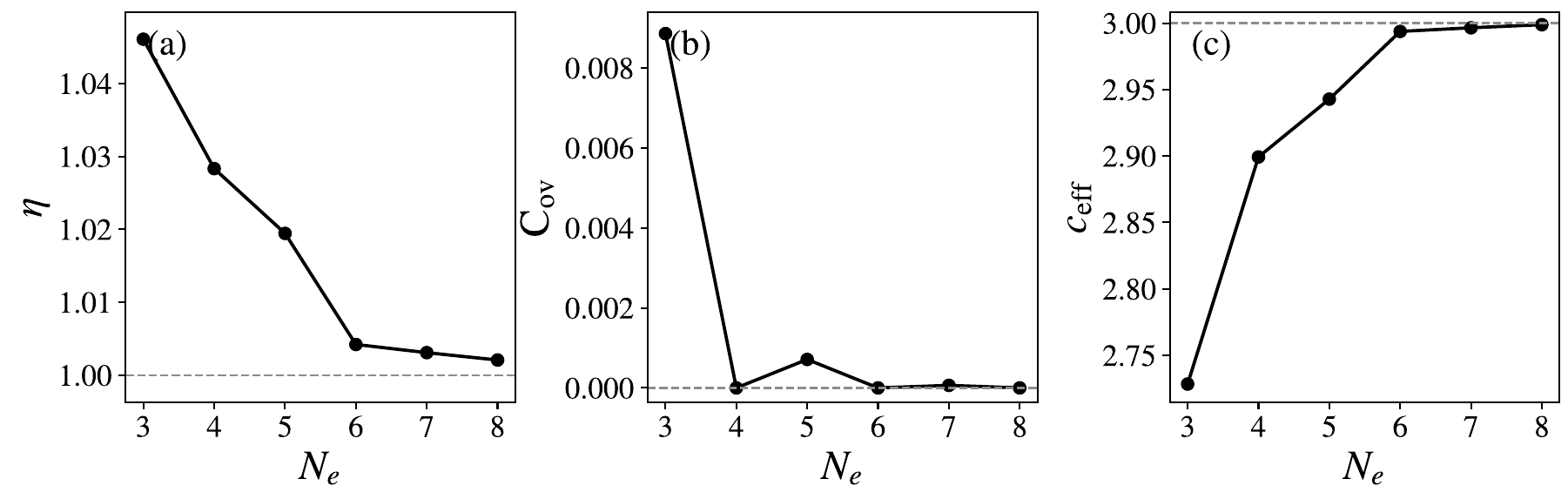}
    \caption{Scaling of the quantities that characterize the saturation of the uncertainty relation \eqref{inequalityAppendix}. All calculations are done for a fixed $\bm{r}=0$.}
    \label{Fig1Appendix}
\end{figure*}

Fig.~\ref{Fig1Appendix}(a) shows the monotonic decrease of $\eta_{\bm{r}}$ calculated at the origin $\bm{r}=0$ as a function of the system size, pointing toward a convergence to 1 in the thermodynamic limit. Fig~\ref{Fig1Appendix}(b) shows that the covariance \eqref{CovarianceExpression} also tends to zero, guaranteeing the equivalence of expressions \eqref{inequalityAppendix} and \eqref{HeisenbergRobertson}. 

A particle of charge $q$ under uniform Berry curvature is expected to satisfy
\begin{equation}
    [\tilde{X}_{\bm{r}},\tilde{Y}_{\bm{r}}] = \frac{\mathrm{i}\hbar}{q B},
\end{equation}
where $B$ is the external magnetic field. For an electron in the LLL, we have $\ell_B^2=\hbar/(eB)$. A $\nu=1/3$ quasihole has charge $q=+e/3$, so we expect
\begin{equation}
    |\bra{\psi(\bm{r})}[\tilde{X}_{\bm{r}},\tilde{Y}_{\bm{r}}]\ket{\psi(\bm{r})}| = 3\ell_B^2,
\end{equation}
where $\ket{\psi(\bm{r})}$ is the single-anyon state pulled from the Spatial Localizer centered at $\bm{R}$. Fig.~\ref{Fig1Appendix} (c) shows the scaling of
\begin{equation}
    c_{\mathrm{eff}} = \frac{|\bra{\psi(\bm{r})}[\tilde{X}_{\bm{r}},\tilde{Y}_{\bm{r}}]\ket{\psi(\bm{r})}|}{\ell_B^2},
\end{equation}
which should converge to the value of 3 in the thermodynamic limit.

\subsection{kagome FCI}

Adopting the model presented in Refs.~\cite{SM:PhysRevLett.106.236803,SM:Lu_2026}, we consider spin-polarized electrons on a kagome lattice with complex NN and real NNN hoppings
\begin{align}
    H_0 &= \sum_{\langle i,j \rangle} [(t_1+\mathrm{i}\nu_{ij}\lambda_1)c^{\dagger}_{i}c_{j}+ \mathrm{h.c.}] + t_2\sum_{\langle\langle i,j \rangle\rangle } (c^{\dagger}_{i}c_{j}+ \mathrm{h.c.}) \nonumber \\
    &= \sum_{s,s'}\sum_{\bm{k}}c_{k,s}^{\dagger}[h_{\bm{k}}]_{s,s'}c_{k,s'},
\end{align}
where $\nu_{ij}=\pm 1$ specifies the orientation of the NN complex hopping pattern. In the Bloch representation, $c^{\dagger}_{\bm{k},s}$ creates an electron with crystal momentum $\bm{k}$ at orbital $s \in \{A,B,C\}$. We take $t_1=-1$, $\lambda_1=1$, and $t_2=0.2$, for which the lowest band is nearly flat and has Chern number $C=1$. As shown in Sec.~\ref{sec:vortexability}, this Chern band is not ideal, \emph{i.e.}, it does not satisfy the trace condition $\operatorname{Tr}g(\bm{k})=|\Omega(\bm{k})|$ relating the quantum metric $g(\bm{k})$ to the Berry curvature $\Omega(\bm{k})$, nor does it satisfy the vortexability criterion of Ref.~\cite{SM:PhysRevB.110.245112}. Consequently, multiplication by a vortex function does not preserve the single-particle Chern-band Hilbert space, precluding the standard analytic construction of exact Laughlin-type trial states for either the FCI ground state or its anyonic excitations~\cite{SM:ledwith2023vortexability}.

\begin{figure*}
    \centering
    \includegraphics[width=0.65\linewidth]{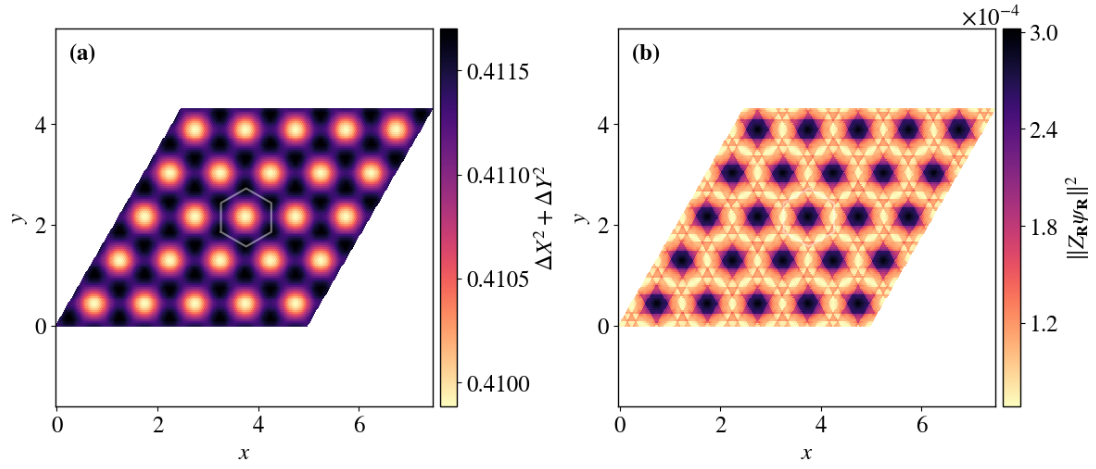}
    \caption{\textbf{Localization texture.} Heatmaps of (a) the projected density variance and (b) the deviation from the exact zero-mode condition $\|Z_{\bm r}\ket{\psi(\bm r)}\|^2=0$ across the lattice.}
    \label{fig:Beautiful1}
\end{figure*}

\begin{figure*}
    \centering
    \includegraphics[width=0.8\linewidth]{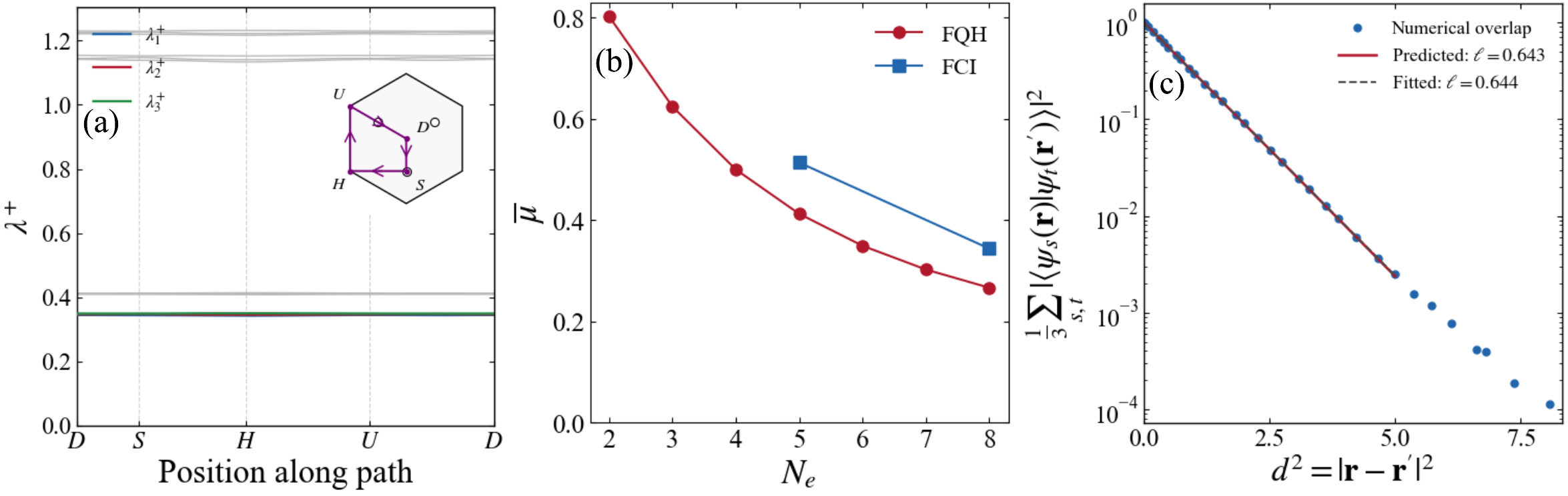}
    \caption{\textbf{Localizer spectrum, finite-size dependence, and coherent-state overlaps. } (a) Positive eigenvalues of the FCI localizer along the real-space path shown in the inset. The three lowest levels are highlighted in blue, red, and green; higher levels are gray. The labels $(D,S,H,U)$ denote high-symmetry points on the unit cell. (b) Spatially averaged localizer indicator $\overline{\mu}$ versus electron number for the FQH and FCI one-quasihole manifolds, with $N_\phi=N_{\mathrm{uc}}=3N_e+1$. (c) Total overlap $\frac13\overline{\sum_{\kappa,\kappa'}|\langle\psi_\kappa(\mathbf r)|\psi_{\kappa'}(\mathbf r')\rangle|^2}$ between anyons $\ket{\psi_\kappa(\bm{r})}$ and $\ket{\psi_{\kappa'}(\bm{r}')}$, where $\kappa,\kappa' \in \{0,1,2\}$ label the three orthonormal anyon states at each position, which span the physical subspaces extracted from the three lowest positive localizer modes. Distances are minimum-image separations on the torus. The solid red curve is the Gaussian prediction $e^{-d^2/(2\ell^2)}$, with $\ell=0.643$ determined independently from the projected-density spread, $\ell^2=\overline{\Delta X^2+\Delta Y^2}$. The dashed curve is a Gaussian fit $Ae^{-d^2/(2\ell_{\mathrm{fit}}^2)}$, yielding $\ell_{\mathrm{fit}}=0.644$.}
    \label{fig:LIFscaling}
\end{figure*}

We partially occupy this band with filling factor $\nu=N_{e}/N_{\mathrm{u.c.}} < 1$, where $N_{\mathrm{u.c.}} = N_{1} \times N_2$ is the number of unit cells of a torus extended along the primitive vectors $\bm{a}_1=(1,0)$ and $\bm{a}_2=\frac{1}{2}(1,\sqrt{3})$. Moreover, we consider a repulsive NN density-density interaction $u=1$. Since, for our parameters, the scale of $u$ is smaller than the single-particle bandgap, we can project the interaction onto the lowest band as
\begin{equation}
    H = H_0 + H_{\mathrm{int}}, \quad H_{\mathrm{int}} = \frac{u}{2}\sum_{\langle i,j\rangle} \overline{n}_{i}\overline{n}_{j}, \label{Interactionkagome}
\end{equation}
where $\overline{n}_{i} \equiv\overline{n}_{s,\bm{R}}$ is the electronic density projected onto the single-particle Chern band, with $i$ labeling the orbital $s$ and unit cell $\bm{R}$. Explicitly,
\begin{equation}
    \overline{n}_{s,\bm{R}} = \frac{1}{N_{\mathrm{u.c.}}}\sum_{\bm{k},\bm{k}'}e^{-\mathrm{i}(\bm{k}-\bm{k}')\cdot\bm{R}}u_{s}^{*}(\bm{k})u_{s}(\bm{k})\gamma^{\dagger}_{\bm{k}}\gamma_{\bm{k}'},
\end{equation}
where $\gamma^{\dagger}_{\bm{k}}$ and $u_{s}(\bm{k})$ are the fermionic creation operator and $s$-component of the Bloch eigenstate of the lowest band, respectively. At system size $N_{1}=N_{2}=5$ and $N_{e}=8$ electrons, the Hamiltonian \eqref{Interactionkagome} has a gapped low-energy band spanned by Bloch quasiholes $\ket{\Psi(\bm{K})}$ at crystal momenta $n_1\bm{q}_1+n_{2}\bm{q}_2$, where $\bm{q}_1 = \frac{\bm{b}_1}{N_1}$, $\bm{q}_2 = \frac{\bm{b_2}}{N_2}$ with $\bm{b}_1 = 2\pi (1,-\frac{1}{\sqrt{3}})$, $\bm{b}_2 = (0,\frac{4\pi}{\sqrt{3}})$, as we can see in Fig.~\ref{fig:5}(b). 

Since the kagome primitive vectors are not orthogonal, an isotropic Spatial Localizer for the kagome lattice needs input from three density operators of the form \eqref{projecteddensity} along the vectors $\bm{q}_1$, $\bm{q}_2$, and $\bm{q}_3 = -\frac{\bm{b}_1+\bm{b}_2}{N_{1}}$. We then embed these position operators into the Clifford Algebra $\mathrm{Cl}_{6,0}(\mathbb{R})$ according to Eq.~\eqref{Localizer} with $d=3$. Similarly to the continuum FQH system, the resulting LIF is almost flat, as shown in Fig.~\ref{fig:5}(g). As a result, we can still find localized anyons centered at any point on the torus, but their charge-distribution profile is now position-dependent, as Figs.~\ref{fig:5}(c-e) and ~\ref{fig:5}(h-j) show. Fig.~\ref{fig:Beautiful1}(a) shows the nearly uniform variance of the charge density $\Delta\tilde{X}_{\bm{r}}^2+\Delta \tilde{Y}_{\bm{r}}^2$ across the lattice, while ~\ref{fig:Beautiful1}(b) shows the quantity $\|Z_{\bm r}\ket{\psi(\bm r)}\|^2$ calculated from the definition
\begin{equation}
    Z_{\bm{r}} = \frac{1}{\sqrt{2|\langle[\tilde{X}_{\bm{r}},\tilde{Y}_{\bm{r}}]\rangle|}}(\tilde{X}_{\bm{r}}+\mathrm{i}\tilde{Y}_{\bm{r}}).
\end{equation}
We should have $\|Z_{\bm r}\ket{\psi(\bm r)}\|^2=0$ for coherent states.

\subsection{Overcompleteness and resolution of the identity of the FCI anyons}\label{sec:klauder}

A family of states $\{\ket{\psi_\kappa(\bm r)}\}$ labeled by a continuous parameter
$\bm r$ forms a set of coherent states in the sense of Klauder~\cite{SM:klauder1963continuous,SM:klauder1985coherent}
if (i) its members are normalized, (ii) they depend continuously on $\bm r$, and
(iii) they resolve the identity with a positive measure. Overcompleteness alone,
i.e., spanning the space with linearly dependent states, is weaker than (iii). Here we test (iii) for
the kagome FCI.

At each position, the Spatial Localizer yields three nearly degenerate positive eigenvalues $\lambda_\kappa(\bm r)$, $\kappa=0,1,2$. Each eigenvalue contributes a single localized anyon $\ket{\psi_\kappa(\bm r)}$. We orthonormalize the three states at each $\bm r$ and define
\begin{equation}
    \mathcal I \equiv \frac{1}{3A_{\mathrm{c}}}\int d^2r\,
    \sum_{\kappa=0}^{2}\ket{\psi_\kappa(\bm r)}\bra{\psi_\kappa(\bm r)},
    \label{eq:ResolutionIdentity}
\end{equation}
where $A_{\mathrm{c}}=|\bm a_1\times\bm a_2|$ is the unit-cell area and the integral runs over the lattice. Condition (iii) requires $\mathcal I = P$, where $P$ is the projector onto the anyon band~\eqref{Projectors}. Since $P$ is diagonal in the momentum basis, we should get $\bra{\Psi(\bm K)}\mathcal I\ket{\Psi(\bm K')}\propto\delta_{\bm K\bm K'}$ exactly. The only nontrivial content of Eq.~\eqref{eq:ResolutionIdentity} is therefore its diagonal, $\bra{\Psi(\bm K)}\mathcal I\ket{\Psi(\bm K)}=\frac{N_\phi}{3}\, \overline{\sum_\kappa|a_{\bm r,\kappa}(\bm K)|^2}$, where the bar denotes the average over $\bm r$. 

\begin{figure}
    \centering
    \includegraphics[width=0.6\linewidth]{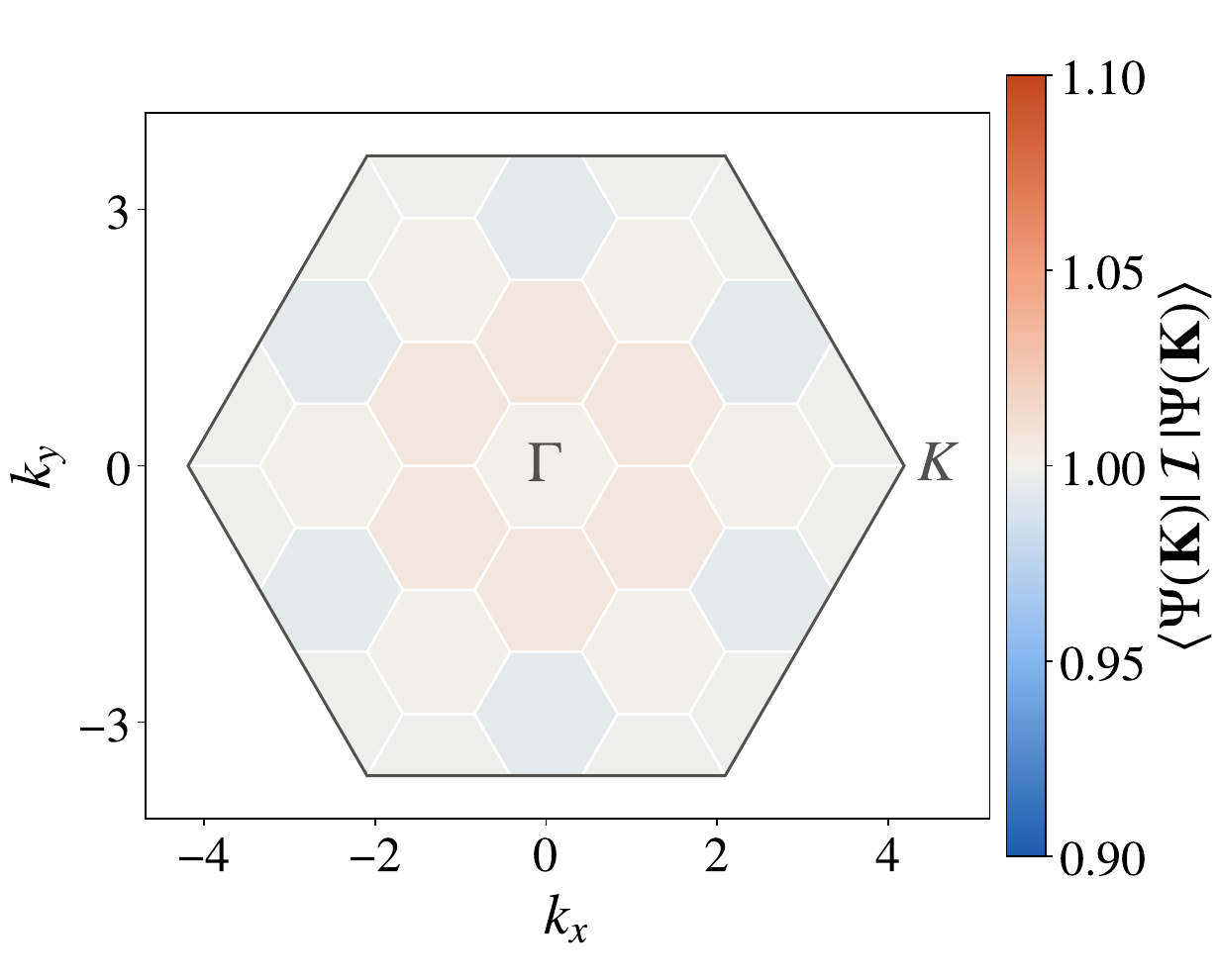}
    \caption{Diagonal matrix elements $\bra{\Psi(\bm K)}\mathcal I\ket{\Psi(\bm K)}$
    of the resolution-of-identity operator~\eqref{eq:ResolutionIdentity} for the
    kagome FCI quasihole band ($5\times5$ torus, $N_e=8$), shown over the hexagonal
    Brillouin zone. Each tile is the Voronoi cell of one many-body momentum $\bm K$;
    tiles on the zone boundary appear as two halves. All values lie between $0.993$
    and $1.007$; off-diagonal elements vanish by translation symmetry.}
    \label{fig:Klauder}
\end{figure}

We evaluate Eq.~\eqref{eq:ResolutionIdentity} on a uniform grid of
sampled positions per unit cell for the $4\times4$ and $5 \times 5$ lattices. The
off-diagonal elements vanish to numerical precision ($<10^{-14}$), and all 25
diagonal elements at $5 \times 5$ lie between $0.993$ and $1.007$ [Fig.~\ref{fig:Klauder}]. The resolution of identity holds to within 1\% at $5 \times 5$ and improves from 4\% at $4 \times 4$, consistent with finite-size mixing within the nearly degenerate triplet.

\begin{figure}
    \centering
    \includegraphics[width=\linewidth]{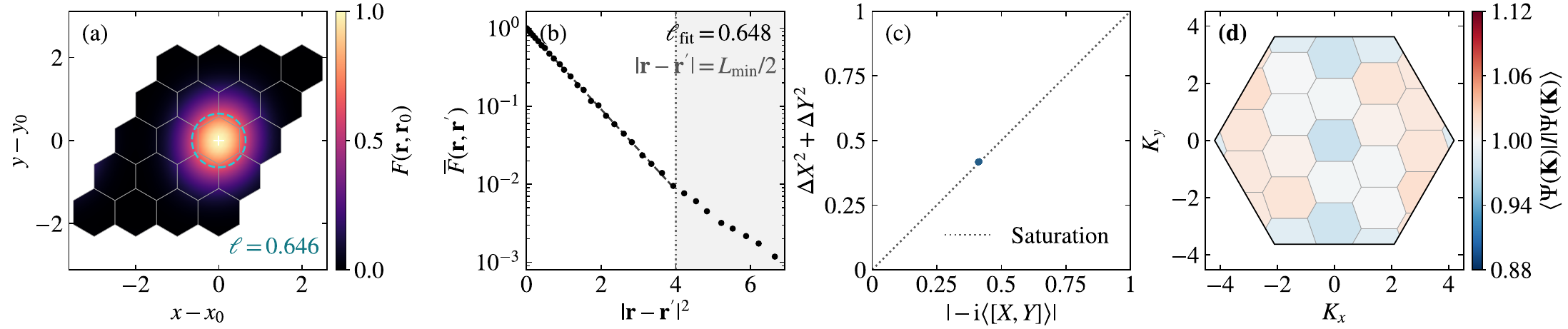}
    \caption{\textbf{Quasielectron coherence on the $4\times5$ kagome FCI torus ($N_e=7$).} (a) Overlap $F(\bm r,\bm r_0)=\frac{1}{3}
\sum_{\kappa,\kappa'=0}^{2}
|\langle\psi_\kappa(\bm r)|\psi_{\kappa'}(\bm r_0)\rangle|^2$
between the quasielectron subspaces centered at $\bm r$ and at the reference
point $\bm r_0$. The cyan circle has radius
$\ell=\sqrt{\overline{\Delta X^2+\Delta Y^2}}=0.646$.
(b) Radially averaged overlap versus squared separation $|\bm r-\bm r'|^2$. The dashed curve is a Gaussian fit with
$\ell_{\mathrm{fit}}=0.648$. The vertical dotted line marks the first
separation at which two periodic images of the quasihole can be equally close,
$|\bm r-\bm r'|=L_{\min}/2$, where $L_{\min}$ is the shortest translation connecting two equivalent points on the torus.
(c) Total density variance versus the magnitude of the commutator expectation for the equally weighted triplet at each sampled center. The dotted diagonal denotes saturation of the sum uncertainty bound. (d) Diagonal matrix elements of \eqref{eq:ResolutionIdentity} in the many-body momentum basis. The values range from $0.9733$ to $1.0192$, and the largest off-diagonal magnitude is $1.1\times10^{-15}$.}
    \label{Quasielectron numerics}
\end{figure}

We also examine the one-quasielectron band on a $4\times5$ torus with
    $N_e=7$. The three nearly degenerate Spatial Localizer modes yield a three-dimensional quasielectron subspace at each sampled center. As shown in Fig.~\ref{Quasielectron numerics}(a,b), the triplet overlap decays as a Gaussian with a fitted magnetic length $\ell_{\mathrm{fit}}=0.648$, close to the independent estimate $\ell=\sqrt{\overline{\Delta X^2+\Delta Y^2}}=0.646$. The triplet nearly saturates the uncertainty bound [Fig.~\ref{Quasielectron numerics}(c)]. Its mean spread $0.4173$, and its mean commutator magnitude is $0.4123$, with $-\mathrm{i}\langle[X,Y]\rangle$ opposite in sign to that of the quasihole. The quasielectron triplets also approximately resolve the identity, with a maximum diagonal deviation of $2.67\%$ and off-diagonal elements at numerical precision [Fig.~\ref{Quasielectron numerics}(d)].

\subsection{Testing vortexability of the kagome Chern band}\label{sec:vortexability}

Here, we show that the $C=1$ band of the kagome model of the main text is not vortexable according to the method developed by Ref.~\cite{SM:PhysRevB.110.245112} for tight-binding models.

Let $\hat{P}$ be the projector onto the candidate vortexable band, and let
\begin{equation}
    \hat{Z} = \sum_{\bm{R},a} z(\bm{R},a)\ket{\bm{R},a}\bra{\bm{R},a}
\end{equation}
be the ``vortex operator'', diagonal in the position basis $\{\ket{\bm{R},a}\}$, where $\bm{R}$ are Bravais lattice vectors and $\bm{r}_{a}$ are the intra-cell positions of the atomic orbitals $a$. 

Any state $\ket{\psi}=\hat{P}\ket{\psi}$ in a vortexable band satisfies
\begin{equation}
    \hat{Z}\ket{\psi}=\hat{P}\hat{Z}\ket{\psi},
\end{equation}
which is equivalent to the operator identity
\begin{equation}
    \hat{Z}\hat{P} = \hat{P}\hat{Z}\hat{P} \Leftrightarrow \hat{Q}\hat{Z}\hat{P}=0,
\end{equation}
where $\hat{Q}=1-\hat{P}$.

Reference~\cite{SM:PhysRevB.110.245112} proposes vortex operators of the form
\begin{align}
    \hat{Z} &= \alpha  \left(\hat{X}+\sum_{a}\tilde{x}_a\hat{P}_a\right) + \beta \left(\hat{Y}+\sum_{a}\tilde{y}_a\hat{P}_a\right) \nonumber \\
    &= \alpha\hat{X}+\beta\hat{Y}+\sum_{a}\gamma_a\hat{P}_a, \label{vortexoperator}
\end{align}
where $\hat{X},\hat{Y}$ are OBC Hermitian position operators in the thermodynamic limit and $\hat{P}_{a}=\ket{a}\bra{a}$ is the projection operator onto sublattice $a$.

Reference~\cite{SM:PhysRevB.110.245112} then defines a vortexability index
\begin{equation}
    I(Z) = \frac{\langle \hat{Z}^{*},\hat{Z}\rangle}{\langle \hat{Z},\hat{Z}^{*}\rangle} \geq 0,
\end{equation}
where $\langle \cdot, \cdot \rangle = \mathrm{Tr}[P\cdot Q\cdot P]/V$, $\hat{Z}$ is the complex conjugate of $\hat{Z}$, and $V$ is the system volume. For a vortexable band, we must have $I(Z)=0$, which follows from $\hat{Q}\hat{P}\hat{Z}=0$. By representing the parameters of $\hat{Z}$ as a vector $\ket{\alpha}=(\alpha,\beta,\gamma_1,\gamma_2,\dots)^{T}$, we can express the vortexability index as
\begin{equation}
    I(Z) = \frac{\bra{\alpha}A\ket{\alpha}}{\bra{\alpha}A^{*}\ket{\alpha}} \geq 0, \label{Iz}
\end{equation}
where 
\[
A_{\mu\nu}
=
\langle \hat{O}_\mu,\hat{O}_\nu\rangle,
\qquad
\bm{\hat{O}}
=
\left(
\hat{X},\,
\hat{Y},\,
P_1,\ldots,P_{N_{\rm orb}}
\right),
\]
or schematically,
\[
A =
\begin{pmatrix}
\langle X,X\rangle
&
\langle X,Y\rangle
&
\langle X,P_b\rangle
\\[2pt]
\langle Y,X\rangle
&
\langle Y,Y\rangle
&
\langle Y,P_b\rangle
\\[2pt]
\langle P_a,X\rangle
&
\langle P_a,Y\rangle
&
\langle P_a,P_b\rangle
\end{pmatrix},
\qquad a,b=1,\ldots,N_{\rm orb}.
\]

Notice that the matrices $A$ and $A^{*}$ are positive-semidefinite and that $\ket{\alpha_0}=(0,0,1,1,1,\dots)^{T}$ is always a trivial zero mode of both matrices: it leads to a vortex operator \eqref{vortexoperator} equal to $\hat{Z} = 1$, which imparts no vorticity. To avoid indeterminacy in \eqref{Iz}, Ref.~\cite{SM:PhysRevB.110.245112} regulates the vorticity index as
\begin{equation}
    I_{\epsilon}(Z) = \frac{\bra{\alpha}(A+\epsilon 1)\ket{\alpha}}{\bra{\alpha}A^{*}+\epsilon 1\ket{\alpha}}
\end{equation}
and calculates it in the $\epsilon \to 0$ limit. Since $A_{\epsilon}^{*} \equiv A^{*}+\epsilon 1$ is positive-semidefinite, we can decompose it as $A^{*}_{\epsilon} = L_{\epsilon}L_{\epsilon}^{\dagger}$ with $L_{\epsilon}$ being a lower triangular matrix. Thus, the index can be rewritten as
\begin{align}
    I_{\epsilon}(Z) &= \frac{\bra{\alpha}A_\epsilon\ket{\alpha}}{\bra{\alpha}L_{\epsilon}L_{\epsilon}^{\dagger}\ket{\alpha}} \nonumber \\
    &= \frac{\bra{\alpha}L_{\epsilon}L_{\epsilon}^{-1}A_\epsilon(L_{\epsilon}^{\dagger})^{-1}L_{\epsilon}^{\dagger}\ket{\alpha}}{\bra{\alpha}L_{\epsilon}L_{\epsilon}^{\dagger}\ket{\alpha}} \nonumber \\
    &\equiv \frac{\bra{v}M_\epsilon\ket{v}}{\langle{v}\ket{v}}
\end{align}
where
\begin{align}
    M_{\epsilon} &= L_{\epsilon}^{-1}A_\epsilon(L_{\epsilon}^{\dagger})^{-1}
\end{align}
and 
\begin{equation}
    \ket{v} = L_{\epsilon}^{\dagger}\ket{\alpha}.
\end{equation}
Therefore, the minimum of $I_{\epsilon}(Z)$ is given by the smallest eigenvalue of $M_{\epsilon}$. From the corresponding eigenvector $\ket{v}$, we can find the optimal set of parameters $\ket{\alpha}=(\alpha,\beta,\gamma_{1},\gamma_2,\dots)$ as
\begin{equation}
    \ket{\alpha}=(L_{\epsilon}^{\dagger})^{-1}\ket{v}.
\end{equation}

\begin{figure}
    \centering
    \includegraphics[width=0.65\linewidth]{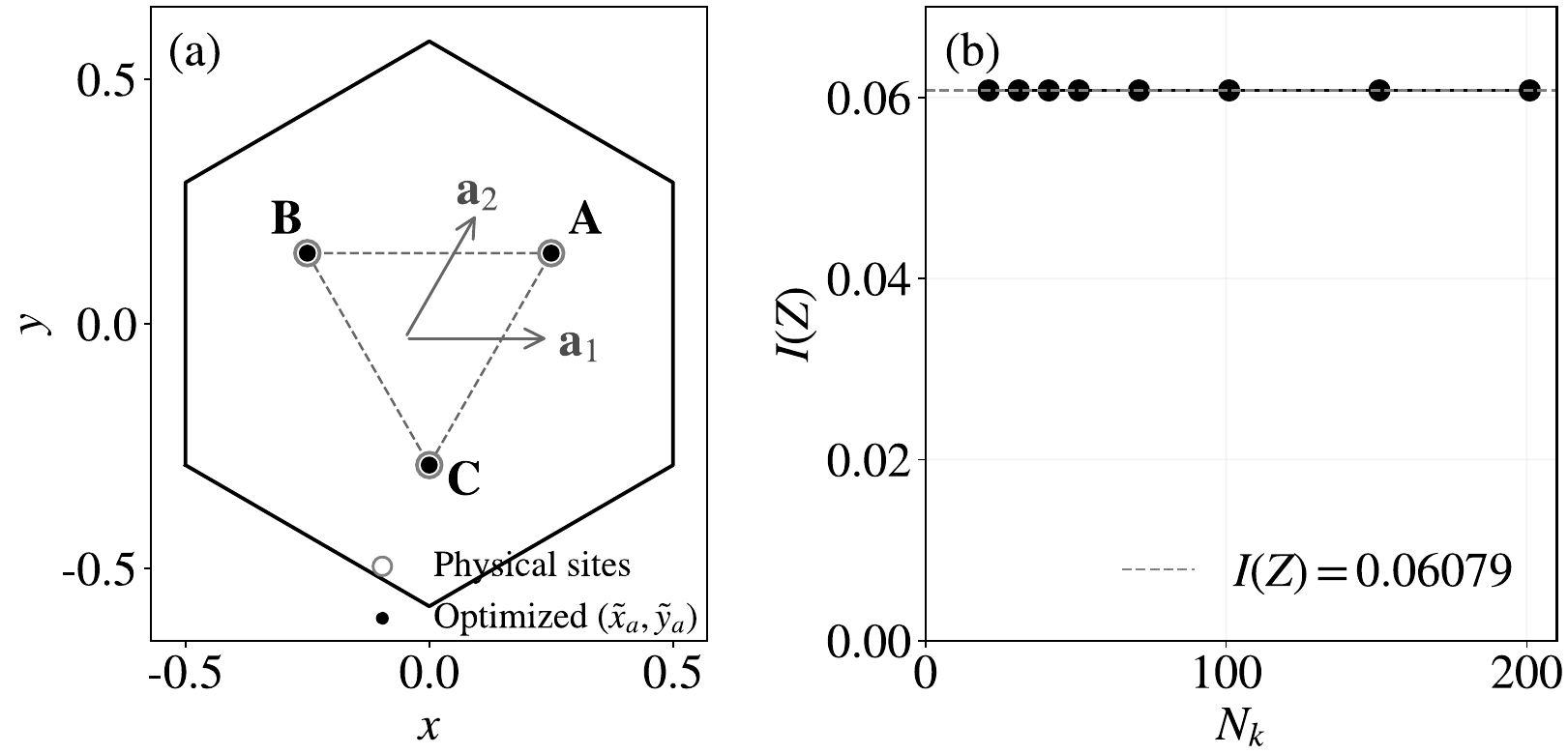}
    \caption{\textbf{Vortexability test of the kagome Chern band}. (a) Orbital positions for the optimal vortex function. (b) Scaling of the vortexability index of the optimal vortex function.}
    \label{FigVortexability}
\end{figure}

Numerically, it is easier to express the matrix A in the momentum representation. Let $\mathcal{O}$ represent the occupied (Chern) band and $\mathcal{U}$ represent the unnocupied bands, such that
\begin{align}
    \hat{P} &= \sum_{\bm{k},\mathcal{O}} \ket{u_{\bm{k},\mathcal{O}}}\bra{u_{\bm{k},\mathcal{O}}} \otimes\ket{\bm{k}}\bra{\bm{k}}, \nonumber \\
    \hat{Q} &= 1 - \hat{P} = \sum_{\bm{k},\mathcal{U}} \ket{u_{\bm{k},\mathcal{U}}}\bra{u_{\bm{k},\mathcal{U}}}\otimes\ket{\bm{k}}\bra{\bm{k}}.
\end{align}
Now, let $\hat{R}_1=\hat{X}$ and $\hat{R}_2=\hat{Y}$. Reference~\cite{SM:PhysRevB.110.245112} derives the expressions
\begin{align}
    \langle R_{i},R_j \rangle &= \sum_{\mathcal{O},\mathcal{U}}\int\frac{d^2k}{(2\pi)^2}\langle\partial_{k_i}u_{\bm{k},\mathcal{O}}|u_{\bm{k},\mathcal{U}}\rangle \langle u_{\bm{k},\mathcal{O}}|\partial_{k_j}u_{\bm{k},\mathcal{U}}\rangle, \nonumber \\
    \langle P_a, R_{i} \rangle &= \sum_{\mathcal{O},\mathcal{U}}\int\frac{d^2k}{(2\pi)^2}u^{*}_{\bm{k},\mathcal{O}}(a)u_{\bm{k},\mathcal{U}}(a) \langle u_{\bm{k},\mathcal{O}}|(-\mathrm{i}\partial_{k_i})|u_{\bm{k},\mathcal{U}}\rangle, \nonumber \\
    \langle P_{a},P_b \rangle &= \sum_{\mathcal{O}}\int\frac{d^2k}{(2\pi)^2}|u_{\bm{k},\mathcal{O}}(a)|^2[\delta_{ab}-|u_{\bm{k},\mathcal{O}}(b)|^2].
\end{align}
Using the expressions above to construct $A$ and $M_{\epsilon}$, we converge to a finite index $I(z)=0.06079$ in the thermodynamic limit, showing that our kagome Chern band is not vortexable according to the method of Ref.~\cite{SM:PhysRevB.110.245112}. Our optimal parameters were $\alpha=-1$, $\beta=\mathrm{i}$, and $\gamma_{1},\gamma_2,\gamma_3$ such that the optimal orbital coordinates were
\begin{align}
    (\tilde{x}_{A},\tilde{y}_A) &= (0.2500,0.1443) \approx \left(\frac{1}{4},\frac{1}{4\sqrt{3}}\right), \nonumber \\
    (\tilde{x}_{B},\tilde{y}_B) &= (-0.2500,0.1443) \approx \left(-\frac{1}{4},\frac{1}{4\sqrt{3}}\right), \nonumber \\
    (\tilde{x}_{C},\tilde{y}_C) &= (0.000,-0.2889) \approx \left(0,-\frac{1}{2\sqrt{3}}\right),
\end{align}
which coincide with the orbital positions of a kagome lattice, as we can see in Fig.~\ref{FigVortexability}.

\subsection{Optimization of the circular spread}\label{sec:spread}

We optimize localization within the anyon band using
normalized superpositions of its orthonormal energy
eigenstates $\ket{\Psi(\bm K)}$,
\begin{equation}
    \ket{\psi(c)}
    =
    \frac{1}{\sqrt{c^\dagger c}}
    \sum_{\bm K}c_{\bm K}\ket{\Psi(\bm K)},
\end{equation}
where $c$ is a vector containing all components $c_{\bm{K}}$. In this basis, we evaluate the expectation values
\begin{equation}
    f_j(c)
    =
    \frac{c^\dagger\tilde{\rho}_j c}{c^\dagger c},
\end{equation}
where $\tilde{\rho}_j$ denotes the matrix of the
projected-density operator associated with direction
$j$. We minimize the circular spread~\cite{SM:PhysRevLett.82.370}
\begin{equation}
    \mathcal{S}(c)
    =
    -\sum_j \log|f_j(c)|^2,
\end{equation}
Its analytical
gradient is
\begin{equation}
    \frac{\partial\mathcal{S}}{\partial c^*}
    =
    -\frac{1}{c^\dagger c}
    \sum_j
    \left[
        \frac{\tilde{\rho}_j}{f_j}
        +\frac{\tilde{\rho}_j^\dagger}{f_j^*}
        -2\mathbb{I}
    \right]c.
    \label{eq:spread_gradient}
\end{equation}
Writing $c=x+\mathrm{i}y$, the gradients supplied to
the optimizer are
$\nabla_x\mathcal{S}
=2\operatorname{Re}(\partial\mathcal{S}/\partial c^*)$
and
$\nabla_y\mathcal{S}
=2\operatorname{Im}(\partial\mathcal{S}/\partial c^*)$.

We initialize each realization with independent
$x_{\bm K},y_{\bm K}\sim\mathcal{N}(0,1)$ and normalize
$c$, producing a random state within the anyon
subspace. We then use \texttt{scipy.optimize.minimize}
with the \texttt{L-BFGS-B} method\cite{SM:byrd1995limited} to minimize the spread.
Starting from the initial guess, the algorithm evaluates
$\mathcal{S}$ and its gradient and uses them, together
with information from previous steps, to generate the
next guess. These evaluations and updates are repeated
until convergence. Since the objective depends only
on the magnitudes of the projected-density expectation
values, it imposes no prescribed localization center.

\section{Anyon parallel transporter, holonomy, and Chern number}\label{AppendixBerryConnection}

Anyon Bloch states belonging to distinct many-body momentum sectors are orthogonal, so their bare overlaps cannot furnish a parallel transporter in momentum space. Projected density operators provide the physical map between neighboring momentum sectors, and their matrix elements therefore define the parallel transport, holonomies, and the anyon Chern number.

\subsection{Parallel transporter}

A central object in our discussion is the orthogonal projector onto the band spanned by the anyon Bloch states $\ket{\Psi(\bm{K})}$,
\begin{equation}
    P = \sum_{\bm{K}} \ket{\Psi(\bm{K})}\bra{\Psi(\bm{K})}
\end{equation}
where
\begin{align}
    \bm{K}&=n_1\bm{q}_1 + n_2 \bm{q}_2,  \label{BZDefinition} \\
    \bm{q}_{j}&=\frac{\bm{b}_{j}}{N_j}, \quad n_{j} \in \{0,1,\dots,N_j-1\} \nonumber
\end{align}
are the points in the Brillouin zone (BZ), which represent the center-of-mass crystal momenta of the many-body wavefunctions in our Hilbert space $\mathcal{H}$. During calculations, it is convenient to work with the half-projector $V : \mathbb{C}^{N_{\mathrm{u.c.}}} \to \mathcal{H}$, defined as
\begin{equation}
    V \equiv \sum_{\bm{K}} \ket{\Psi(\bm{K})}\bra{\bm{K}}, \label{HalfProj}
\end{equation}
where $\{\ket{\bm{K}}\}$ is an orthonormal basis of the auxiliary space $\mathbb{C}^{N_{\mathrm{u.c.}}}$.

The other central objects in our discussion are the $\bm{q}_{j}$-Fourier components of the normalized density operator,
\begin{equation}
    \rho_{\bm{q}_{j}} = \frac{1}{N_e}\sum_{\bm{R},s} e^{-\mathrm{i}\bm{q}_{j}\cdot(\bm{R}+\bm{r}_s)}n_{\bm{R},s} = \frac{1}{N_e}\sum_{\bm{K},s}e^{-\mathrm{i}\bm{q}_{j}\cdot\bm{r}_{s}}c^{\dagger}_{\bm{K},s}c_{\bm{K}+\bm{q}_{j},s}, \label{Doperator}
\end{equation}
where
\begin{equation}
    c_{\bm{R},s} = \frac{1}{\sqrt{N_{\mathrm{u.c.}}}}\sum_{\bm{K}}e^{\mathrm{i}\bm{K}\cdot\bm{R}}c_{\bm{K},s},
\end{equation}
and $s \in \{1,\dots,N_{\mathrm{orb}}\}$ is the orbital index. Using the half-projectors, we can express the density operator projected onto the anyon band as
\begin{align}
    \Tilde{\rho}_{\bm{q}_{j}} &= V^{\dagger}\rho_{\bm{q}_{j}}V \nonumber \\
    &= \sum_{\bm{K},\bm{K}'} [\Tilde{\rho}_{\bm{q}_j}]_{\bm{K}\bm{K}'}\ket{\bm{K}}\bra{\bm{K}'}.
\end{align}
Notice that the above object acts on the auxiliary space $\mathbb{C}^{N_{\mathrm{u.c.}}}$ of dimension $N_{\mathrm{u.c.}}=N_{1}N_{2}$. Its matrix elements are
\begin{equation}
    [\Tilde{\rho}_{\bm{q}_j}]_{\bm{K}\bm{K}'} = \bra{\Psi(\bm{K})}\rho_{\bm{q}_{j}}\ket{\Psi(\bm{K}')}. \label{matrixelements}
\end{equation}

Let $T_{j}$ be the translator by one unit cell along $\bm{a}_{j}$, so $T_{j}\hat{n}_{\bm{R},s}T_{j}^{-1} = \hat{n}_{\bm{R}+\bm{a}_{j},s}$. The density operator \eqref{Doperator} transforms as
\begin{align}
    T_{j}\rho_{\bm{q}_{j}}T_{j}^{-1} &= \frac{1}{N_e}\sum_{\bm{R},s}e^{-\mathrm{i}\bm{q}_{j}\cdot (\bm{R}+\bm{r}_s)}T_{j}\hat{n}_{\bm{R},s}T_{j}^{-1} \nonumber \\
    &=\frac{1}{N_e} \sum_{\bm{R},s}e^{-\mathrm{i}\bm{q}_{j}\cdot (\bm{R}+\bm{r}_s)}\hat{n}_{\bm{R}+\bm{a}_{j},s} \nonumber \\
    &= e^{\mathrm{i}\bm{q}_{j}\cdot\bm{a}_{j}}\rho_{\bm{q}_{j}}.
\end{align}
Meanwhile, the action of the translator on an anyon Bloch state gives
\begin{equation}
    T_{j}\ket{\Psi(\bm{K})} = e^{-\mathrm{i}\bm{K}\cdot\bm{a}_{j}}\ket{\Psi(\bm{K})}.
\end{equation}
Therefore, the matrix elements \eqref{matrixelements} become
\begin{align}
    [\Tilde{\rho}_{\bm{q}_{j}}]_{\bm{K}\bm{K}'} &= \bra{\Psi(\bm{K})}T^{-1}_{j}T_{j}\rho_{\bm{q}_{j}}T^{-1}_{j}T_{j}\ket{\Psi(\bm{K}')} \nonumber \\
    &= e^{\mathrm{i}(\bm{K}-\bm{K}'+\bm{q}_{j})\cdot\bm{a}_{j}}\bra{\Psi(\bm{K})}\rho_{\bm{q}_{j}}\ket{\Psi(\bm{K}')} \nonumber \\
    &=e^{\mathrm{i}(\bm{K}-\bm{K}'+\bm{q}_{j})\cdot\bm{a}_{j}}[\Tilde{\rho}_{\bm{q}_{j}}]_{\bm{K}\bm{K}'}
\end{align}
The equality above can only be true if
\begin{equation}
    \bm{K}' = \bm{K}+\bm{q}_{j},
\end{equation}
modulo a reciprocal lattice vector. Thus, the reduced position operator projected onto the anyon band reads
\begin{equation}
    \tilde{\rho}_{\bm{q}_{j}} = \sum_{\bm{k}}t_{j}(\bm{K})\ket{\bm{K}}\bra{\bm{K}+\bm{q}_{j}}, \label{reducedbasis}
\end{equation}
where
\begin{equation}
    t_{j}(\bm{K}) = \bra{\Psi(\bm{K})}\rho_{\bm{q}_{j}}\ket{\Psi(\bm{K}+\bm{q}_{j})}. \label{MatrixElement}
\end{equation}
The expression \eqref{reducedbasis} describes momentum translators of anyon Bloch states and thus possesses the essential characteristics of a parallel transporter along the anyon fiber bundle. In particular, the fact that $t_{j}(\bm{K}) \neq 0$ for all $\bm{K}$ implies that $\tilde{\rho}_{j}$ has full rank in the auziliary subspace $\mathbb{C}^{N_{\phi}}$.

\subsection{Holonomy and Chern number}

Acting with \eqref{reducedbasis} on $\ket{\bm{K}_0}$ yields
\begin{equation}
\tilde{\rho}_{\bm{q}_{j}}\ket{\bm{K}_0} = t_{j}(\bm{K}_0)\ket{\bm{K}_0-\bm{q}_{j}}.
\end{equation}
Therefore, acting $N_{j}$ times, where $N_{j}$ is the number of unit cells along $\bm{a}_{j}$, transports $\ket{\bm{k}_0}$ along a closed path in the $j$ direction,
\begin{align}
    (\tilde{\rho}_{\bm{q}_{j}})^{N_{j}}\ket{\bm{K}_0} &= t_{j}(\bm{K}_0)\dots t_{j}(\bm{K}_0-\bm{b}_{j}+\bm{q}_{j})\ket{\bm{K}_0-\bm{b}_{j}} \nonumber \\
    &= t_{j}(\bm{K}_0)\dots t_{j}(\bm{K}_0-\bm{b}_{j}+\bm{q}_{j})\ket{\bm{K}_0},
\end{align}
where $\bm{b}_{j}$ is a reciprocal lattice vector. Therefore, the quantity
\begin{equation}
    W_{j}(\bm{K}_0) \equiv \prod_{\bm{K}=\bm{K}_0-\bm{b}_{j}+\bm{q}_{j}}^{\bm{k}_0}t_{j}(\bm{K}) 
\end{equation}
is a holonomy of the anyon band along $\bm{q}_{j}$ at base point $\bm{k}_0$. For a generic closed path $\mathcal{C}$, we have
\begin{equation}
    W[\mathcal{C}] = \prod_{(\bm{k},j) \in \mathcal{C}}t_{j}(\bm{K}), \label{Holonomy}
\end{equation}
which, for the case of a single anyon band, is a complex number.

To calculate the Chern number, we consider the unitary part of the parallel transport along
$\bm q_{j}$,
\begin{equation}
    e^{\mathrm{i}\theta_{j}(\bm{K})}
    \equiv
    \frac{
        t_{j}(\bm{K})
    }{
        |t_{j}(\bm{K})|
    }, \quad -\pi < \theta_{j}(\bm{K}) \leq \pi.
    \label{UnitaryLinks}
\end{equation}
Using these links, we define the Wilson loop around the
plaquette based at $\bm{K}$ as
\begin{equation}
\begin{split}
    W_{12}(\bm{k})
    ={}&
    e^{\mathrm{i}\theta_{1}(\bm{K})}e^{\mathrm{i}\theta_{2}(\bm{K}+\bm{q}_1)}e^{-\mathrm{i}\theta_{1}(\bm{K}+\bm{q}_2)}e^{-\mathrm{i}\theta_{2}(\bm{K})} \label{FHSWilsonLoop}
\end{split}
\end{equation}
The
discrete Berry flux through the plaquette is defined as~\cite{SM:fukui2005chern}
\begin{equation}
    \theta(\bm{K})
    =
    \operatorname{Arg}
    W_{12}(\bm{K}),
    \qquad
    -\pi<\theta(\bm{K})\leq\pi.
    \label{DiscreteBerryFlux}
\end{equation}
The Chern number is then
\begin{equation}
    C_{\mathrm{anyon}}
    =
    -\frac{1}{2\pi}
    \sum_{\bm{K}}
    \theta(\bm{K}).
    \label{AnyonBandChernNumber}
\end{equation}
Using the states $\ket{\Psi(\bm{K})}$ directly obtained from the exact diagonalization of the interacting kagome Hamiltonian for a $5 \times 5$ lattice with aspect ratio 1 and $N_{e}=8$ electrons, we obtained $C_{\mathrm{anyon}}=3$ to machine precision.

\subsection{Analytic calculation of the anyon Chern number in the FQH system}

Let $c_{j}^{\dagger}$ be the creation operator of an electron in the $j$-th Landau orbital of the LLL, where $j \in \{0,1,\dots,N_\phi-1\}$. Our anyon ``Bloch states'' $\ket{\Psi(K)}$ are eigenstates of the Hamiltonian \eqref{PseudoPotential} in the $N_e$-electron sector. Their many-body wavefunctions have the form
\begin{equation}
    \ket{\Psi(K)} = \sum_{\{j_n\}} \phi(j_1,\dots,j_{N_e})c^{\dagger}_{j_1}\cdots c^{\dagger}_{j_{N_e}}\ket{0},
\end{equation}
where we adopt the sum notation
\begin{equation}
    \sum_{\{j_n\}}[\cdots] \equiv \sum_{\substack{
    j_1<\dots<j_{N_e}\\
    \sum_nj_n= K \ \text{(mod } N_{\phi}), 
}} [\cdots]
\end{equation}
with $K \equiv \sum_{n}j_{n} \ \text{(mod } N_{\phi}) $ being the magnetic momentum quantum number the number $N_{\phi}$ of magnetic unit cells is given by $pN_{\phi} -qN_e=s$ with $s = -1$ $(+1)$ for a quasielectron (quasihole). The periodic density operators projected onto the anyon band are defined in \eqref{FQHEdensityoperators}. We can represent them, for $a\in\{x,y\}$, in the magnetic momentum basis as
\begin{equation}
    \tilde{\rho}_{a} = \sum_{K,K'} [\tilde{\rho}_{a}]_{KK'}\ket{K}\bra{K'}, \label{umatrixelements}
\end{equation}
where the matrix elements can be directly evaluated from the anyon Bloch states $\ket{\Psi(K)}$ and the unprojected periodic density operator $\rho_{a}$ as
\begin{equation}
    [\tilde{\rho}_{a}]_{KK'} = \bra{\Psi(K)}\rho_{a}\ket{\Psi(K')}. \label{Overlap}
\end{equation}

To determine the form of \eqref{Overlap}, we use the fact that the Hamiltonian \eqref{PseudoPotential} is invariant under magnetic translations. First, consider the magnetic momentum operator and the magnetic translator in the $y$ direction as
\begin{equation}
    \hat{K} = \sum_{j}j \ \hat{n}_{j}, \quad \hat{T}_y = e^{\mathrm{i}\frac{2\pi}{N_{\phi}}\hat{K}}.
\end{equation}
Using
\begin{equation}
    [\hat{K},c_{j}^{\dagger}] = jc_{j}^{\dagger}, \quad [\hat{K},c_{j}] = -jc_{j},
\end{equation}
We get
\begin{equation}
    \hat{T}_yc_{j}^{\dagger}\hat{T}_y^{\dagger} = \omega^{j}c^{\dagger}_j, \quad \hat{T}_yc_{j}\hat{T}_y^{\dagger} = \omega^{-j}c_{j}, \label{Ty}
\end{equation}
where
\begin{equation}
    \omega \equiv e^{\mathrm{i}2\pi/N_{\phi}}.
\end{equation}

Therefore, using the fact that $\hat{T}_y\ket{0}=\ket{0}$, we have
\begin{align}
    \hat{T}_{y}\ket{\Psi(K)} &= \sum_{\{j_n\}}\phi(\{j_n\})\hat{T}_yc_{j_1}^{\dagger}\hat{T}_{y}^{\dagger}\cdots\hat{T}_{y}c_{j_{N_{e}}}\hat{T}_{y}^{\dagger}\hat{T}_{y}\ket{0} \nonumber \\
    &=\sum_{\{j_n\}}\phi(\{j_n\})\omega^{j_1+\cdots+j_{N_e}}c^{\dagger}_{j_1}\cdots c^{\dagger}_{j_{N_e}} \nonumber \\
    &= \omega^{K}\ket{\Psi(K)}. \label{Intermediate1}
\end{align}
Moreover,
\begin{align}
    [\hat{K},\rho_x] = \sum_{j,\ell}j \ w^{\ell}[\hat{n}_j,\hat{n}_\ell] = 0,
\end{align}
so $\hat{T}_{y}\rho_x\hat{T}_{y}^{\dagger}=\rho_x$ and
\begin{align}
    \bra{\Psi(K)}\rho_{x}\ket{\Psi(K')} &= \bra{\Psi(K)}\hat{T}_{y}^{\dagger}\hat{T}_{y}\rho_{x}\hat{T}_{y}^{\dagger}\hat{T}_{y}\ket{\Psi(K')} \nonumber \\
    &= \omega^{K'-K}\bra{\Psi(K)}\rho_{x}\ket{\Psi(K')}
\end{align}
The last equality is only possible if $K'=K \ (\text{mod } N_{\phi}) $. Therefore,
\begin{equation}
   \tilde{\rho}_x = \sum_{K}t_{x}(K)\ket{K}\bra{K}, \quad t_{x}(K) \equiv \bra{\Psi(K)}\rho_{x}\ket{\Psi(K)}. \label{Ux}
\end{equation}

Now, consider the magnetic translators in the $x$ direction, defined as
\begin{equation}
    \hat{T}_{x}c_{j}^{\dagger}\hat{T}_{x}=c^{\dagger}_{j+1}, \quad \hat{T}_{x}c_{j}\hat{T}_{x}=c_{j+1}.
\end{equation}
Acting with this translator on a Bloch anyon and noting that $\hat{T}_{y}\ket{0}=\ket{0}$ yields
\begin{align}
    \hat{T}_{x}\ket{\Psi(K)} &= \hat{T}_x \sum_{\substack{
    j_1<\dots<j_{N_e}\\
    \sum_nj_n= K \ \text{(mod } N_{\phi})}}\phi(\{j_n\})c^{\dagger}_{j_1}\cdots c^{\dagger}_{j_{N_e}}\ket{0} \nonumber \\
    &=\sum_{\substack{
    j_1<\dots<j_{N_e}\\
    \sum_nj_n= K \ \text{(mod } N_{\phi})}}\phi(\{j_n\})c^{\dagger}_{j_1+1}\cdots c^{\dagger}_{j_{N_e+1}}\ket{0}.
\end{align}
Defining $j'_{n}=j_{n}+1$, $\tilde{\phi}(j_{1}',\dots,j'_{N_e})=\phi(j'_{1}-1,\dots,j'_{N_e}-1)$, and noting that $j'_{1}+\dots j'_{N_e} = K+N_{e} \ (\text{mod } N_{\phi})$, we can rewrite the above expression as
\begin{align}
    \hat{T}_{x}\ket{\Psi(K)} &= \sum_{\substack{
    j'_1<\dots<j'_{N_e}\\
    \sum_nj'_n= K + N_{e} \ \text{(mod } N_{\phi})}}\tilde{\phi}(j'_{1},\dots,j'_{N_{e}})c^{\dagger}_{j'_1}\cdots c^{\dagger}_{j'_{N_e}}\ket{0} \nonumber \\
    &\equiv \ket{\tilde{\Psi}(K+N_{e})},
\end{align}
where $\ket{\tilde{\Psi}(K+N_{e})}$ is a vector with magnetic momentum $K+N_{e}$. Since the projector onto the anyon band is invariant under magnetic translations $\hat{T}_y$, the vector $\ket{\tilde{\Psi}(K+N_{e})}$ lies in the anyon band. Given that the anyon band only has one state per magnetic momentum sector, we must conclude that $\ket{\tilde{\Psi}(K+N_{e})}$ is equal to the anyon Bloch state in the $K+N_{e}$ sector in our original basis up to a phase factor,
\begin{equation}
    \hat{T}_{x}\ket{\Psi(K)} = e^{\mathrm{i}\alpha(K)}\ket{\Psi(K+N_{e})}. \label{IntermediatePsi}
\end{equation}

We can also show
\begin{align}
    \hat{T}_{x}\rho_{x}\hat{T}^{\dagger}_{x} &= \sum_{j}w^{j}\hat{T}_{x}\hat{n}_{j}\hat{T}^{\dagger}_{x} \nonumber \\
    &=\sum_{j}w^{j}\hat{n}_{j+1} \nonumber \\
    &= \sum_{j}w^{j-1}\hat{n}_{j} \nonumber \\
    &= w^{-1}\rho_{x}.
\end{align}

Therefore, using \eqref{IntermediatePsi}, we have
\begin{align}
    t_{x}(K) &= \bra{\Psi(K)}\rho_{x}\ket{\Psi(K)} \nonumber \\
    &= \bra{\Psi(K)}\hat{T}^{\dagger}_{x}\hat{T}_{x}\rho_{x}\hat{T}_{x}^{\dagger}\hat{T}_{x}\ket{\Psi(K)} \nonumber \\
    &= \bra{\Psi(K+N_{e})}e^{-\mathrm{i}\alpha(K)}w^{-1}\rho_{x}e^{\mathrm{i}\alpha(K)}\ket{\Psi(K+N_e)} \nonumber \\
    &= e^{-\mathrm{i}\frac{2\pi}{N_{\phi}}}t_{x}(K+N_{e}).
\end{align}

By iterating this translation of $t_{x}(K)$ $r$ times, we obtain
\begin{equation}
    t_{x}(K+rN_{e}) = e^{\mathrm{i}\frac{2\pi r}{N_{\phi}}}t_{x}(K), \quad r \in \mathbb{Z}. \label{GeneralTranslation}
\end{equation}
For the quasihole sector, we have $pN_{\phi}-qN_{e}=1$. Therefore, $-pN_{e} = 1 - N_{\phi}$, and
\begin{equation}
    -qN_{e} = 1 \ (\text{mod } N_{\phi}).
\end{equation}
Since $K$ is only defined modulo $N_{\phi}$, we can take $r=-q$ in \eqref{GeneralTranslation} and write
\begin{align}
    t^{\mathrm{qh}}_{x}(K+1)&= t_{x}^{\mathrm{qh}}(K-qN_{e})  \nonumber \\
    &= e^{-\mathrm{i}\frac{2\pi q}{N_{\phi}}}t_{x}^{\mathrm{qh}}(K). \label{TranslationQuasihole}
\end{align}
For the quasielectron sector, we have $pN_{\phi}-qN_{e}=-1$ and thus $qN_{e} = 1 \ (\mathrm{mod} \ N_{\phi})$, which gives us
\begin{equation}
    t^{\mathrm{qe}}_{x}(K+1)=e^{\mathrm{i}\frac{2\pi q}{N_{\phi}}}t_{x}^{\mathrm{qe}}(K). \label{TranslationQuasielectron}
\end{equation}

Now, let us investigate the matrix elements of the parallel transport \eqref{umatrixelements} in the $y$ direction. The definitions \eqref{FQHEdensityoperators} and the identities \eqref{Ty} imply
\begin{align}
    \hat{T}_{y}\rho_{y}\hat{T}_{y}^{\dagger} &= \sum_{j}\hat{T}_{y}c^{\dagger}_{j+1}\hat{T}^{\dagger}_{y}\hat{T}_{y}c_{j}\hat{T}_{y}^{\dagger} \nonumber \\
    &=\sum_{j}\omega^{j+1}\omega^{-j}c_{j+1}^{\dagger}c_{j} \nonumber \\
    &= e^{\mathrm{i}\frac{2\pi}{N_{\phi}}}\rho_{y}.
\end{align}
Using \eqref{Intermediate1}, we have 
\begin{align}
    \bra{\Psi(K)}\rho_y\ket{\Psi(K')} &= \bra{\Psi(K)}\hat{T}_{y}^{\dagger}\hat{T}_{y}\rho_y\hat{T}_{y}^{\dagger}\hat{T}_{y}\ket{\Psi(K')} \nonumber \\
    &=\exp\left[\mathrm{i}\frac{2\pi}{N_{\phi}}(K'-K+1)\right]\bra{\Psi(K)}\rho_y\ket{\Psi(K')}
\end{align}
This can only hold if
\begin{equation}
    K'=K-1 \ (\text{mod} \ N_{\phi}).
\end{equation}
Therefore,
\begin{equation}
    \tilde{\rho}_{y} = \sum_{K}t_{y}(K)\ket{K+1}\bra{K}, \label{Uy}
\end{equation}
where
\begin{equation}
    t_{y}(K) = \bra{\Psi(K+1)}\rho_{y}\ket{\Psi(K)}.
\end{equation}

The results \eqref{Ux} and \eqref{Uy} imply
\begin{align}
    \tilde{\rho}_x\tilde{\rho}_{y}\ket{K} &= t_{y}(K)\tilde{\rho}_{x}\ket{K+1} = t_{x}(K+1)t_{y}(K)\ket{K+1}, \\
    \tilde{\rho}_y\tilde{\rho}_{x}\ket{K} &= t_{x}(K)\tilde{\rho}_{y}\ket{K} = t_{x}(K)t_{y}(K)\ket{K+1}.
\end{align}
Therefore, we have
\begin{equation}
    \tilde{\rho}_x\tilde{\rho}_{y}\ket{K} = \frac{t_x(K+1)}{t_{x}(K)}\tilde{\rho}_y\tilde{\rho}_{x}\ket{K}.
\end{equation}
But the results \eqref{TranslationQuasihole} and \eqref{TranslationQuasielectron} imply
\begin{equation}
    \tilde{\rho}_x\tilde{\rho}_y\ket{K} = e^{\mp \mathrm{i}2\pi q/N_{\phi}}\tilde{\rho}_y\tilde{\rho}_{x}\ket{K},
\end{equation}
where $-$ $(+)$ corresponds to quasiholes (quasielectrons). This Weyl algebra emerges on the anyon manifold. To each magnetic unit cell, we associate an elementary Wilson plaquette,
\begin{equation}
    W = (\tilde{\rho}_x\tilde{\rho}_y)^{-1}(\tilde{\rho}_{y}\tilde{\rho}_x)
\end{equation}
The flux picked up by each plaquette is
\begin{align}
    \mathrm{Im}\bra{K} \log W \ket{K} = \pm\frac{2\pi q}{N_{\phi}}.
\end{align}
The sum of the contributions of all plaquettes is the anyon Chern number,
\begin{align}
    C_{\mathrm{anyon}} &= \frac{1}{2\pi} \mathrm{Im} \  \mathrm{Tr} \log W \nonumber \\
    &= \frac{1}{2\pi}\mathrm{Im}\sum_{K=0}^{N_{\phi}-1}\bra{K} \log W \ket{K} \nonumber \\
    &= \pm q,
\end{align}
where $+$ $(-)$ corresponds to the Chern number of a quasihole (quasielectron).

\section{Gauge-invariant vorticity of the coherent-state coefficients}\label{sec:vorticity}

Under transformations of the anyon Bloch gauge $\ket{\Psi(\bm{K})} \to e^{\mathrm{i}\chi_{\bm{K}}} \ket{\Psi(\bm{K})}$, the coefficients in Eq.~\eqref{CoherentBloch} must transform as $a_{\bm r}(\bm K) \to e^{-\mathrm{i}\chi_{\bm{K}}}a_{\bm r}(\bm K)$ to preserve the coherent state, while the parallel transport links~\eqref{projecteddensityMain} change as $[\tilde{\rho}_{j}]_{\bm{K}\bm{K}'} \to e^{\mathrm{i}(\chi_{\bm{K}'}-\chi_{\bm{K}})}[\tilde{\rho}_{j}]_{\bm{K}\bm{K}'}$. Thus, for neighboring points $\bm{K}$ and $\bm{K}'=\bm{K}+\bm{q}_{j}$ in the BZ, we can define the gauge-invariant phases $\alpha_{\bm{r}}({\bm{K},\bm{K}'})=\mathrm{Arg}\{a_{\bm{r}}(\bm{K})^{*}[\tilde{\rho}_{j}]_{\bm{K}\bm{K}'}a_{\bm{r}}(\bm{K}')\}$. Summed counterclockwise along the boundary of a region $R$ of the discretized BZ, these phases yield a total circulation
\begin{equation}
    \sum_{(\bm{K} \to \bm{K}') \in \partial R} \alpha_{\bm{r}}({\bm{K},\bm{K}'}) = 2\pi n_R + \Phi_R,
\end{equation}
where $\Phi_R=\sum_{\bm{K}\in R}\theta_{\bm{K}}$ is the nonquantized Berry flux [Eq.~\eqref{PlaquettePhases}] through $R$, coming from the $[\tilde{\rho}_{j}]_{\bm{K}\bm{K}'}$ factors, and $n_R$ is the vorticity associated with the coefficients $a_{\bm{r}}(\bm{K})$ inside $R$~\cite{SM:fukui2005chern}. For a counterclockwise loop around the highlighted region in Fig.~\ref{fig:Beautiful}(d), we obtain $n_{\mathrm{loop}}=3=C_{\mathrm{anyon}}$.


\begin{thebibliography}{86}%
\makeatletter
\providecommand \@ifxundefined [1]{%
 \@ifx{#1\undefined}
}%
\providecommand \@ifnum [1]{%
 \ifnum #1\expandafter \@firstoftwo
 \else \expandafter \@secondoftwo
 \fi
}%
\providecommand \@ifx [1]{%
 \ifx #1\expandafter \@firstoftwo
 \else \expandafter \@secondoftwo
 \fi
}%
\providecommand \natexlab [1]{#1}%
\providecommand \enquote  [1]{``#1''}%
\providecommand \bibnamefont  [1]{#1}%
\providecommand \bibfnamefont [1]{#1}%
\providecommand \citenamefont [1]{#1}%
\providecommand \href@noop [0]{\@secondoftwo}%
\providecommand \href [0]{\begingroup \@sanitize@url \@href}%
\providecommand \@href[1]{\@@startlink{#1}\@@href}%
\providecommand \@@href[1]{\endgroup#1\@@endlink}%
\providecommand \@sanitize@url [0]{\catcode `\\12\catcode `\$12\catcode
  `\&12\catcode `\#12\catcode `\^12\catcode `\_12\catcode `\%12\relax}%
\providecommand \@@startlink[1]{}%
\providecommand \@@endlink[0]{}%
\providecommand \url  [0]{\begingroup\@sanitize@url \@url }%
\providecommand \@url [1]{\endgroup\@href {#1}{\urlprefix }}%
\providecommand \urlprefix  [0]{URL }%
\providecommand \Eprint [0]{\href }%
\providecommand \doibase [0]{https://doi.org/}%
\providecommand \selectlanguage [0]{\@gobble}%
\providecommand \bibinfo  [0]{\@secondoftwo}%
\providecommand \bibfield  [0]{\@secondoftwo}%
\providecommand \translation [1]{[#1]}%
\providecommand \BibitemOpen [0]{}%
\providecommand \bibitemStop [0]{}%
\providecommand \bibitemNoStop [0]{.\EOS\space}%
\providecommand \EOS [0]{\spacefactor3000\relax}%
\providecommand \BibitemShut  [1]{\csname bibitem#1\endcsname}%
\let\auto@bib@innerbib\@empty
\bibitem [{\citenamefont {Leinaas}\ and\ \citenamefont
  {Myrheim}(1977)}]{leinaas1977theory}%
  \BibitemOpen
  \bibfield  {author} {\bibinfo {author} {\bibfnamefont {J.~M.}\ \bibnamefont
  {Leinaas}}\ and\ \bibinfo {author} {\bibfnamefont {J.}~\bibnamefont
  {Myrheim}},\ }\bibfield  {title} {\bibinfo {title} {On the theory of
  identical particles},\ }\href {https://doi.org/10.1007/BF02727953} {\bibfield
   {journal} {\bibinfo  {journal} {Il Nuovo Cimento B (1971-1996)}\ }\textbf
  {\bibinfo {volume} {37}},\ \bibinfo {pages} {1} (\bibinfo {year}
  {1977})}\BibitemShut {NoStop}%
\bibitem [{\citenamefont {Wilczek}(1982)}]{wilczek1982quantum}%
  \BibitemOpen
  \bibfield  {author} {\bibinfo {author} {\bibfnamefont {F.}~\bibnamefont
  {Wilczek}},\ }\bibfield  {title} {\bibinfo {title} {Quantum mechanics of
  fractional-spin particles},\ }\href
  {https://doi.org/10.1103/PhysRevLett.49.957} {\bibfield  {journal} {\bibinfo
  {journal} {Phys. Rev. Lett.}\ }\textbf {\bibinfo {volume} {49}},\ \bibinfo
  {pages} {957} (\bibinfo {year} {1982})}\BibitemShut {NoStop}%
\bibitem [{\citenamefont {Wen}(1990)}]{wen1990topological}%
  \BibitemOpen
  \bibfield  {author} {\bibinfo {author} {\bibfnamefont {X.-G.}\ \bibnamefont
  {Wen}},\ }\bibfield  {title} {\bibinfo {title} {Topological orders in rigid
  states},\ }\href@noop {} {\bibfield  {journal} {\bibinfo  {journal}
  {International Journal of Modern Physics B}\ }\textbf {\bibinfo {volume}
  {4}},\ \bibinfo {pages} {239} (\bibinfo {year} {1990})}\BibitemShut {NoStop}%
\bibitem [{\citenamefont {Moore}\ and\ \citenamefont
  {Seiberg}(1989)}]{moore1989classical}%
  \BibitemOpen
  \bibfield  {author} {\bibinfo {author} {\bibfnamefont {G.}~\bibnamefont
  {Moore}}\ and\ \bibinfo {author} {\bibfnamefont {N.}~\bibnamefont
  {Seiberg}},\ }\bibfield  {title} {\bibinfo {title} {Classical and quantum
  conformal field theory},\ }\href@noop {} {\bibfield  {journal} {\bibinfo
  {journal} {Communications in Mathematical Physics}\ }\textbf {\bibinfo
  {volume} {123}},\ \bibinfo {pages} {177} (\bibinfo {year}
  {1989})}\BibitemShut {NoStop}%
\bibitem [{\citenamefont {Tsui}\ \emph {et~al.}(1982)\citenamefont {Tsui},
  \citenamefont {Stormer},\ and\ \citenamefont
  {Gossard}}]{TsuiStormerGossard82}%
  \BibitemOpen
  \bibfield  {author} {\bibinfo {author} {\bibfnamefont {D.~C.}\ \bibnamefont
  {Tsui}}, \bibinfo {author} {\bibfnamefont {H.~L.}\ \bibnamefont {Stormer}},\
  and\ \bibinfo {author} {\bibfnamefont {A.~C.}\ \bibnamefont {Gossard}},\
  }\bibfield  {title} {\bibinfo {title} {Two-dimensional magnetotransport in
  the extreme quantum limit},\ }\href
  {https://doi.org/10.1103/PhysRevLett.48.1559} {\bibfield  {journal} {\bibinfo
   {journal} {Phys. Rev. Lett.}\ }\textbf {\bibinfo {volume} {48}},\ \bibinfo
  {pages} {1559} (\bibinfo {year} {1982})}\BibitemShut {NoStop}%
\bibitem [{\citenamefont {Laughlin}(1983)}]{laughlin1983}%
  \BibitemOpen
  \bibfield  {author} {\bibinfo {author} {\bibfnamefont {R.~B.}\ \bibnamefont
  {Laughlin}},\ }\bibfield  {title} {\bibinfo {title} {Anomalous quantum hall
  effect: An incompressible quantum fluid with fractionally charged
  excitations},\ }\href {https://doi.org/10.1103/PhysRevLett.50.1395}
  {\bibfield  {journal} {\bibinfo  {journal} {Phys. Rev. Lett.}\ }\textbf
  {\bibinfo {volume} {50}},\ \bibinfo {pages} {1395} (\bibinfo {year}
  {1983})}\BibitemShut {NoStop}%
\bibitem [{\citenamefont {Haldane}(1983)}]{Haldane1983}%
  \BibitemOpen
  \bibfield  {author} {\bibinfo {author} {\bibfnamefont {F.~D.~M.}\
  \bibnamefont {Haldane}},\ }\bibfield  {title} {\bibinfo {title} {Fractional
  quantization of the {Hall} effect: A hierarchy of incompressible quantum
  fluid states},\ }\href {https://doi.org/10.1103/PhysRevLett.51.605}
  {\bibfield  {journal} {\bibinfo  {journal} {Phys. Rev. Lett.}\ }\textbf
  {\bibinfo {volume} {51}},\ \bibinfo {pages} {605} (\bibinfo {year}
  {1983})}\BibitemShut {NoStop}%
\bibitem [{\citenamefont {Halperin}(1984)}]{halperin1984}%
  \BibitemOpen
  \bibfield  {author} {\bibinfo {author} {\bibfnamefont {B.~I.}\ \bibnamefont
  {Halperin}},\ }\bibfield  {title} {\bibinfo {title} {Statistics of
  quasiparticles and the hierarchy of fractional quantized hall states},\
  }\href {https://doi.org/10.1103/PhysRevLett.52.1583} {\bibfield  {journal}
  {\bibinfo  {journal} {Phys. Rev. Lett.}\ }\textbf {\bibinfo {volume} {52}},\
  \bibinfo {pages} {1583} (\bibinfo {year} {1984})}\BibitemShut {NoStop}%
\bibitem [{\citenamefont {Arovas}\ \emph {et~al.}(1984)\citenamefont {Arovas},
  \citenamefont {Schrieffer},\ and\ \citenamefont
  {Wilczek}}]{arovas1984fractional}%
  \BibitemOpen
  \bibfield  {author} {\bibinfo {author} {\bibfnamefont {D.}~\bibnamefont
  {Arovas}}, \bibinfo {author} {\bibfnamefont {J.~R.}\ \bibnamefont
  {Schrieffer}},\ and\ \bibinfo {author} {\bibfnamefont {F.}~\bibnamefont
  {Wilczek}},\ }\bibfield  {title} {\bibinfo {title} {Fractional statistics and
  the quantum {Hall} effect},\ }\href
  {https://doi.org/10.1103/PhysRevLett.53.722} {\bibfield  {journal} {\bibinfo
  {journal} {Phys. Rev. Lett.}\ }\textbf {\bibinfo {volume} {53}},\ \bibinfo
  {pages} {722} (\bibinfo {year} {1984})}\BibitemShut {NoStop}%
\bibitem [{\citenamefont {Saminadayar}\ \emph {et~al.}(1997)\citenamefont
  {Saminadayar}, \citenamefont {Glattli}, \citenamefont {Jin},\ and\
  \citenamefont {Etienne}}]{saminadayar1997observation}%
  \BibitemOpen
  \bibfield  {author} {\bibinfo {author} {\bibfnamefont {L.}~\bibnamefont
  {Saminadayar}}, \bibinfo {author} {\bibfnamefont {D.~C.}\ \bibnamefont
  {Glattli}}, \bibinfo {author} {\bibfnamefont {Y.}~\bibnamefont {Jin}},\ and\
  \bibinfo {author} {\bibfnamefont {B.}~\bibnamefont {Etienne}},\ }\bibfield
  {title} {\bibinfo {title} {Observation of the $e/3$ fractionally charged
  {Laughlin} quasiparticle},\ }\href
  {https://doi.org/10.1103/PhysRevLett.79.2526} {\bibfield  {journal} {\bibinfo
   {journal} {Phys. Rev. Lett.}\ }\textbf {\bibinfo {volume} {79}},\ \bibinfo
  {pages} {2526} (\bibinfo {year} {1997})}\BibitemShut {NoStop}%
\bibitem [{\citenamefont {de~Picciotto}\ \emph {et~al.}(1998)\citenamefont
  {de~Picciotto}, \citenamefont {Reznikov}, \citenamefont {Heiblum},
  \citenamefont {Umansky}, \citenamefont {Bunin},\ and\ \citenamefont
  {Mahalu}}]{de1998direct}%
  \BibitemOpen
  \bibfield  {author} {\bibinfo {author} {\bibfnamefont {R.}~\bibnamefont
  {de~Picciotto}}, \bibinfo {author} {\bibfnamefont {M.}~\bibnamefont
  {Reznikov}}, \bibinfo {author} {\bibfnamefont {M.}~\bibnamefont {Heiblum}},
  \bibinfo {author} {\bibfnamefont {V.}~\bibnamefont {Umansky}}, \bibinfo
  {author} {\bibfnamefont {G.}~\bibnamefont {Bunin}},\ and\ \bibinfo {author}
  {\bibfnamefont {D.}~\bibnamefont {Mahalu}},\ }\bibfield  {title} {\bibinfo
  {title} {Direct observation of a fractional charge},\ }\href
  {https://doi.org/10.1016/S0921-4526(98)00139-2} {\bibfield  {journal}
  {\bibinfo  {journal} {Physica B: Condensed Matter}\ }\textbf {\bibinfo
  {volume} {249--251}},\ \bibinfo {pages} {395} (\bibinfo {year}
  {1998})}\BibitemShut {NoStop}%
\bibitem [{\citenamefont {Bartolomei}\ \emph {et~al.}(2020)\citenamefont
  {Bartolomei}, \citenamefont {Kumar}, \citenamefont {Bisognin}, \citenamefont
  {Marguerite}, \citenamefont {Berroir}, \citenamefont {Bocquillon},
  \citenamefont {Placais}, \citenamefont {Cavanna}, \citenamefont {Dong},
  \citenamefont {Gennser}, \citenamefont {Jin},\ and\ \citenamefont
  {F{\`e}ve}}]{bartolomei2020fractional}%
  \BibitemOpen
  \bibfield  {author} {\bibinfo {author} {\bibfnamefont {H.}~\bibnamefont
  {Bartolomei}}, \bibinfo {author} {\bibfnamefont {M.}~\bibnamefont {Kumar}},
  \bibinfo {author} {\bibfnamefont {R.}~\bibnamefont {Bisognin}}, \bibinfo
  {author} {\bibfnamefont {A.}~\bibnamefont {Marguerite}}, \bibinfo {author}
  {\bibfnamefont {J.-M.}\ \bibnamefont {Berroir}}, \bibinfo {author}
  {\bibfnamefont {E.}~\bibnamefont {Bocquillon}}, \bibinfo {author}
  {\bibfnamefont {B.}~\bibnamefont {Placais}}, \bibinfo {author} {\bibfnamefont
  {A.}~\bibnamefont {Cavanna}}, \bibinfo {author} {\bibfnamefont
  {Q.}~\bibnamefont {Dong}}, \bibinfo {author} {\bibfnamefont {U.}~\bibnamefont
  {Gennser}}, \bibinfo {author} {\bibfnamefont {Y.}~\bibnamefont {Jin}},\ and\
  \bibinfo {author} {\bibfnamefont {G.}~\bibnamefont {F{\`e}ve}},\ }\bibfield
  {title} {\bibinfo {title} {Fractional statistics in anyon collisions},\
  }\href {https://doi.org/10.1126/science.aaz5601} {\bibfield  {journal}
  {\bibinfo  {journal} {Science}\ }\textbf {\bibinfo {volume} {368}},\ \bibinfo
  {pages} {173} (\bibinfo {year} {2020})}\BibitemShut {NoStop}%
\bibitem [{\citenamefont {Nakamura}\ \emph {et~al.}(2020)\citenamefont
  {Nakamura}, \citenamefont {Liang}, \citenamefont {Gardner},\ and\
  \citenamefont {Manfra}}]{nakamura2020direct}%
  \BibitemOpen
  \bibfield  {author} {\bibinfo {author} {\bibfnamefont {J.}~\bibnamefont
  {Nakamura}}, \bibinfo {author} {\bibfnamefont {S.}~\bibnamefont {Liang}},
  \bibinfo {author} {\bibfnamefont {G.~C.}\ \bibnamefont {Gardner}},\ and\
  \bibinfo {author} {\bibfnamefont {M.~J.}\ \bibnamefont {Manfra}},\ }\bibfield
   {title} {\bibinfo {title} {Direct observation of anyonic braiding
  statistics},\ }\href {https://doi.org/10.1038/s41567-020-1019-1} {\bibfield
  {journal} {\bibinfo  {journal} {Nature Physics}\ }\textbf {\bibinfo {volume}
  {16}},\ \bibinfo {pages} {931} (\bibinfo {year} {2020})}\BibitemShut
  {NoStop}%
\bibitem [{\citenamefont {Neupert}\ \emph {et~al.}(2011)\citenamefont
  {Neupert}, \citenamefont {Santos}, \citenamefont {Chamon},\ and\
  \citenamefont {Mudry}}]{neupert2011fractional}%
  \BibitemOpen
  \bibfield  {author} {\bibinfo {author} {\bibfnamefont {T.}~\bibnamefont
  {Neupert}}, \bibinfo {author} {\bibfnamefont {L.}~\bibnamefont {Santos}},
  \bibinfo {author} {\bibfnamefont {C.}~\bibnamefont {Chamon}},\ and\ \bibinfo
  {author} {\bibfnamefont {C.}~\bibnamefont {Mudry}},\ }\bibfield  {title}
  {\bibinfo {title} {Fractional quantum hall states at zero magnetic field},\
  }\href {https://doi.org/10.1103/PhysRevLett.106.236804} {\bibfield  {journal}
  {\bibinfo  {journal} {Phys. Rev. Lett.}\ }\textbf {\bibinfo {volume} {106}},\
  \bibinfo {pages} {236804} (\bibinfo {year} {2011})}\BibitemShut {NoStop}%
\bibitem [{\citenamefont {Sheng}\ \emph {et~al.}(2011)\citenamefont {Sheng},
  \citenamefont {Gu}, \citenamefont {Sun},\ and\ \citenamefont
  {Sheng}}]{sheng2011fractional}%
  \BibitemOpen
  \bibfield  {author} {\bibinfo {author} {\bibfnamefont {D.~N.}\ \bibnamefont
  {Sheng}}, \bibinfo {author} {\bibfnamefont {Z.-C.}\ \bibnamefont {Gu}},
  \bibinfo {author} {\bibfnamefont {K.}~\bibnamefont {Sun}},\ and\ \bibinfo
  {author} {\bibfnamefont {L.}~\bibnamefont {Sheng}},\ }\bibfield  {title}
  {\bibinfo {title} {Fractional quantum hall effect in the absence of landau
  levels},\ }\href {https://doi.org/10.1038/ncomms1380} {\bibfield  {journal}
  {\bibinfo  {journal} {Nature Communications}\ }\textbf {\bibinfo {volume}
  {2}},\ \bibinfo {pages} {389} (\bibinfo {year} {2011})}\BibitemShut {NoStop}%
\bibitem [{\citenamefont {Tang}\ \emph {et~al.}(2011)\citenamefont {Tang},
  \citenamefont {Mei},\ and\ \citenamefont {Wen}}]{tang2011fractional}%
  \BibitemOpen
  \bibfield  {author} {\bibinfo {author} {\bibfnamefont {E.}~\bibnamefont
  {Tang}}, \bibinfo {author} {\bibfnamefont {J.-W.}\ \bibnamefont {Mei}},\ and\
  \bibinfo {author} {\bibfnamefont {X.-G.}\ \bibnamefont {Wen}},\ }\bibfield
  {title} {\bibinfo {title} {High-temperature fractional quantum hall states},\
  }\href {https://doi.org/10.1103/PhysRevLett.106.236802} {\bibfield  {journal}
  {\bibinfo  {journal} {Phys. Rev. Lett.}\ }\textbf {\bibinfo {volume} {106}},\
  \bibinfo {pages} {236802} (\bibinfo {year} {2011})}\BibitemShut {NoStop}%
\bibitem [{\citenamefont {Regnault}\ and\ \citenamefont
  {Bernevig}(2011)}]{regnault2011fractional}%
  \BibitemOpen
  \bibfield  {author} {\bibinfo {author} {\bibfnamefont {N.}~\bibnamefont
  {Regnault}}\ and\ \bibinfo {author} {\bibfnamefont {B.~A.}\ \bibnamefont
  {Bernevig}},\ }\bibfield  {title} {\bibinfo {title} {Fractional chern
  insulator},\ }\href {https://doi.org/10.1103/PhysRevX.1.021014} {\bibfield
  {journal} {\bibinfo  {journal} {Phys. Rev. X}\ }\textbf {\bibinfo {volume}
  {1}},\ \bibinfo {pages} {021014} (\bibinfo {year} {2011})}\BibitemShut
  {NoStop}%
\bibitem [{\citenamefont {Cai}\ \emph {et~al.}(2023)\citenamefont {Cai},
  \citenamefont {Anderson}, \citenamefont {Wang}, \citenamefont {Zhang},
  \citenamefont {Liu}, \citenamefont {Holtzmann}, \citenamefont {Zhang},
  \citenamefont {Fan}, \citenamefont {Taniguchi}, \citenamefont {Watanabe},
  \citenamefont {Ran}, \citenamefont {Cao}, \citenamefont {Fu}, \citenamefont
  {Xiao}, \citenamefont {Yao},\ and\ \citenamefont {Xu}}]{cai2023signatures}%
  \BibitemOpen
  \bibfield  {author} {\bibinfo {author} {\bibfnamefont {J.}~\bibnamefont
  {Cai}}, \bibinfo {author} {\bibfnamefont {E.}~\bibnamefont {Anderson}},
  \bibinfo {author} {\bibfnamefont {C.}~\bibnamefont {Wang}}, \bibinfo {author}
  {\bibfnamefont {X.}~\bibnamefont {Zhang}}, \bibinfo {author} {\bibfnamefont
  {X.}~\bibnamefont {Liu}}, \bibinfo {author} {\bibfnamefont {W.}~\bibnamefont
  {Holtzmann}}, \bibinfo {author} {\bibfnamefont {Y.}~\bibnamefont {Zhang}},
  \bibinfo {author} {\bibfnamefont {F.}~\bibnamefont {Fan}}, \bibinfo {author}
  {\bibfnamefont {T.}~\bibnamefont {Taniguchi}}, \bibinfo {author}
  {\bibfnamefont {K.}~\bibnamefont {Watanabe}}, \bibinfo {author}
  {\bibfnamefont {Y.}~\bibnamefont {Ran}}, \bibinfo {author} {\bibfnamefont
  {T.}~\bibnamefont {Cao}}, \bibinfo {author} {\bibfnamefont {L.}~\bibnamefont
  {Fu}}, \bibinfo {author} {\bibfnamefont {D.}~\bibnamefont {Xiao}}, \bibinfo
  {author} {\bibfnamefont {W.}~\bibnamefont {Yao}},\ and\ \bibinfo {author}
  {\bibfnamefont {X.}~\bibnamefont {Xu}},\ }\bibfield  {title} {\bibinfo
  {title} {Signatures of fractional quantum anomalous hall states in twisted
  mote2},\ }\href {https://doi.org/10.1038/s41586-023-06289-w} {\bibfield
  {journal} {\bibinfo  {journal} {Nature}\ }\textbf {\bibinfo {volume} {622}},\
  \bibinfo {pages} {63} (\bibinfo {year} {2023})}\BibitemShut {NoStop}%
\bibitem [{\citenamefont {Zeng}\ \emph {et~al.}(2023)\citenamefont {Zeng},
  \citenamefont {Xia}, \citenamefont {Kang}, \citenamefont {Zhu}, \citenamefont
  {Kn{\"u}ppel}, \citenamefont {Vaswani}, \citenamefont {Watanabe},
  \citenamefont {Taniguchi}, \citenamefont {Mak},\ and\ \citenamefont
  {Shan}}]{zeng2023thermodynamic}%
  \BibitemOpen
  \bibfield  {author} {\bibinfo {author} {\bibfnamefont {Y.}~\bibnamefont
  {Zeng}}, \bibinfo {author} {\bibfnamefont {Z.}~\bibnamefont {Xia}}, \bibinfo
  {author} {\bibfnamefont {K.}~\bibnamefont {Kang}}, \bibinfo {author}
  {\bibfnamefont {J.}~\bibnamefont {Zhu}}, \bibinfo {author} {\bibfnamefont
  {P.}~\bibnamefont {Kn{\"u}ppel}}, \bibinfo {author} {\bibfnamefont
  {C.}~\bibnamefont {Vaswani}}, \bibinfo {author} {\bibfnamefont
  {K.}~\bibnamefont {Watanabe}}, \bibinfo {author} {\bibfnamefont
  {T.}~\bibnamefont {Taniguchi}}, \bibinfo {author} {\bibfnamefont {K.~F.}\
  \bibnamefont {Mak}},\ and\ \bibinfo {author} {\bibfnamefont {J.}~\bibnamefont
  {Shan}},\ }\bibfield  {title} {\bibinfo {title} {Thermodynamic evidence of
  fractional {Chern} insulator in moir{\'e} {MoTe$_2$}},\ }\href
  {https://doi.org/10.1038/s41586-023-06452-3} {\bibfield  {journal} {\bibinfo
  {journal} {Nature}\ }\textbf {\bibinfo {volume} {622}},\ \bibinfo {pages}
  {69} (\bibinfo {year} {2023})}\BibitemShut {NoStop}%
\bibitem [{\citenamefont {Park}\ \emph {et~al.}(2023)\citenamefont {Park},
  \citenamefont {Cai}, \citenamefont {Anderson}, \citenamefont {Zhang},
  \citenamefont {Zhu}, \citenamefont {Liu}, \citenamefont {Wang}, \citenamefont
  {Holtzmann}, \citenamefont {Hu}, \citenamefont {Liu}, \citenamefont
  {Taniguchi}, \citenamefont {Watanabe}, \citenamefont {Chu}, \citenamefont
  {Cao}, \citenamefont {Fu}, \citenamefont {Yao}, \citenamefont {Chang},
  \citenamefont {Cobden}, \citenamefont {Xiao},\ and\ \citenamefont
  {Xu}}]{park2023observation}%
  \BibitemOpen
  \bibfield  {author} {\bibinfo {author} {\bibfnamefont {H.}~\bibnamefont
  {Park}}, \bibinfo {author} {\bibfnamefont {J.}~\bibnamefont {Cai}}, \bibinfo
  {author} {\bibfnamefont {E.}~\bibnamefont {Anderson}}, \bibinfo {author}
  {\bibfnamefont {Y.}~\bibnamefont {Zhang}}, \bibinfo {author} {\bibfnamefont
  {J.}~\bibnamefont {Zhu}}, \bibinfo {author} {\bibfnamefont {X.}~\bibnamefont
  {Liu}}, \bibinfo {author} {\bibfnamefont {C.}~\bibnamefont {Wang}}, \bibinfo
  {author} {\bibfnamefont {W.}~\bibnamefont {Holtzmann}}, \bibinfo {author}
  {\bibfnamefont {C.}~\bibnamefont {Hu}}, \bibinfo {author} {\bibfnamefont
  {Z.}~\bibnamefont {Liu}}, \bibinfo {author} {\bibfnamefont {T.}~\bibnamefont
  {Taniguchi}}, \bibinfo {author} {\bibfnamefont {K.}~\bibnamefont {Watanabe}},
  \bibinfo {author} {\bibfnamefont {J.-H.}\ \bibnamefont {Chu}}, \bibinfo
  {author} {\bibfnamefont {T.}~\bibnamefont {Cao}}, \bibinfo {author}
  {\bibfnamefont {L.}~\bibnamefont {Fu}}, \bibinfo {author} {\bibfnamefont
  {W.}~\bibnamefont {Yao}}, \bibinfo {author} {\bibfnamefont {C.-Z.}\
  \bibnamefont {Chang}}, \bibinfo {author} {\bibfnamefont {D.}~\bibnamefont
  {Cobden}}, \bibinfo {author} {\bibfnamefont {D.}~\bibnamefont {Xiao}},\ and\
  \bibinfo {author} {\bibfnamefont {X.}~\bibnamefont {Xu}},\ }\bibfield
  {title} {\bibinfo {title} {Observation of fractionally quantized anomalous
  {Hall} effect},\ }\href {https://doi.org/10.1038/s41586-023-06536-0}
  {\bibfield  {journal} {\bibinfo  {journal} {Nature}\ }\textbf {\bibinfo
  {volume} {622}},\ \bibinfo {pages} {74} (\bibinfo {year} {2023})}\BibitemShut
  {NoStop}%
\bibitem [{\citenamefont {Xu}\ \emph {et~al.}(2023)\citenamefont {Xu},
  \citenamefont {Sun}, \citenamefont {Jia}, \citenamefont {Liu}, \citenamefont
  {Xu}, \citenamefont {Li}, \citenamefont {Gu}, \citenamefont {Watanabe},
  \citenamefont {Taniguchi}, \citenamefont {Tong}, \citenamefont {Jia},
  \citenamefont {Shi}, \citenamefont {Jiang}, \citenamefont {Zhang},
  \citenamefont {Liu},\ and\ \citenamefont {Li}}]{xu2023observation}%
  \BibitemOpen
  \bibfield  {author} {\bibinfo {author} {\bibfnamefont {F.}~\bibnamefont
  {Xu}}, \bibinfo {author} {\bibfnamefont {Z.}~\bibnamefont {Sun}}, \bibinfo
  {author} {\bibfnamefont {T.}~\bibnamefont {Jia}}, \bibinfo {author}
  {\bibfnamefont {C.}~\bibnamefont {Liu}}, \bibinfo {author} {\bibfnamefont
  {C.}~\bibnamefont {Xu}}, \bibinfo {author} {\bibfnamefont {C.}~\bibnamefont
  {Li}}, \bibinfo {author} {\bibfnamefont {Y.}~\bibnamefont {Gu}}, \bibinfo
  {author} {\bibfnamefont {K.}~\bibnamefont {Watanabe}}, \bibinfo {author}
  {\bibfnamefont {T.}~\bibnamefont {Taniguchi}}, \bibinfo {author}
  {\bibfnamefont {B.}~\bibnamefont {Tong}}, \bibinfo {author} {\bibfnamefont
  {J.}~\bibnamefont {Jia}}, \bibinfo {author} {\bibfnamefont {Z.}~\bibnamefont
  {Shi}}, \bibinfo {author} {\bibfnamefont {S.}~\bibnamefont {Jiang}}, \bibinfo
  {author} {\bibfnamefont {Y.}~\bibnamefont {Zhang}}, \bibinfo {author}
  {\bibfnamefont {X.}~\bibnamefont {Liu}},\ and\ \bibinfo {author}
  {\bibfnamefont {T.}~\bibnamefont {Li}},\ }\bibfield  {title} {\bibinfo
  {title} {Observation of integer and fractional quantum anomalous {Hall}
  effects in twisted bilayer {MoTe$_2$}},\ }\href
  {https://doi.org/10.1103/PhysRevX.13.031037} {\bibfield  {journal} {\bibinfo
  {journal} {Phys. Rev. X}\ }\textbf {\bibinfo {volume} {13}},\ \bibinfo
  {pages} {031037} (\bibinfo {year} {2023})}\BibitemShut {NoStop}%
\bibitem [{\citenamefont {Lu}\ \emph {et~al.}(2024)\citenamefont {Lu},
  \citenamefont {Han}, \citenamefont {Yao}, \citenamefont {Reddy},
  \citenamefont {Yang}, \citenamefont {Seo}, \citenamefont {Watanabe},
  \citenamefont {Taniguchi}, \citenamefont {Fu},\ and\ \citenamefont
  {Ju}}]{Lu2024FQAH}%
  \BibitemOpen
  \bibfield  {author} {\bibinfo {author} {\bibfnamefont {Z.}~\bibnamefont
  {Lu}}, \bibinfo {author} {\bibfnamefont {T.}~\bibnamefont {Han}}, \bibinfo
  {author} {\bibfnamefont {Y.}~\bibnamefont {Yao}}, \bibinfo {author}
  {\bibfnamefont {A.~P.}\ \bibnamefont {Reddy}}, \bibinfo {author}
  {\bibfnamefont {J.}~\bibnamefont {Yang}}, \bibinfo {author} {\bibfnamefont
  {J.}~\bibnamefont {Seo}}, \bibinfo {author} {\bibfnamefont {K.}~\bibnamefont
  {Watanabe}}, \bibinfo {author} {\bibfnamefont {T.}~\bibnamefont {Taniguchi}},
  \bibinfo {author} {\bibfnamefont {L.}~\bibnamefont {Fu}},\ and\ \bibinfo
  {author} {\bibfnamefont {L.}~\bibnamefont {Ju}},\ }\bibfield  {title}
  {\bibinfo {title} {Fractional quantum anomalous {Hall} effect in multilayer
  graphene},\ }\href {https://doi.org/10.1038/s41586-023-07010-7} {\bibfield
  {journal} {\bibinfo  {journal} {Nature}\ }\textbf {\bibinfo {volume} {626}},\
  \bibinfo {pages} {759} (\bibinfo {year} {2024})}\BibitemShut {NoStop}%
\bibitem [{\citenamefont {Deng}\ \emph {et~al.}(2026)\citenamefont {Deng},
  \citenamefont {Sun}, \citenamefont {Pimenov}, \citenamefont {Taniguchi},
  \citenamefont {Watanabe}, \citenamefont {Mueller},\ and\ \citenamefont
  {Liu}}]{deng2026realspaceimagingquantumhall}%
  \BibitemOpen
  \bibfield  {author} {\bibinfo {author} {\bibfnamefont {J.}~\bibnamefont
  {Deng}}, \bibinfo {author} {\bibfnamefont {Y.}~\bibnamefont {Sun}}, \bibinfo
  {author} {\bibfnamefont {D.}~\bibnamefont {Pimenov}}, \bibinfo {author}
  {\bibfnamefont {T.}~\bibnamefont {Taniguchi}}, \bibinfo {author}
  {\bibfnamefont {K.}~\bibnamefont {Watanabe}}, \bibinfo {author}
  {\bibfnamefont {E.~J.}\ \bibnamefont {Mueller}},\ and\ \bibinfo {author}
  {\bibfnamefont {X.}~\bibnamefont {Liu}},\ }\href
  {https://arxiv.org/abs/2606.25036} {\bibinfo {title} {Real-space imaging of
  quantum hall quasiparticles}} (\bibinfo {year} {2026}),\ \Eprint
  {https://arxiv.org/abs/2606.25036} {arXiv:2606.25036 [cond-mat.mes-hall]}
  \BibitemShut {NoStop}%
\bibitem [{\citenamefont {Park}\ \emph {et~al.}(2026)\citenamefont {Park},
  \citenamefont {Voinea}, \citenamefont {Tsui}, \citenamefont {Pu},
  \citenamefont {Watanabe}, \citenamefont {Taniguchi}, \citenamefont {Cooper},
  \citenamefont {Zaletel}, \citenamefont {Papić},\ and\ \citenamefont
  {Yazdani}}]{park2026localspectroscopyanyonsbound}%
  \BibitemOpen
  \bibfield  {author} {\bibinfo {author} {\bibfnamefont {J.~M.}\ \bibnamefont
  {Park}}, \bibinfo {author} {\bibfnamefont {C.}~\bibnamefont {Voinea}},
  \bibinfo {author} {\bibfnamefont {Y.-C.}\ \bibnamefont {Tsui}}, \bibinfo
  {author} {\bibfnamefont {S.}~\bibnamefont {Pu}}, \bibinfo {author}
  {\bibfnamefont {K.}~\bibnamefont {Watanabe}}, \bibinfo {author}
  {\bibfnamefont {T.}~\bibnamefont {Taniguchi}}, \bibinfo {author}
  {\bibfnamefont {N.~R.}\ \bibnamefont {Cooper}}, \bibinfo {author}
  {\bibfnamefont {M.~P.}\ \bibnamefont {Zaletel}}, \bibinfo {author}
  {\bibfnamefont {Z.}~\bibnamefont {Papić}},\ and\ \bibinfo {author}
  {\bibfnamefont {A.}~\bibnamefont {Yazdani}},\ }\href
  {https://arxiv.org/abs/2606.25024} {\bibinfo {title} {Local spectroscopy of
  anyons bound to charge traps}} (\bibinfo {year} {2026}),\ \Eprint
  {https://arxiv.org/abs/2606.25024} {arXiv:2606.25024 [cond-mat.mes-hall]}
  \BibitemShut {NoStop}%
\bibitem [{\citenamefont {Kitaev}(2003)}]{kitaev2003fault}%
  \BibitemOpen
  \bibfield  {author} {\bibinfo {author} {\bibfnamefont {A.~Y.}\ \bibnamefont
  {Kitaev}},\ }\bibfield  {title} {\bibinfo {title} {Fault-tolerant quantum
  computation by anyons},\ }\href
  {https://doi.org/10.1016/S0003-4916(02)00018-0} {\bibfield  {journal}
  {\bibinfo  {journal} {Annals of Physics}\ }\textbf {\bibinfo {volume}
  {303}},\ \bibinfo {pages} {2} (\bibinfo {year} {2003})}\BibitemShut {NoStop}%
\bibitem [{\citenamefont {Nayak}\ \emph {et~al.}(2008)\citenamefont {Nayak},
  \citenamefont {Simon}, \citenamefont {Stern}, \citenamefont {Freedman},\ and\
  \citenamefont {Das~Sarma}}]{nayak2008nonabelian}%
  \BibitemOpen
  \bibfield  {author} {\bibinfo {author} {\bibfnamefont {C.}~\bibnamefont
  {Nayak}}, \bibinfo {author} {\bibfnamefont {S.~H.}\ \bibnamefont {Simon}},
  \bibinfo {author} {\bibfnamefont {A.}~\bibnamefont {Stern}}, \bibinfo
  {author} {\bibfnamefont {M.}~\bibnamefont {Freedman}},\ and\ \bibinfo
  {author} {\bibfnamefont {S.}~\bibnamefont {Das~Sarma}},\ }\bibfield  {title}
  {\bibinfo {title} {Non-abelian anyons and topological quantum computation},\
  }\href {https://doi.org/10.1103/RevModPhys.80.1083} {\bibfield  {journal}
  {\bibinfo  {journal} {Reviews of Modern Physics}\ }\textbf {\bibinfo {volume}
  {80}},\ \bibinfo {pages} {1083} (\bibinfo {year} {2008})}\BibitemShut
  {NoStop}%
\bibitem [{\citenamefont {Kj{\o}nsberg}\ and\ \citenamefont
  {Leinaas}(1997)}]{kjonsberg1997anyon}%
  \BibitemOpen
  \bibfield  {author} {\bibinfo {author} {\bibfnamefont {H.}~\bibnamefont
  {Kj{\o}nsberg}}\ and\ \bibinfo {author} {\bibfnamefont {J.~M.}\ \bibnamefont
  {Leinaas}},\ }\bibfield  {title} {\bibinfo {title} {On the anyon description
  of the laughlin hole states},\ }\href@noop {} {\bibfield  {journal} {\bibinfo
   {journal} {International Journal of Modern Physics A}\ }\textbf {\bibinfo
  {volume} {12}},\ \bibinfo {pages} {1975} (\bibinfo {year}
  {1997})}\BibitemShut {NoStop}%
\bibitem [{\citenamefont {Yan}\ \emph {et~al.}(2026)\citenamefont {Yan},
  \citenamefont {Li}, \citenamefont {Soejima},\ and\ \citenamefont
  {Khalaf}}]{shn5-vwgt}%
  \BibitemOpen
  \bibfield  {author} {\bibinfo {author} {\bibfnamefont {Z.}~\bibnamefont
  {Yan}}, \bibinfo {author} {\bibfnamefont {Q.}~\bibnamefont {Li}}, \bibinfo
  {author} {\bibfnamefont {T.}~\bibnamefont {Soejima}},\ and\ \bibinfo {author}
  {\bibfnamefont {E.}~\bibnamefont {Khalaf}},\ }\bibfield  {title} {\bibinfo
  {title} {Anyon dispersion in {Aharonov}-{Casher} bands and implications for
  twisted {MoTe$_2$}},\ }\href {https://doi.org/10.1103/shn5-vwgt} {\bibfield
  {journal} {\bibinfo  {journal} {Phys. Rev. B}\ }\textbf {\bibinfo {volume}
  {113}},\ \bibinfo {pages} {235155} (\bibinfo {year} {2026})}\BibitemShut
  {NoStop}%
\bibitem [{\citenamefont {Schleith}\ \emph {et~al.}(2026)\citenamefont
  {Schleith}, \citenamefont {Soejima},\ and\ \citenamefont
  {Khalaf}}]{gzdq-wkxj}%
  \BibitemOpen
  \bibfield  {author} {\bibinfo {author} {\bibfnamefont {M.-L.}\ \bibnamefont
  {Schleith}}, \bibinfo {author} {\bibfnamefont {T.}~\bibnamefont {Soejima}},\
  and\ \bibinfo {author} {\bibfnamefont {E.}~\bibnamefont {Khalaf}},\
  }\bibfield  {title} {\bibinfo {title} {Anyon dispersion from nonuniform
  magnetic field on the sphere},\ }\href {https://doi.org/10.1103/gzdq-wkxj}
  {\bibfield  {journal} {\bibinfo  {journal} {Phys. Rev. Lett.}\ }\textbf
  {\bibinfo {volume} {137}},\ \bibinfo {pages} {046506} (\bibinfo {year}
  {2026})}\BibitemShut {NoStop}%
\bibitem [{\citenamefont {Jain}(1989)}]{jain1989}%
  \BibitemOpen
  \bibfield  {author} {\bibinfo {author} {\bibfnamefont {J.~K.}\ \bibnamefont
  {Jain}},\ }\bibfield  {title} {\bibinfo {title} {Composite-fermion approach
  for the fractional quantum hall effect},\ }\href
  {https://doi.org/10.1103/PhysRevLett.63.199} {\bibfield  {journal} {\bibinfo
  {journal} {Phys. Rev. Lett.}\ }\textbf {\bibinfo {volume} {63}},\ \bibinfo
  {pages} {199} (\bibinfo {year} {1989})}\BibitemShut {NoStop}%
\bibitem [{\citenamefont {Jain}(2007)}]{jain2007composite}%
  \BibitemOpen
  \bibfield  {author} {\bibinfo {author} {\bibfnamefont {J.}~\bibnamefont
  {Jain}},\ }\href@noop {} {\bibinfo {title} {Composite fermions cambridge
  univ}} (\bibinfo {year} {2007})\BibitemShut {NoStop}%
\bibitem [{\citenamefont {Hansson}\ \emph {et~al.}(2009)\citenamefont
  {Hansson}, \citenamefont {Hermanns}, \citenamefont {Regnault},\ and\
  \citenamefont {Viefers}}]{Hansson2009}%
  \BibitemOpen
  \bibfield  {author} {\bibinfo {author} {\bibfnamefont {T.~H.}\ \bibnamefont
  {Hansson}}, \bibinfo {author} {\bibfnamefont {M.}~\bibnamefont {Hermanns}},
  \bibinfo {author} {\bibfnamefont {N.}~\bibnamefont {Regnault}},\ and\
  \bibinfo {author} {\bibfnamefont {S.}~\bibnamefont {Viefers}},\ }\bibfield
  {title} {\bibinfo {title} {Conformal field theory approach to {Abelian} and
  non-{Abelian} quantum {Hall} quasielectrons},\ }\href
  {https://doi.org/10.1103/PhysRevLett.102.166805} {\bibfield  {journal}
  {\bibinfo  {journal} {Phys. Rev. Lett.}\ }\textbf {\bibinfo {volume} {102}},\
  \bibinfo {pages} {166805} (\bibinfo {year} {2009})}\BibitemShut {NoStop}%
\bibitem [{\citenamefont {Nardin}\ and\ \citenamefont
  {Mazza}(2023)}]{PhysRevB.108.L201106}%
  \BibitemOpen
  \bibfield  {author} {\bibinfo {author} {\bibfnamefont {A.}~\bibnamefont
  {Nardin}}\ and\ \bibinfo {author} {\bibfnamefont {L.}~\bibnamefont {Mazza}},\
  }\bibfield  {title} {\bibinfo {title} {Laughlin's quasielectron as a nonlocal
  composite fermion},\ }\href {https://doi.org/10.1103/PhysRevB.108.L201106}
  {\bibfield  {journal} {\bibinfo  {journal} {Phys. Rev. B}\ }\textbf {\bibinfo
  {volume} {108}},\ \bibinfo {pages} {L201106} (\bibinfo {year}
  {2023})}\BibitemShut {NoStop}%
\bibitem [{\citenamefont {Iyer}\ \emph {et~al.}(2026)\citenamefont {Iyer},
  \citenamefont {Feuerpfeil}, \citenamefont {Cr{\'e}pel}, \citenamefont
  {Regnault},\ and\ \citenamefont {Mora}}]{iyer2026dispersionanyonblochbands}%
  \BibitemOpen
  \bibfield  {author} {\bibinfo {author} {\bibfnamefont {K.}~\bibnamefont
  {Iyer}}, \bibinfo {author} {\bibfnamefont {A.}~\bibnamefont {Feuerpfeil}},
  \bibinfo {author} {\bibfnamefont {V.}~\bibnamefont {Cr{\'e}pel}}, \bibinfo
  {author} {\bibfnamefont {N.}~\bibnamefont {Regnault}},\ and\ \bibinfo
  {author} {\bibfnamefont {C.}~\bibnamefont {Mora}},\ }\href
  {https://arxiv.org/abs/2604.24859} {\bibinfo {title} {Dispersion of anyon
  bloch bands}} (\bibinfo {year} {2026}),\ \Eprint
  {https://arxiv.org/abs/2604.24859} {arXiv:2604.24859 [cond-mat.mes-hall]}
  \BibitemShut {NoStop}%
\bibitem [{\citenamefont {Roy}(2014)}]{PhysRevB.90.165139}%
  \BibitemOpen
  \bibfield  {author} {\bibinfo {author} {\bibfnamefont {R.}~\bibnamefont
  {Roy}},\ }\bibfield  {title} {\bibinfo {title} {Band geometry of fractional
  topological insulators},\ }\href {https://doi.org/10.1103/PhysRevB.90.165139}
  {\bibfield  {journal} {\bibinfo  {journal} {Phys. Rev. B}\ }\textbf {\bibinfo
  {volume} {90}},\ \bibinfo {pages} {165139} (\bibinfo {year}
  {2014})}\BibitemShut {NoStop}%
\bibitem [{\citenamefont {Jackson}\ \emph {et~al.}(2015)\citenamefont
  {Jackson}, \citenamefont {M{\"o}ller},\ and\ \citenamefont
  {Roy}}]{jackson2015geometric}%
  \BibitemOpen
  \bibfield  {author} {\bibinfo {author} {\bibfnamefont {T.~S.}\ \bibnamefont
  {Jackson}}, \bibinfo {author} {\bibfnamefont {G.}~\bibnamefont
  {M{\"o}ller}},\ and\ \bibinfo {author} {\bibfnamefont {R.}~\bibnamefont
  {Roy}},\ }\bibfield  {title} {\bibinfo {title} {Geometric stability of
  topological lattice phases},\ }\href@noop {} {\bibfield  {journal} {\bibinfo
  {journal} {Nature communications}\ }\textbf {\bibinfo {volume} {6}},\
  \bibinfo {pages} {8629} (\bibinfo {year} {2015})}\BibitemShut {NoStop}%
\bibitem [{\citenamefont {Ledwith}\ \emph {et~al.}(2023)\citenamefont
  {Ledwith}, \citenamefont {Vishwanath},\ and\ \citenamefont
  {Parker}}]{ledwith2023vortexability}%
  \BibitemOpen
  \bibfield  {author} {\bibinfo {author} {\bibfnamefont {P.~J.}\ \bibnamefont
  {Ledwith}}, \bibinfo {author} {\bibfnamefont {A.}~\bibnamefont
  {Vishwanath}},\ and\ \bibinfo {author} {\bibfnamefont {D.~E.}\ \bibnamefont
  {Parker}},\ }\bibfield  {title} {\bibinfo {title} {Vortexability: A unifying
  criterion for ideal fractional chern insulators},\ }\href
  {https://doi.org/10.1103/PhysRevB.108.205144} {\bibfield  {journal} {\bibinfo
   {journal} {Phys. Rev. B}\ }\textbf {\bibinfo {volume} {108}},\ \bibinfo
  {pages} {205144} (\bibinfo {year} {2023})}\BibitemShut {NoStop}%
\bibitem [{\citenamefont {Macaluso}\ \emph {et~al.}(2020)\citenamefont
  {Macaluso}, \citenamefont {Comparin}, \citenamefont {Umucal{\i}lar},
  \citenamefont {Gerster}, \citenamefont {Montangero}, \citenamefont {Rizzi},\
  and\ \citenamefont {Carusotto}}]{macaluso2020charge}%
  \BibitemOpen
  \bibfield  {author} {\bibinfo {author} {\bibfnamefont {E.}~\bibnamefont
  {Macaluso}}, \bibinfo {author} {\bibfnamefont {T.}~\bibnamefont {Comparin}},
  \bibinfo {author} {\bibfnamefont {R.~O.}\ \bibnamefont {Umucal{\i}lar}},
  \bibinfo {author} {\bibfnamefont {M.}~\bibnamefont {Gerster}}, \bibinfo
  {author} {\bibfnamefont {S.}~\bibnamefont {Montangero}}, \bibinfo {author}
  {\bibfnamefont {M.}~\bibnamefont {Rizzi}},\ and\ \bibinfo {author}
  {\bibfnamefont {I.}~\bibnamefont {Carusotto}},\ }\bibfield  {title} {\bibinfo
  {title} {Charge and statistics of lattice quasiholes from density
  measurements: A tree tensor network study},\ }\href@noop {} {\bibfield
  {journal} {\bibinfo  {journal} {Physical Review Research}\ }\textbf {\bibinfo
  {volume} {2}},\ \bibinfo {pages} {013145} (\bibinfo {year}
  {2020})}\BibitemShut {NoStop}%
\bibitem [{\citenamefont {Ra{\v{c}}i{\=u}nas}\ \emph
  {et~al.}(2018)\citenamefont {Ra{\v{c}}i{\=u}nas}, \citenamefont {{\"U}nal},
  \citenamefont {Anisimovas},\ and\ \citenamefont
  {Eckardt}}]{ravciunas2018creating}%
  \BibitemOpen
  \bibfield  {author} {\bibinfo {author} {\bibfnamefont {M.}~\bibnamefont
  {Ra{\v{c}}i{\=u}nas}}, \bibinfo {author} {\bibfnamefont {F.~N.}\ \bibnamefont
  {{\"U}nal}}, \bibinfo {author} {\bibfnamefont {E.}~\bibnamefont
  {Anisimovas}},\ and\ \bibinfo {author} {\bibfnamefont {A.}~\bibnamefont
  {Eckardt}},\ }\bibfield  {title} {\bibinfo {title} {Creating, probing, and
  manipulating fractionally charged excitations of fractional {Chern}
  insulators in optical lattices},\ }\href
  {https://doi.org/10.1103/PhysRevA.98.063621} {\bibfield  {journal} {\bibinfo
  {journal} {Phys. Rev. A}\ }\textbf {\bibinfo {volume} {98}},\ \bibinfo
  {pages} {063621} (\bibinfo {year} {2018})}\BibitemShut {NoStop}%
\bibitem [{\citenamefont {Liu}\ \emph {et~al.}(2025)\citenamefont {Liu},
  \citenamefont {Li}, \citenamefont {Shi},\ and\ \citenamefont
  {Wu}}]{Liu2025FCIquasiparticle}%
  \BibitemOpen
  \bibfield  {author} {\bibinfo {author} {\bibfnamefont {Z.}~\bibnamefont
  {Liu}}, \bibinfo {author} {\bibfnamefont {B.}~\bibnamefont {Li}}, \bibinfo
  {author} {\bibfnamefont {Y.}~\bibnamefont {Shi}},\ and\ \bibinfo {author}
  {\bibfnamefont {F.}~\bibnamefont {Wu}},\ }\bibfield  {title} {\bibinfo
  {title} {Characterization of fractional chern insulator quasiparticles in
  twisted homobilayer mote$_2$},\ }\href {https://doi.org/10.1103/nddl-729x}
  {\bibfield  {journal} {\bibinfo  {journal} {Phys. Rev. B}\ }\textbf {\bibinfo
  {volume} {112}},\ \bibinfo {pages} {245104} (\bibinfo {year} {2025})},\
  \Eprint {https://arxiv.org/abs/2507.04056} {arXiv:2507.04056} \BibitemShut
  {NoStop}%
\bibitem [{\citenamefont {Palm}\ \emph {et~al.}(2026)\citenamefont {Palm},
  \citenamefont {Mostaan}, \citenamefont {Goldman},\ and\ \citenamefont
  {Grusdt}}]{palm2026interferometric}%
  \BibitemOpen
  \bibfield  {author} {\bibinfo {author} {\bibfnamefont {F.~A.}\ \bibnamefont
  {Palm}}, \bibinfo {author} {\bibfnamefont {N.}~\bibnamefont {Mostaan}},
  \bibinfo {author} {\bibfnamefont {N.}~\bibnamefont {Goldman}},\ and\ \bibinfo
  {author} {\bibfnamefont {F.}~\bibnamefont {Grusdt}},\ }\bibfield  {title}
  {\bibinfo {title} {Interferometric braiding of anyons in chern insulators},\
  }\href@noop {} {\bibfield  {journal} {\bibinfo  {journal} {Physical Review
  B}\ }\textbf {\bibinfo {volume} {113}},\ \bibinfo {pages} {155419} (\bibinfo
  {year} {2026})}\BibitemShut {NoStop}%
\bibitem [{\citenamefont {Wang}\ \emph {et~al.}(2026)\citenamefont {Wang},
  \citenamefont {Vashisht}, \citenamefont {Palm}, \citenamefont {Grusdt},
  \citenamefont {Vanderstraeten},\ and\ \citenamefont
  {Goldman}}]{wang2026anyon}%
  \BibitemOpen
  \bibfield  {author} {\bibinfo {author} {\bibfnamefont {B.}~\bibnamefont
  {Wang}}, \bibinfo {author} {\bibfnamefont {A.}~\bibnamefont {Vashisht}},
  \bibinfo {author} {\bibfnamefont {F.~A.}\ \bibnamefont {Palm}}, \bibinfo
  {author} {\bibfnamefont {F.}~\bibnamefont {Grusdt}}, \bibinfo {author}
  {\bibfnamefont {L.}~\bibnamefont {Vanderstraeten}},\ and\ \bibinfo {author}
  {\bibfnamefont {N.}~\bibnamefont {Goldman}},\ }\bibfield  {title} {\bibinfo
  {title} {Anyon-impurity bound states in quantum-engineered fractional chern
  insulators},\ }\href@noop {} {\bibfield  {journal} {\bibinfo  {journal}
  {arXiv preprint arXiv:2608.06233}\ } (\bibinfo {year} {2026})}\BibitemShut
  {NoStop}%
\bibitem [{\citenamefont {Wagner}\ and\ \citenamefont
  {Neupert}(2026)}]{cfm5-wgz1}%
  \BibitemOpen
  \bibfield  {author} {\bibinfo {author} {\bibfnamefont {G.}~\bibnamefont
  {Wagner}}\ and\ \bibinfo {author} {\bibfnamefont {T.}~\bibnamefont
  {Neupert}},\ }\bibfield  {title} {\bibinfo {title} {Sensing the binding and
  unbinding of anyons at impurities},\ }\href
  {https://doi.org/10.1103/cfm5-wgz1} {\bibfield  {journal} {\bibinfo
  {journal} {Phys. Rev. Res.}\ }\textbf {\bibinfo {volume} {8}},\ \bibinfo
  {pages} {013263} (\bibinfo {year} {2026})}\BibitemShut {NoStop}%
\bibitem [{\citenamefont {Adlong}\ \emph {et~al.}(2026)\citenamefont {Adlong},
  \citenamefont {Christianen}, \citenamefont {Imamoglu},\ and\ \citenamefont
  {Wagner}}]{adlong2026anyonsimpuritiesfractionalcharge}%
  \BibitemOpen
  \bibfield  {author} {\bibinfo {author} {\bibfnamefont {H.~S.}\ \bibnamefont
  {Adlong}}, \bibinfo {author} {\bibfnamefont {A.}~\bibnamefont {Christianen}},
  \bibinfo {author} {\bibfnamefont {A.}~\bibnamefont {Imamoglu}},\ and\
  \bibinfo {author} {\bibfnamefont {G.}~\bibnamefont {Wagner}},\ }\href
  {https://arxiv.org/abs/2610.20522} {\bibinfo {title} {Anyons at impurities:
  Fractional charge measurement and "wrong sign" binding}} (\bibinfo {year}
  {2026}),\ \Eprint {https://arxiv.org/abs/2610.20522} {arXiv:2610.20522
  [cond-mat.str-el]} \BibitemShut {NoStop}%
\bibitem [{\citenamefont {Lu}\ and\ \citenamefont
  {Santos}(2026{\natexlab{a}})}]{Lu_2026}%
  \BibitemOpen
  \bibfield  {author} {\bibinfo {author} {\bibfnamefont {T.}~\bibnamefont
  {Lu}}\ and\ \bibinfo {author} {\bibfnamefont {L.~H.}\ \bibnamefont
  {Santos}},\ }\bibfield  {title} {\bibinfo {title} {Exciton-anyon binding in
  fractional chern insulators: Spectral fingerprints},\ }\bibfield  {journal}
  {\bibinfo  {journal} {Physical Review Letters}\ }\textbf {\bibinfo {volume}
  {136}},\ \href {https://doi.org/10.1103/8bp8-g8g4} {10.1103/8bp8-g8g4}
  (\bibinfo {year} {2026}{\natexlab{a}})\BibitemShut {NoStop}%
\bibitem [{\citenamefont {Lu}\ and\ \citenamefont
  {Santos}(2026{\natexlab{b}})}]{lu2026anyonboundstatesmote2}%
  \BibitemOpen
  \bibfield  {author} {\bibinfo {author} {\bibfnamefont {T.}~\bibnamefont
  {Lu}}\ and\ \bibinfo {author} {\bibfnamefont {L.~H.}\ \bibnamefont
  {Santos}},\ }\href {https://arxiv.org/abs/2610.22473} {\bibinfo {title}
  {Anyon bound states in mote$_2$ fractional chern insulators}} (\bibinfo
  {year} {2026}{\natexlab{b}}),\ \Eprint {https://arxiv.org/abs/2610.22473}
  {arXiv:2610.22473 [cond-mat.str-el]} \BibitemShut {NoStop}%
\bibitem [{\citenamefont {Thouless}(1984)}]{thouless1984wannier}%
  \BibitemOpen
  \bibfield  {author} {\bibinfo {author} {\bibfnamefont {D.}~\bibnamefont
  {Thouless}},\ }\bibfield  {title} {\bibinfo {title} {Wannier functions for
  magnetic sub-bands},\ }\href@noop {} {\bibfield  {journal} {\bibinfo
  {journal} {Journal of Physics C: Solid State Physics}\ }\textbf {\bibinfo
  {volume} {17}},\ \bibinfo {pages} {L325} (\bibinfo {year}
  {1984})}\BibitemShut {NoStop}%
\bibitem [{\citenamefont {Brouder}\ \emph {et~al.}(2007)\citenamefont
  {Brouder}, \citenamefont {Panati}, \citenamefont {Calandra}, \citenamefont
  {Mourougane},\ and\ \citenamefont {Marzari}}]{brouder2007exponential}%
  \BibitemOpen
  \bibfield  {author} {\bibinfo {author} {\bibfnamefont {C.}~\bibnamefont
  {Brouder}}, \bibinfo {author} {\bibfnamefont {G.}~\bibnamefont {Panati}},
  \bibinfo {author} {\bibfnamefont {M.}~\bibnamefont {Calandra}}, \bibinfo
  {author} {\bibfnamefont {C.}~\bibnamefont {Mourougane}},\ and\ \bibinfo
  {author} {\bibfnamefont {N.}~\bibnamefont {Marzari}},\ }\bibfield  {title}
  {\bibinfo {title} {Exponential localization of wannier functions in
  insulators},\ }\href@noop {} {\bibfield  {journal} {\bibinfo  {journal}
  {Physical review letters}\ }\textbf {\bibinfo {volume} {98}},\ \bibinfo
  {pages} {046402} (\bibinfo {year} {2007})}\BibitemShut {NoStop}%
\bibitem [{\citenamefont {Gerhard}\ \emph {et~al.}(2026)\citenamefont
  {Gerhard}, \citenamefont {Wang}, \citenamefont {Cerjan},\ and\ \citenamefont
  {Benalcazar}}]{SpatLoc}%
  \BibitemOpen
  \bibfield  {author} {\bibinfo {author} {\bibfnamefont {H.}~\bibnamefont
  {Gerhard}}, \bibinfo {author} {\bibfnamefont {Y.}~\bibnamefont {Wang}},
  \bibinfo {author} {\bibfnamefont {A.}~\bibnamefont {Cerjan}},\ and\ \bibinfo
  {author} {\bibfnamefont {W.~A.}\ \bibnamefont {Benalcazar}},\ }\bibfield
  {title} {\bibinfo {title} {A spatial localizer for electrons in insulators},\
  }\href@noop {} {\bibfield  {journal} {\bibinfo  {journal} {arXiv:2603.13206}\
  } (\bibinfo {year} {2026})},\ \Eprint {https://arxiv.org/abs/2603.13206}
  {arXiv:2603.13206 [cond-mat.mtrl-sci]} \BibitemShut {NoStop}%
\bibitem [{\citenamefont {Gerhard}\ and\ \citenamefont
  {Benalcazar}(2026)}]{gerhard2026spatiallocalizerconstituentresolvedexciton}%
  \BibitemOpen
  \bibfield  {author} {\bibinfo {author} {\bibfnamefont {H.}~\bibnamefont
  {Gerhard}}\ and\ \bibinfo {author} {\bibfnamefont {W.~A.}\ \bibnamefont
  {Benalcazar}},\ }\href {https://arxiv.org/abs/2609.01728} {\bibinfo {title}
  {A spatial localizer for constituent-resolved exciton wannier functions}}
  (\bibinfo {year} {2026}),\ \Eprint {https://arxiv.org/abs/2609.01728}
  {arXiv:2609.01728 [cond-mat.mtrl-sci]} \BibitemShut {NoStop}%
\bibitem [{\citenamefont {Sun}\ \emph {et~al.}(2011)\citenamefont {Sun},
  \citenamefont {Gu}, \citenamefont {Katsura},\ and\ \citenamefont
  {Das~Sarma}}]{PhysRevLett.106.236803}%
  \BibitemOpen
  \bibfield  {author} {\bibinfo {author} {\bibfnamefont {K.}~\bibnamefont
  {Sun}}, \bibinfo {author} {\bibfnamefont {Z.}~\bibnamefont {Gu}}, \bibinfo
  {author} {\bibfnamefont {H.}~\bibnamefont {Katsura}},\ and\ \bibinfo {author}
  {\bibfnamefont {S.}~\bibnamefont {Das~Sarma}},\ }\bibfield  {title} {\bibinfo
  {title} {Nearly flatbands with nontrivial topology},\ }\href
  {https://doi.org/10.1103/PhysRevLett.106.236803} {\bibfield  {journal}
  {\bibinfo  {journal} {Phys. Rev. Lett.}\ }\textbf {\bibinfo {volume} {106}},\
  \bibinfo {pages} {236803} (\bibinfo {year} {2011})}\BibitemShut {NoStop}%
\bibitem [{\citenamefont {Nakahara}(2018)}]{nakahara2018geometry}%
  \BibitemOpen
  \bibfield  {author} {\bibinfo {author} {\bibfnamefont {M.}~\bibnamefont
  {Nakahara}},\ }\href@noop {} {\emph {\bibinfo {title} {Geometry, topology and
  physics}}}\ (\bibinfo  {publisher} {CRC press},\ \bibinfo {year}
  {2018})\BibitemShut {NoStop}%
\bibitem [{\citenamefont {Okuma}(2024)}]{PhysRevB.110.245112}%
  \BibitemOpen
  \bibfield  {author} {\bibinfo {author} {\bibfnamefont {N.}~\bibnamefont
  {Okuma}},\ }\bibfield  {title} {\bibinfo {title} {Constructing vortex
  functions and basis states of chern insulators: Ideal condition, inequality
  from index theorem, and coherentlike states on the von neumann lattice},\
  }\href {https://doi.org/10.1103/PhysRevB.110.245112} {\bibfield  {journal}
  {\bibinfo  {journal} {Phys. Rev. B}\ }\textbf {\bibinfo {volume} {110}},\
  \bibinfo {pages} {245112} (\bibinfo {year} {2024})}\BibitemShut {NoStop}%
\bibitem [{Note1()}]{Note1}%
  \BibitemOpen
  \bibinfo {note} {Throughout, $\protect \bm {K}$ denotes magnetic momentum in
  the continuum FQH system and crystal momentum in the FCI.}\BibitemShut
  {Stop}%
\bibitem [{SM()}]{SM}%
  \BibitemOpen
  \href@noop {} {}\bibinfo {note} {See Supplemental Material (appended below) for the periodic density operators, the comparison
  with Resta's position operator, the degeneracy of Laughlin quasiholes on the
  torus, details of the many-body Spatial Localizer and its uncertainty bounds,
  exact-diagonalization details and additional numerical results, the
  vortexability test of the kagome band, the derivation of the anyon parallel
  transport and Chern number, and the gauge-invariant vorticity of the
  coherent-state coefficients.}\BibitemShut {Stop}%
\bibitem [{Note2()}]{Note2}%
  \BibitemOpen
  \bibinfo {note} {Concretely, in our calculations we adopt the Weyl-Brauer
  representation of the Clifford elements~\cite
  {brauer1935spinors}}\BibitemShut {NoStop}%
\bibitem [{\citenamefont {Greiter}\ \emph {et~al.}(2016)\citenamefont
  {Greiter}, \citenamefont {Schnells},\ and\ \citenamefont
  {Thomale}}]{PhysRevB.93.245156}%
  \BibitemOpen
  \bibfield  {author} {\bibinfo {author} {\bibfnamefont {M.}~\bibnamefont
  {Greiter}}, \bibinfo {author} {\bibfnamefont {V.}~\bibnamefont {Schnells}},\
  and\ \bibinfo {author} {\bibfnamefont {R.}~\bibnamefont {Thomale}},\
  }\bibfield  {title} {\bibinfo {title} {Laughlin states and their
  quasiparticle excitations on the torus},\ }\href
  {https://doi.org/10.1103/PhysRevB.93.245156} {\bibfield  {journal} {\bibinfo
  {journal} {Phys. Rev. B}\ }\textbf {\bibinfo {volume} {93}},\ \bibinfo
  {pages} {245156} (\bibinfo {year} {2016})}\BibitemShut {NoStop}%
\bibitem [{\citenamefont {Nielsen}\ \emph {et~al.}(2013)\citenamefont
  {Nielsen}, \citenamefont {Sierra},\ and\ \citenamefont
  {Cirac}}]{nielsen2013local}%
  \BibitemOpen
  \bibfield  {author} {\bibinfo {author} {\bibfnamefont {A.~E.~B.}\
  \bibnamefont {Nielsen}}, \bibinfo {author} {\bibfnamefont {G.}~\bibnamefont
  {Sierra}},\ and\ \bibinfo {author} {\bibfnamefont {J.~I.}\ \bibnamefont
  {Cirac}},\ }\bibfield  {title} {\bibinfo {title} {Local models of fractional
  quantum {Hall} states in lattices and physical implementation},\ }\href
  {https://doi.org/10.1038/ncomms3864} {\bibfield  {journal} {\bibinfo
  {journal} {Nature Communications}\ }\textbf {\bibinfo {volume} {4}},\
  \bibinfo {pages} {2864} (\bibinfo {year} {2013})}\BibitemShut {NoStop}%
\bibitem [{\citenamefont {Liu}\ \emph {et~al.}(2015)\citenamefont {Liu},
  \citenamefont {Bhatt},\ and\ \citenamefont {Regnault}}]{PhysRevB.91.045126}%
  \BibitemOpen
  \bibfield  {author} {\bibinfo {author} {\bibfnamefont {Z.}~\bibnamefont
  {Liu}}, \bibinfo {author} {\bibfnamefont {R.~N.}\ \bibnamefont {Bhatt}},\
  and\ \bibinfo {author} {\bibfnamefont {N.}~\bibnamefont {Regnault}},\
  }\bibfield  {title} {\bibinfo {title} {Characterization of quasiholes in
  fractional chern insulators},\ }\href
  {https://doi.org/10.1103/PhysRevB.91.045126} {\bibfield  {journal} {\bibinfo
  {journal} {Phys. Rev. B}\ }\textbf {\bibinfo {volume} {91}},\ \bibinfo
  {pages} {045126} (\bibinfo {year} {2015})}\BibitemShut {NoStop}%
\bibitem [{\citenamefont {Robertson}(1929)}]{robertson1929uncertainty}%
  \BibitemOpen
  \bibfield  {author} {\bibinfo {author} {\bibfnamefont {H.~P.}\ \bibnamefont
  {Robertson}},\ }\bibfield  {title} {\bibinfo {title} {The uncertainty
  principle},\ }\href@noop {} {\bibfield  {journal} {\bibinfo  {journal}
  {Physical Review}\ }\textbf {\bibinfo {volume} {34}},\ \bibinfo {pages} {163}
  (\bibinfo {year} {1929})}\BibitemShut {NoStop}%
\bibitem [{\citenamefont {Rezayi}\ and\ \citenamefont
  {Haldane}(1985)}]{PhysRevB.32.6924}%
  \BibitemOpen
  \bibfield  {author} {\bibinfo {author} {\bibfnamefont {E.~H.}\ \bibnamefont
  {Rezayi}}\ and\ \bibinfo {author} {\bibfnamefont {F.~D.~M.}\ \bibnamefont
  {Haldane}},\ }\bibfield  {title} {\bibinfo {title} {Incompressible states of
  the fractionally quantized hall effect in the presence of impurities: A
  finite-size study},\ }\href {https://doi.org/10.1103/PhysRevB.32.6924}
  {\bibfield  {journal} {\bibinfo  {journal} {Phys. Rev. B}\ }\textbf {\bibinfo
  {volume} {32}},\ \bibinfo {pages} {6924(R)} (\bibinfo {year}
  {1985})}\BibitemShut {NoStop}%
\bibitem [{\citenamefont {Trugman}\ and\ \citenamefont
  {Kivelson}(1985)}]{Trugman1985}%
  \BibitemOpen
  \bibfield  {author} {\bibinfo {author} {\bibfnamefont {S.~A.}\ \bibnamefont
  {Trugman}}\ and\ \bibinfo {author} {\bibfnamefont {S.}~\bibnamefont
  {Kivelson}},\ }\bibfield  {title} {\bibinfo {title} {Exact results for the
  fractional quantum {Hall} effect with general interactions},\ }\href
  {https://doi.org/10.1103/PhysRevB.31.5280} {\bibfield  {journal} {\bibinfo
  {journal} {Phys. Rev. B}\ }\textbf {\bibinfo {volume} {31}},\ \bibinfo
  {pages} {5280} (\bibinfo {year} {1985})}\BibitemShut {NoStop}%
\bibitem [{\citenamefont {Lee}\ \emph {et~al.}(2015)\citenamefont {Lee},
  \citenamefont {Papi\'{c}},\ and\ \citenamefont {Thomale}}]{Papic2015}%
  \BibitemOpen
  \bibfield  {author} {\bibinfo {author} {\bibfnamefont {C.~H.}\ \bibnamefont
  {Lee}}, \bibinfo {author} {\bibfnamefont {Z.}~\bibnamefont {Papi\'{c}}},\
  and\ \bibinfo {author} {\bibfnamefont {R.}~\bibnamefont {Thomale}},\
  }\bibfield  {title} {\bibinfo {title} {Geometric construction of quantum
  {Hall} clustering {Hamiltonians}},\ }\href
  {https://doi.org/10.1103/PhysRevX.5.041003} {\bibfield  {journal} {\bibinfo
  {journal} {Phys. Rev. X}\ }\textbf {\bibinfo {volume} {5}},\ \bibinfo {pages}
  {041003} (\bibinfo {year} {2015})}\BibitemShut {NoStop}%
\bibitem [{\citenamefont {Yoshioka}(1984)}]{yoshioka1984ground}%
  \BibitemOpen
  \bibfield  {author} {\bibinfo {author} {\bibfnamefont {D.}~\bibnamefont
  {Yoshioka}},\ }\bibfield  {title} {\bibinfo {title} {Ground state of the
  two-dimensional charged particles in a strong magnetic field and the
  fractional quantum hall effect},\ }\href@noop {} {\bibfield  {journal}
  {\bibinfo  {journal} {Physical Review B}\ }\textbf {\bibinfo {volume} {29}},\
  \bibinfo {pages} {6833} (\bibinfo {year} {1984})}\BibitemShut {NoStop}%
\bibitem [{\citenamefont {Yoshioka}\ \emph {et~al.}(1983)\citenamefont
  {Yoshioka}, \citenamefont {Halperin},\ and\ \citenamefont
  {Lee}}]{PhysRevLett.50.1219}%
  \BibitemOpen
  \bibfield  {author} {\bibinfo {author} {\bibfnamefont {D.}~\bibnamefont
  {Yoshioka}}, \bibinfo {author} {\bibfnamefont {B.~I.}\ \bibnamefont
  {Halperin}},\ and\ \bibinfo {author} {\bibfnamefont {P.~A.}\ \bibnamefont
  {Lee}},\ }\bibfield  {title} {\bibinfo {title} {Ground state of
  two-dimensional electrons in strong magnetic fields and $\frac{1}{3}$
  quantized hall effect},\ }\href {https://doi.org/10.1103/PhysRevLett.50.1219}
  {\bibfield  {journal} {\bibinfo  {journal} {Phys. Rev. Lett.}\ }\textbf
  {\bibinfo {volume} {50}},\ \bibinfo {pages} {1219} (\bibinfo {year}
  {1983})}\BibitemShut {NoStop}%
\bibitem [{\citenamefont {Jain}(1990)}]{PhysRevB.41.7653}%
  \BibitemOpen
  \bibfield  {author} {\bibinfo {author} {\bibfnamefont {J.~K.}\ \bibnamefont
  {Jain}},\ }\bibfield  {title} {\bibinfo {title} {Theory of the fractional
  quantum hall effect},\ }\href {https://doi.org/10.1103/PhysRevB.41.7653}
  {\bibfield  {journal} {\bibinfo  {journal} {Phys. Rev. B}\ }\textbf {\bibinfo
  {volume} {41}},\ \bibinfo {pages} {7653} (\bibinfo {year}
  {1990})}\BibitemShut {NoStop}%
\bibitem [{\citenamefont {Dev}\ and\ \citenamefont
  {Jain}(1992)}]{PhysRevB.45.1223}%
  \BibitemOpen
  \bibfield  {author} {\bibinfo {author} {\bibfnamefont {G.}~\bibnamefont
  {Dev}}\ and\ \bibinfo {author} {\bibfnamefont {J.~K.}\ \bibnamefont {Jain}},\
  }\bibfield  {title} {\bibinfo {title} {Jastrow-slater trial wave functions
  for the fractional quantum hall effect: Results for few-particle systems},\
  }\href {https://doi.org/10.1103/PhysRevB.45.1223} {\bibfield  {journal}
  {\bibinfo  {journal} {Phys. Rev. B}\ }\textbf {\bibinfo {volume} {45}},\
  \bibinfo {pages} {1223} (\bibinfo {year} {1992})}\BibitemShut {NoStop}%
\bibitem [{\citenamefont {Jain}\ and\ \citenamefont
  {Kamilla}(1997)}]{doi:10.1142/S0217979297001301}%
  \BibitemOpen
  \bibfield  {author} {\bibinfo {author} {\bibfnamefont {J.~K.}\ \bibnamefont
  {Jain}}\ and\ \bibinfo {author} {\bibfnamefont {R.~K.}\ \bibnamefont
  {Kamilla}},\ }\bibfield  {title} {\bibinfo {title} {Composite fermions in the
  hilbert space of the lowest electronic landau level},\ }\href
  {https://doi.org/10.1142/S0217979297001301} {\bibfield  {journal} {\bibinfo
  {journal} {International Journal of Modern Physics B}\ }\textbf {\bibinfo
  {volume} {11}},\ \bibinfo {pages} {2621} (\bibinfo {year} {1997})},\ \Eprint
  {https://arxiv.org/abs/https://doi.org/10.1142/S0217979297001301}
  {https://doi.org/10.1142/S0217979297001301} \BibitemShut {NoStop}%
\bibitem [{\citenamefont {Klauder}(1963)}]{klauder1963continuous}%
  \BibitemOpen
  \bibfield  {author} {\bibinfo {author} {\bibfnamefont {J.~R.}\ \bibnamefont
  {Klauder}},\ }\bibfield  {title} {\bibinfo {title} {Continuous-representation
  theory. ii. generalized relation between quantum and classical dynamics},\
  }\href@noop {} {\bibfield  {journal} {\bibinfo  {journal} {Journal of
  Mathematical Physics}\ }\textbf {\bibinfo {volume} {4}},\ \bibinfo {pages}
  {1058} (\bibinfo {year} {1963})}\BibitemShut {NoStop}%
\bibitem [{\citenamefont {Klauder}\ and\ \citenamefont
  {Skagerstam}(1985)}]{klauder1985coherent}%
  \BibitemOpen
  \bibfield  {author} {\bibinfo {author} {\bibfnamefont {J.~R.}\ \bibnamefont
  {Klauder}}\ and\ \bibinfo {author} {\bibfnamefont {B.-S.}\ \bibnamefont
  {Skagerstam}},\ }\href@noop {} {\emph {\bibinfo {title} {Coherent states:
  applications in physics and mathematical physics}}}\ (\bibinfo  {publisher}
  {World scientific},\ \bibinfo {year} {1985})\BibitemShut {NoStop}%
\bibitem [{\citenamefont {Davenport}\ \emph {et~al.}(2026)\citenamefont
  {Davenport}, \citenamefont {Hwang}, \citenamefont {Knolle},\ and\
  \citenamefont
  {Schindler}}]{davenport2026compositequantumgeometrysemiclassical}%
  \BibitemOpen
  \bibfield  {author} {\bibinfo {author} {\bibfnamefont {H.}~\bibnamefont
  {Davenport}}, \bibinfo {author} {\bibfnamefont {Y.}~\bibnamefont {Hwang}},
  \bibinfo {author} {\bibfnamefont {J.}~\bibnamefont {Knolle}},\ and\ \bibinfo
  {author} {\bibfnamefont {F.}~\bibnamefont {Schindler}},\ }\href
  {https://arxiv.org/abs/2606.12525} {\bibinfo {title} {Composite quantum
  geometry and semiclassical dynamics}} (\bibinfo {year} {2026}),\ \Eprint
  {https://arxiv.org/abs/2606.12525} {arXiv:2606.12525 [cond-mat.mes-hall]}
  \BibitemShut {NoStop}%
\bibitem [{\citenamefont {Hwang}\ \emph
  {et~al.}(2026{\natexlab{a}})\citenamefont {Hwang}, \citenamefont
  {Davenport},\ and\ \citenamefont {Schindler}}]{hwang2026exciton}%
  \BibitemOpen
  \bibfield  {author} {\bibinfo {author} {\bibfnamefont {Y.}~\bibnamefont
  {Hwang}}, \bibinfo {author} {\bibfnamefont {H.}~\bibnamefont {Davenport}},\
  and\ \bibinfo {author} {\bibfnamefont {F.}~\bibnamefont {Schindler}},\
  }\bibfield  {title} {\bibinfo {title} {Exciton alchemy: Chern excitons from
  trivial bands},\ }\href@noop {} {\bibfield  {journal} {\bibinfo  {journal}
  {arXiv preprint arXiv:2608.07372}\ } (\bibinfo {year}
  {2026}{\natexlab{a}})}\BibitemShut {NoStop}%
\bibitem [{\citenamefont {Paiva}\ \emph {et~al.}(2024)\citenamefont {Paiva},
  \citenamefont {Holder},\ and\ \citenamefont
  {Ilan}}]{paiva2024shiftpolarizationexcitonsquantum}%
  \BibitemOpen
  \bibfield  {author} {\bibinfo {author} {\bibfnamefont {C.}~\bibnamefont
  {Paiva}}, \bibinfo {author} {\bibfnamefont {T.}~\bibnamefont {Holder}},\ and\
  \bibinfo {author} {\bibfnamefont {R.}~\bibnamefont {Ilan}},\ }\href
  {https://arxiv.org/abs/2408.10300} {\bibinfo {title} {Shift and polarization
  of excitons from quantum geometry}} (\bibinfo {year} {2024}),\ \Eprint
  {https://arxiv.org/abs/2408.10300} {arXiv:2408.10300 [cond-mat.mes-hall]}
  \BibitemShut {NoStop}%
\bibitem [{\citenamefont {Hwang}\ \emph
  {et~al.}(2026{\natexlab{b}})\citenamefont {Hwang}, \citenamefont
  {Davenport},\ and\ \citenamefont
  {Schindler}}]{hwang2026stablewavefunctionzerosindicate}%
  \BibitemOpen
  \bibfield  {author} {\bibinfo {author} {\bibfnamefont {Y.}~\bibnamefont
  {Hwang}}, \bibinfo {author} {\bibfnamefont {H.}~\bibnamefont {Davenport}},\
  and\ \bibinfo {author} {\bibfnamefont {F.}~\bibnamefont {Schindler}},\ }\href
  {https://arxiv.org/abs/2604.21643} {\bibinfo {title} {Stable wave-function
  zeros indicate exciton topology}} (\bibinfo {year} {2026}{\natexlab{b}}),\
  \Eprint {https://arxiv.org/abs/2604.21643} {arXiv:2604.21643
  [cond-mat.mes-hall]} \BibitemShut {NoStop}%
\bibitem [{\citenamefont {Li}\ \emph {et~al.}(2024)\citenamefont {Li},
  \citenamefont {Dong}, \citenamefont {Ledwith},\ and\ \citenamefont
  {Khalaf}}]{li2024constraints}%
  \BibitemOpen
  \bibfield  {author} {\bibinfo {author} {\bibfnamefont {Q.}~\bibnamefont
  {Li}}, \bibinfo {author} {\bibfnamefont {J.}~\bibnamefont {Dong}}, \bibinfo
  {author} {\bibfnamefont {P.~J.}\ \bibnamefont {Ledwith}},\ and\ \bibinfo
  {author} {\bibfnamefont {E.}~\bibnamefont {Khalaf}},\ }\href
  {https://arxiv.org/abs/2407.02561} {\bibinfo {title} {Constraints on real
  space representations of chern bands}} (\bibinfo {year} {2024}),\ \Eprint
  {https://arxiv.org/abs/2407.02561} {arXiv:2407.02561 [cond-mat.str-el]}
  \BibitemShut {NoStop}%
\bibitem [{\citenamefont {Fukui}\ \emph {et~al.}(2005)\citenamefont {Fukui},
  \citenamefont {Hatsugai},\ and\ \citenamefont {Suzuki}}]{fukui2005chern}%
  \BibitemOpen
  \bibfield  {author} {\bibinfo {author} {\bibfnamefont {T.}~\bibnamefont
  {Fukui}}, \bibinfo {author} {\bibfnamefont {Y.}~\bibnamefont {Hatsugai}},\
  and\ \bibinfo {author} {\bibfnamefont {H.}~\bibnamefont {Suzuki}},\
  }\bibfield  {title} {\bibinfo {title} {Chern numbers in discretized brillouin
  zone: Efficient method of computing (spin) hall conductances},\ }\href@noop
  {} {\bibfield  {journal} {\bibinfo  {journal} {Journal of the Physical
  Society of Japan}\ }\textbf {\bibinfo {volume} {74}},\ \bibinfo {pages}
  {1674} (\bibinfo {year} {2005})}\BibitemShut {NoStop}%
\bibitem [{\citenamefont {Vaidya}\ \emph {et~al.}(2024)\citenamefont {Vaidya},
  \citenamefont {Rechtsman},\ and\ \citenamefont
  {Benalcazar}}]{PhysRevLett.132.116602}%
  \BibitemOpen
  \bibfield  {author} {\bibinfo {author} {\bibfnamefont {S.}~\bibnamefont
  {Vaidya}}, \bibinfo {author} {\bibfnamefont {M.~C.}\ \bibnamefont
  {Rechtsman}},\ and\ \bibinfo {author} {\bibfnamefont {W.~A.}\ \bibnamefont
  {Benalcazar}},\ }\bibfield  {title} {\bibinfo {title} {Polarization and weak
  topology in chern insulators},\ }\href
  {https://doi.org/10.1103/PhysRevLett.132.116602} {\bibfield  {journal}
  {\bibinfo  {journal} {Phys. Rev. Lett.}\ }\textbf {\bibinfo {volume} {132}},\
  \bibinfo {pages} {116602} (\bibinfo {year} {2024})}\BibitemShut {NoStop}%
\bibitem [{\citenamefont {Benalcazar}\ \emph {et~al.}(2014)\citenamefont
  {Benalcazar}, \citenamefont {Teo},\ and\ \citenamefont
  {Hughes}}]{PhysRevB.89.224503}%
  \BibitemOpen
  \bibfield  {author} {\bibinfo {author} {\bibfnamefont {W.~A.}\ \bibnamefont
  {Benalcazar}}, \bibinfo {author} {\bibfnamefont {J.~C.~Y.}\ \bibnamefont
  {Teo}},\ and\ \bibinfo {author} {\bibfnamefont {T.~L.}\ \bibnamefont
  {Hughes}},\ }\bibfield  {title} {\bibinfo {title} {Classification of
  two-dimensional topological crystalline superconductors and majorana bound
  states at disclinations},\ }\href
  {https://doi.org/10.1103/PhysRevB.89.224503} {\bibfield  {journal} {\bibinfo
  {journal} {Phys. Rev. B}\ }\textbf {\bibinfo {volume} {89}},\ \bibinfo
  {pages} {224503} (\bibinfo {year} {2014})}\BibitemShut {NoStop}%
\bibitem [{\citenamefont {Ran}\ \emph {et~al.}(2009)\citenamefont {Ran},
  \citenamefont {Zhang},\ and\ \citenamefont {Vishwanath}}]{ran2009one}%
  \BibitemOpen
  \bibfield  {author} {\bibinfo {author} {\bibfnamefont {Y.}~\bibnamefont
  {Ran}}, \bibinfo {author} {\bibfnamefont {Y.}~\bibnamefont {Zhang}},\ and\
  \bibinfo {author} {\bibfnamefont {A.}~\bibnamefont {Vishwanath}},\ }\bibfield
   {title} {\bibinfo {title} {One-dimensional topologically protected modes in
  topological insulators with lattice dislocations},\ }\href@noop {} {\bibfield
   {journal} {\bibinfo  {journal} {Nature Physics}\ }\textbf {\bibinfo {volume}
  {5}},\ \bibinfo {pages} {298} (\bibinfo {year} {2009})}\BibitemShut {NoStop}%
\bibitem [{\citenamefont {Teo}\ and\ \citenamefont
  {Kane}(2010)}]{PhysRevB.82.115120}%
  \BibitemOpen
  \bibfield  {author} {\bibinfo {author} {\bibfnamefont {J.~C.~Y.}\
  \bibnamefont {Teo}}\ and\ \bibinfo {author} {\bibfnamefont {C.~L.}\
  \bibnamefont {Kane}},\ }\bibfield  {title} {\bibinfo {title} {Topological
  defects and gapless modes in insulators and superconductors},\ }\href
  {https://doi.org/10.1103/PhysRevB.82.115120} {\bibfield  {journal} {\bibinfo
  {journal} {Phys. Rev. B}\ }\textbf {\bibinfo {volume} {82}},\ \bibinfo
  {pages} {115120} (\bibinfo {year} {2010})}\BibitemShut {NoStop}%
\bibitem [{\citenamefont {Bombin}(2010)}]{PhysRevLett.105.030403}%
  \BibitemOpen
  \bibfield  {author} {\bibinfo {author} {\bibfnamefont {H.}~\bibnamefont
  {Bombin}},\ }\bibfield  {title} {\bibinfo {title} {Topological order with a
  twist: Ising anyons from an abelian model},\ }\href
  {https://doi.org/10.1103/PhysRevLett.105.030403} {\bibfield  {journal}
  {\bibinfo  {journal} {Phys. Rev. Lett.}\ }\textbf {\bibinfo {volume} {105}},\
  \bibinfo {pages} {030403} (\bibinfo {year} {2010})}\BibitemShut {NoStop}%
\bibitem [{\citenamefont {Gattu}\ and\ \citenamefont {Jain}(2025)}]{scl5-8pv6}%
  \BibitemOpen
  \bibfield  {author} {\bibinfo {author} {\bibfnamefont {M.}~\bibnamefont
  {Gattu}}\ and\ \bibinfo {author} {\bibfnamefont {J.~K.}\ \bibnamefont
  {Jain}},\ }\bibfield  {title} {\bibinfo {title} {Molecular anyons in the
  fractional quantum hall effect},\ }\href {https://doi.org/10.1103/scl5-8pv6}
  {\bibfield  {journal} {\bibinfo  {journal} {Phys. Rev. Lett.}\ }\textbf
  {\bibinfo {volume} {135}},\ \bibinfo {pages} {236601} (\bibinfo {year}
  {2025})}\BibitemShut {NoStop}%
\bibitem [{\citenamefont {Pu}\ \emph {et~al.}(2026)\citenamefont {Pu},
  \citenamefont {Peng},\ and\ \citenamefont
  {Adam}}]{pu2026compositefermionsidealchern}%
  \BibitemOpen
  \bibfield  {author} {\bibinfo {author} {\bibfnamefont {S.}~\bibnamefont
  {Pu}}, \bibinfo {author} {\bibfnamefont {L.}~\bibnamefont {Peng}},\ and\
  \bibinfo {author} {\bibfnamefont {S.}~\bibnamefont {Adam}},\ }\href
  {https://arxiv.org/abs/2610.01217} {\bibinfo {title} {Composite fermions in
  ideal chern bands}} (\bibinfo {year} {2026}),\ \Eprint
  {https://arxiv.org/abs/2610.01217} {arXiv:2610.01217 [cond-mat.mes-hall]}
  \BibitemShut {NoStop}%
\bibitem [{\citenamefont {Xiao}\ \emph {et~al.}(2026)\citenamefont {Xiao},
  \citenamefont {Wang},\ and\ \citenamefont
  {Ryu}}]{xiao2026anyonchernbandsopposite}%
  \BibitemOpen
  \bibfield  {author} {\bibinfo {author} {\bibfnamefont {Z.}~\bibnamefont
  {Xiao}}, \bibinfo {author} {\bibfnamefont {M.}~\bibnamefont {Wang}},\ and\
  \bibinfo {author} {\bibfnamefont {S.}~\bibnamefont {Ryu}},\ }\href
  {https://arxiv.org/abs/2610.30743} {\bibinfo {title} {Anyon chern bands of
  opposite sign and reversed drift}} (\bibinfo {year} {2026}),\ \Eprint
  {https://arxiv.org/abs/2610.30743} {arXiv:2610.30743 [cond-mat.mes-hall]}
  \BibitemShut {NoStop}%
\bibitem [{\citenamefont {Sarkar}\ and\ \citenamefont
  {Wan}(2026)}]{sarkar2026latticeinducedsplittingconnectivitytopology}%
  \BibitemOpen
  \bibfield  {author} {\bibinfo {author} {\bibfnamefont {S.}~\bibnamefont
  {Sarkar}}\ and\ \bibinfo {author} {\bibfnamefont {X.}~\bibnamefont {Wan}},\
  }\href {https://arxiv.org/abs/2611.02268} {\bibinfo {title} {Lattice-induced
  splitting, connectivity, and topology of anyon bands}} (\bibinfo {year}
  {2026}),\ \Eprint {https://arxiv.org/abs/2611.02268} {arXiv:2611.02268
  [cond-mat.str-el]} \BibitemShut {NoStop}%
\bibitem [{\citenamefont {Brauer}\ and\ \citenamefont
  {Weyl}(1935)}]{brauer1935spinors}%
  \BibitemOpen
  \bibfield  {author} {\bibinfo {author} {\bibfnamefont {R.}~\bibnamefont
  {Brauer}}\ and\ \bibinfo {author} {\bibfnamefont {H.}~\bibnamefont {Weyl}},\
  }\bibfield  {title} {\bibinfo {title} {Spinors in n dimensions},\ }\href@noop
  {} {\bibfield  {journal} {\bibinfo  {journal} {American Journal of
  Mathematics}\ }\textbf {\bibinfo {volume} {57}},\ \bibinfo {pages} {425}
  (\bibinfo {year} {1935})}\BibitemShut {NoStop}%
\end{thebibliography}

\begin{thebibliography}{25}%
\makeatletter
\providecommand \@ifxundefined [1]{%
 \@ifx{#1\undefined}
}%
\providecommand \@ifnum [1]{%
 \ifnum #1\expandafter \@firstoftwo
 \else \expandafter \@secondoftwo
 \fi
}%
\providecommand \@ifx [1]{%
 \ifx #1\expandafter \@firstoftwo
 \else \expandafter \@secondoftwo
 \fi
}%
\providecommand \natexlab [1]{#1}%
\providecommand \enquote  [1]{``#1''}%
\providecommand \bibnamefont  [1]{#1}%
\providecommand \bibfnamefont [1]{#1}%
\providecommand \citenamefont [1]{#1}%
\providecommand \href@noop [0]{\@secondoftwo}%
\providecommand \href [0]{\begingroup \@sanitize@url \@href}%
\providecommand \@href[1]{\@@startlink{#1}\@@href}%
\providecommand \@@href[1]{\endgroup#1\@@endlink}%
\providecommand \@sanitize@url [0]{\catcode `\\12\catcode `\$12\catcode
  `\&12\catcode `\#12\catcode `\^12\catcode `\_12\catcode `\%12\relax}%
\providecommand \@@startlink[1]{}%
\providecommand \@@endlink[0]{}%
\providecommand \url  [0]{\begingroup\@sanitize@url \@url }%
\providecommand \@url [1]{\endgroup\@href {#1}{\urlprefix }}%
\providecommand \urlprefix  [0]{URL }%
\providecommand \Eprint [0]{\href }%
\providecommand \doibase [0]{https://doi.org/}%
\providecommand \selectlanguage [0]{\@gobble}%
\providecommand \bibinfo  [0]{\@secondoftwo}%
\providecommand \bibfield  [0]{\@secondoftwo}%
\providecommand \translation [1]{[#1]}%
\providecommand \BibitemOpen [0]{}%
\providecommand \bibitemStop [0]{}%
\providecommand \bibitemNoStop [0]{.\EOS\space}%
\providecommand \EOS [0]{\spacefactor3000\relax}%
\providecommand \BibitemShut  [1]{\csname bibitem#1\endcsname}%
\let\auto@bib@innerbib\@empty
\bibitem [{\citenamefont {Resta}(1998)}]{SM:Resta1998}%
  \BibitemOpen
  \bibfield  {author} {\bibinfo {author} {\bibfnamefont {R.}~\bibnamefont
  {Resta}},\ }\bibfield  {title} {\bibinfo {title} {Quantum-mechanical position
  operator in extended systems},\ }\href
  {https://doi.org/10.1103/PhysRevLett.80.1800} {\bibfield  {journal} {\bibinfo
   {journal} {Phys. Rev. Lett.}\ }\textbf {\bibinfo {volume} {80}},\ \bibinfo
  {pages} {1800} (\bibinfo {year} {1998})}\BibitemShut {NoStop}%
\bibitem [{\citenamefont {Souza}\ \emph {et~al.}(2000)\citenamefont {Souza},
  \citenamefont {Wilkens},\ and\ \citenamefont
  {Martin}}]{SM:SouzaWilkensMartin2000}%
  \BibitemOpen
  \bibfield  {author} {\bibinfo {author} {\bibfnamefont {I.}~\bibnamefont
  {Souza}}, \bibinfo {author} {\bibfnamefont {T.}~\bibnamefont {Wilkens}},\
  and\ \bibinfo {author} {\bibfnamefont {R.~M.}\ \bibnamefont {Martin}},\
  }\bibfield  {title} {\bibinfo {title} {Polarization and localization in
  insulators: Generating function approach},\ }\href
  {https://doi.org/10.1103/PhysRevB.62.1666} {\bibfield  {journal} {\bibinfo
  {journal} {Phys. Rev. B}\ }\textbf {\bibinfo {volume} {62}},\ \bibinfo
  {pages} {1666} (\bibinfo {year} {2000})}\BibitemShut {NoStop}%
\bibitem [{\citenamefont {Resta}\ and\ \citenamefont
  {Sorella}(1999)}]{SM:PhysRevLett.82.370}%
  \BibitemOpen
  \bibfield  {author} {\bibinfo {author} {\bibfnamefont {R.}~\bibnamefont
  {Resta}}\ and\ \bibinfo {author} {\bibfnamefont {S.}~\bibnamefont
  {Sorella}},\ }\bibfield  {title} {\bibinfo {title} {Electron localization in
  the insulating state},\ }\href {https://doi.org/10.1103/PhysRevLett.82.370}
  {\bibfield  {journal} {\bibinfo  {journal} {Phys. Rev. Lett.}\ }\textbf
  {\bibinfo {volume} {82}},\ \bibinfo {pages} {370} (\bibinfo {year}
  {1999})}\BibitemShut {NoStop}%
\bibitem [{\citenamefont {Thouless}(1984)}]{SM:thouless1984wannier}%
  \BibitemOpen
  \bibfield  {author} {\bibinfo {author} {\bibfnamefont {D.}~\bibnamefont
  {Thouless}},\ }\bibfield  {title} {\bibinfo {title} {Wannier functions for
  magnetic sub-bands},\ }\href@noop {} {\bibfield  {journal} {\bibinfo
  {journal} {Journal of Physics C: Solid State Physics}\ }\textbf {\bibinfo
  {volume} {17}},\ \bibinfo {pages} {L325} (\bibinfo {year}
  {1984})}\BibitemShut {NoStop}%
\bibitem [{\citenamefont {Brouder}\ \emph {et~al.}(2007)\citenamefont
  {Brouder}, \citenamefont {Panati}, \citenamefont {Calandra}, \citenamefont
  {Mourougane},\ and\ \citenamefont {Marzari}}]{SM:brouder2007exponential}%
  \BibitemOpen
  \bibfield  {author} {\bibinfo {author} {\bibfnamefont {C.}~\bibnamefont
  {Brouder}}, \bibinfo {author} {\bibfnamefont {G.}~\bibnamefont {Panati}},
  \bibinfo {author} {\bibfnamefont {M.}~\bibnamefont {Calandra}}, \bibinfo
  {author} {\bibfnamefont {C.}~\bibnamefont {Mourougane}},\ and\ \bibinfo
  {author} {\bibfnamefont {N.}~\bibnamefont {Marzari}},\ }\bibfield  {title}
  {\bibinfo {title} {Exponential localization of wannier functions in
  insulators},\ }\href@noop {} {\bibfield  {journal} {\bibinfo  {journal}
  {Physical review letters}\ }\textbf {\bibinfo {volume} {98}},\ \bibinfo
  {pages} {046402} (\bibinfo {year} {2007})}\BibitemShut {NoStop}%
\bibitem [{\citenamefont {Greiter}\ \emph {et~al.}(2016)\citenamefont
  {Greiter}, \citenamefont {Schnells},\ and\ \citenamefont
  {Thomale}}]{SM:PhysRevB.93.245156}%
  \BibitemOpen
  \bibfield  {author} {\bibinfo {author} {\bibfnamefont {M.}~\bibnamefont
  {Greiter}}, \bibinfo {author} {\bibfnamefont {V.}~\bibnamefont {Schnells}},\
  and\ \bibinfo {author} {\bibfnamefont {R.}~\bibnamefont {Thomale}},\
  }\bibfield  {title} {\bibinfo {title} {Laughlin states and their
  quasiparticle excitations on the torus},\ }\href
  {https://doi.org/10.1103/PhysRevB.93.245156} {\bibfield  {journal} {\bibinfo
  {journal} {Phys. Rev. B}\ }\textbf {\bibinfo {volume} {93}},\ \bibinfo
  {pages} {245156} (\bibinfo {year} {2016})}\BibitemShut {NoStop}%
\bibitem [{\citenamefont {Mumford}(1983)}]{SM:mumford1983tata}%
  \BibitemOpen
  \bibfield  {author} {\bibinfo {author} {\bibfnamefont {D.}~\bibnamefont
  {Mumford}},\ }\href@noop {} {\emph {\bibinfo {title} {Tata lectures on theta
  I}}}\ (\bibinfo  {publisher} {Springer},\ \bibinfo {year} {1983})\BibitemShut
  {NoStop}%
\bibitem [{\citenamefont {Gerhard}\ \emph {et~al.}(2026)\citenamefont
  {Gerhard}, \citenamefont {Wang}, \citenamefont {Cerjan},\ and\ \citenamefont
  {Benalcazar}}]{SM:SpatLoc}%
  \BibitemOpen
  \bibfield  {author} {\bibinfo {author} {\bibfnamefont {H.}~\bibnamefont
  {Gerhard}}, \bibinfo {author} {\bibfnamefont {Y.}~\bibnamefont {Wang}},
  \bibinfo {author} {\bibfnamefont {A.}~\bibnamefont {Cerjan}},\ and\ \bibinfo
  {author} {\bibfnamefont {W.~A.}\ \bibnamefont {Benalcazar}},\ }\bibfield
  {title} {\bibinfo {title} {A spatial localizer for electrons in insulators},\
  }\href@noop {} {\bibfield  {journal} {\bibinfo  {journal} {arXiv:2603.13206}\
  } (\bibinfo {year} {2026})},\ \Eprint {https://arxiv.org/abs/2603.13206}
  {arXiv:2603.13206 [cond-mat.mtrl-sci]} \BibitemShut {NoStop}%
\bibitem [{\citenamefont {Robertson}(1929)}]{SM:robertson1929uncertainty}%
  \BibitemOpen
  \bibfield  {author} {\bibinfo {author} {\bibfnamefont {H.~P.}\ \bibnamefont
  {Robertson}},\ }\bibfield  {title} {\bibinfo {title} {The uncertainty
  principle},\ }\href@noop {} {\bibfield  {journal} {\bibinfo  {journal}
  {Physical Review}\ }\textbf {\bibinfo {volume} {34}},\ \bibinfo {pages} {163}
  (\bibinfo {year} {1929})}\BibitemShut {NoStop}%
\bibitem [{\citenamefont {Laughlin}(1983)}]{SM:laughlin1983}%
  \BibitemOpen
  \bibfield  {author} {\bibinfo {author} {\bibfnamefont {R.~B.}\ \bibnamefont
  {Laughlin}},\ }\bibfield  {title} {\bibinfo {title} {Anomalous quantum hall
  effect: An incompressible quantum fluid with fractionally charged
  excitations},\ }\href {https://doi.org/10.1103/PhysRevLett.50.1395}
  {\bibfield  {journal} {\bibinfo  {journal} {Phys. Rev. Lett.}\ }\textbf
  {\bibinfo {volume} {50}},\ \bibinfo {pages} {1395} (\bibinfo {year}
  {1983})}\BibitemShut {NoStop}%
\bibitem [{\citenamefont {Haldane}(1983)}]{SM:Haldane1983}%
  \BibitemOpen
  \bibfield  {author} {\bibinfo {author} {\bibfnamefont {F.~D.~M.}\
  \bibnamefont {Haldane}},\ }\bibfield  {title} {\bibinfo {title} {Fractional
  quantization of the {Hall} effect: A hierarchy of incompressible quantum
  fluid states},\ }\href {https://doi.org/10.1103/PhysRevLett.51.605}
  {\bibfield  {journal} {\bibinfo  {journal} {Phys. Rev. Lett.}\ }\textbf
  {\bibinfo {volume} {51}},\ \bibinfo {pages} {605} (\bibinfo {year}
  {1983})}\BibitemShut {NoStop}%
\bibitem [{\citenamefont {Rezayi}\ and\ \citenamefont
  {Haldane}(1985)}]{SM:PhysRevB.32.6924}%
  \BibitemOpen
  \bibfield  {author} {\bibinfo {author} {\bibfnamefont {E.~H.}\ \bibnamefont
  {Rezayi}}\ and\ \bibinfo {author} {\bibfnamefont {F.~D.~M.}\ \bibnamefont
  {Haldane}},\ }\bibfield  {title} {\bibinfo {title} {Incompressible states of
  the fractionally quantized hall effect in the presence of impurities: A
  finite-size study},\ }\href {https://doi.org/10.1103/PhysRevB.32.6924}
  {\bibfield  {journal} {\bibinfo  {journal} {Phys. Rev. B}\ }\textbf {\bibinfo
  {volume} {32}},\ \bibinfo {pages} {6924(R)} (\bibinfo {year}
  {1985})}\BibitemShut {NoStop}%
\bibitem [{\citenamefont {Trugman}\ and\ \citenamefont
  {Kivelson}(1985)}]{SM:Trugman1985}%
  \BibitemOpen
  \bibfield  {author} {\bibinfo {author} {\bibfnamefont {S.~A.}\ \bibnamefont
  {Trugman}}\ and\ \bibinfo {author} {\bibfnamefont {S.}~\bibnamefont
  {Kivelson}},\ }\bibfield  {title} {\bibinfo {title} {Exact results for the
  fractional quantum {Hall} effect with general interactions},\ }\href
  {https://doi.org/10.1103/PhysRevB.31.5280} {\bibfield  {journal} {\bibinfo
  {journal} {Phys. Rev. B}\ }\textbf {\bibinfo {volume} {31}},\ \bibinfo
  {pages} {5280} (\bibinfo {year} {1985})}\BibitemShut {NoStop}%
\bibitem [{\citenamefont {Lee}\ \emph {et~al.}(2015)\citenamefont {Lee},
  \citenamefont {Papi\'{c}},\ and\ \citenamefont {Thomale}}]{SM:Papic2015}%
  \BibitemOpen
  \bibfield  {author} {\bibinfo {author} {\bibfnamefont {C.~H.}\ \bibnamefont
  {Lee}}, \bibinfo {author} {\bibfnamefont {Z.}~\bibnamefont {Papi\'{c}}},\
  and\ \bibinfo {author} {\bibfnamefont {R.}~\bibnamefont {Thomale}},\
  }\bibfield  {title} {\bibinfo {title} {Geometric construction of quantum
  {Hall} clustering {Hamiltonians}},\ }\href
  {https://doi.org/10.1103/PhysRevX.5.041003} {\bibfield  {journal} {\bibinfo
  {journal} {Phys. Rev. X}\ }\textbf {\bibinfo {volume} {5}},\ \bibinfo {pages}
  {041003} (\bibinfo {year} {2015})}\BibitemShut {NoStop}%
\bibitem [{\citenamefont {Jain}(1989)}]{SM:jain1989}%
  \BibitemOpen
  \bibfield  {author} {\bibinfo {author} {\bibfnamefont {J.~K.}\ \bibnamefont
  {Jain}},\ }\bibfield  {title} {\bibinfo {title} {Composite-fermion approach
  for the fractional quantum hall effect},\ }\href
  {https://doi.org/10.1103/PhysRevLett.63.199} {\bibfield  {journal} {\bibinfo
  {journal} {Phys. Rev. Lett.}\ }\textbf {\bibinfo {volume} {63}},\ \bibinfo
  {pages} {199} (\bibinfo {year} {1989})}\BibitemShut {NoStop}%
\bibitem [{\citenamefont {Yoshioka}(1984)}]{SM:yoshioka1984ground}%
  \BibitemOpen
  \bibfield  {author} {\bibinfo {author} {\bibfnamefont {D.}~\bibnamefont
  {Yoshioka}},\ }\bibfield  {title} {\bibinfo {title} {Ground state of the
  two-dimensional charged particles in a strong magnetic field and the
  fractional quantum hall effect},\ }\href@noop {} {\bibfield  {journal}
  {\bibinfo  {journal} {Physical Review B}\ }\textbf {\bibinfo {volume} {29}},\
  \bibinfo {pages} {6833} (\bibinfo {year} {1984})}\BibitemShut {NoStop}%
\bibitem [{\citenamefont {Yoshioka}\ \emph {et~al.}(1983)\citenamefont
  {Yoshioka}, \citenamefont {Halperin},\ and\ \citenamefont
  {Lee}}]{SM:PhysRevLett.50.1219}%
  \BibitemOpen
  \bibfield  {author} {\bibinfo {author} {\bibfnamefont {D.}~\bibnamefont
  {Yoshioka}}, \bibinfo {author} {\bibfnamefont {B.~I.}\ \bibnamefont
  {Halperin}},\ and\ \bibinfo {author} {\bibfnamefont {P.~A.}\ \bibnamefont
  {Lee}},\ }\bibfield  {title} {\bibinfo {title} {Ground state of
  two-dimensional electrons in strong magnetic fields and $\frac{1}{3}$
  quantized hall effect},\ }\href {https://doi.org/10.1103/PhysRevLett.50.1219}
  {\bibfield  {journal} {\bibinfo  {journal} {Phys. Rev. Lett.}\ }\textbf
  {\bibinfo {volume} {50}},\ \bibinfo {pages} {1219} (\bibinfo {year}
  {1983})}\BibitemShut {NoStop}%
\bibitem [{\citenamefont {Sun}\ \emph {et~al.}(2011)\citenamefont {Sun},
  \citenamefont {Gu}, \citenamefont {Katsura},\ and\ \citenamefont
  {Das~Sarma}}]{SM:PhysRevLett.106.236803}%
  \BibitemOpen
  \bibfield  {author} {\bibinfo {author} {\bibfnamefont {K.}~\bibnamefont
  {Sun}}, \bibinfo {author} {\bibfnamefont {Z.}~\bibnamefont {Gu}}, \bibinfo
  {author} {\bibfnamefont {H.}~\bibnamefont {Katsura}},\ and\ \bibinfo {author}
  {\bibfnamefont {S.}~\bibnamefont {Das~Sarma}},\ }\bibfield  {title} {\bibinfo
  {title} {Nearly flatbands with nontrivial topology},\ }\href
  {https://doi.org/10.1103/PhysRevLett.106.236803} {\bibfield  {journal}
  {\bibinfo  {journal} {Phys. Rev. Lett.}\ }\textbf {\bibinfo {volume} {106}},\
  \bibinfo {pages} {236803} (\bibinfo {year} {2011})}\BibitemShut {NoStop}%
\bibitem [{\citenamefont {Lu}\ and\ \citenamefont {Santos}(2026)}]{SM:Lu_2026}%
  \BibitemOpen
  \bibfield  {author} {\bibinfo {author} {\bibfnamefont {T.}~\bibnamefont
  {Lu}}\ and\ \bibinfo {author} {\bibfnamefont {L.~H.}\ \bibnamefont
  {Santos}},\ }\bibfield  {title} {\bibinfo {title} {Exciton-anyon binding in
  fractional chern insulators: Spectral fingerprints},\ }\bibfield  {journal}
  {\bibinfo  {journal} {Physical Review Letters}\ }\textbf {\bibinfo {volume}
  {136}},\ \href {https://doi.org/10.1103/8bp8-g8g4} {10.1103/8bp8-g8g4}
  (\bibinfo {year} {2026})\BibitemShut {NoStop}%
\bibitem [{\citenamefont {Okuma}(2024)}]{SM:PhysRevB.110.245112}%
  \BibitemOpen
  \bibfield  {author} {\bibinfo {author} {\bibfnamefont {N.}~\bibnamefont
  {Okuma}},\ }\bibfield  {title} {\bibinfo {title} {Constructing vortex
  functions and basis states of chern insulators: Ideal condition, inequality
  from index theorem, and coherentlike states on the von neumann lattice},\
  }\href {https://doi.org/10.1103/PhysRevB.110.245112} {\bibfield  {journal}
  {\bibinfo  {journal} {Phys. Rev. B}\ }\textbf {\bibinfo {volume} {110}},\
  \bibinfo {pages} {245112} (\bibinfo {year} {2024})}\BibitemShut {NoStop}%
\bibitem [{\citenamefont {Ledwith}\ \emph {et~al.}(2023)\citenamefont
  {Ledwith}, \citenamefont {Vishwanath},\ and\ \citenamefont
  {Parker}}]{SM:ledwith2023vortexability}%
  \BibitemOpen
  \bibfield  {author} {\bibinfo {author} {\bibfnamefont {P.~J.}\ \bibnamefont
  {Ledwith}}, \bibinfo {author} {\bibfnamefont {A.}~\bibnamefont
  {Vishwanath}},\ and\ \bibinfo {author} {\bibfnamefont {D.~E.}\ \bibnamefont
  {Parker}},\ }\bibfield  {title} {\bibinfo {title} {Vortexability: A unifying
  criterion for ideal fractional chern insulators},\ }\href
  {https://doi.org/10.1103/PhysRevB.108.205144} {\bibfield  {journal} {\bibinfo
   {journal} {Phys. Rev. B}\ }\textbf {\bibinfo {volume} {108}},\ \bibinfo
  {pages} {205144} (\bibinfo {year} {2023})}\BibitemShut {NoStop}%
\bibitem [{\citenamefont {Klauder}(1963)}]{SM:klauder1963continuous}%
  \BibitemOpen
  \bibfield  {author} {\bibinfo {author} {\bibfnamefont {J.~R.}\ \bibnamefont
  {Klauder}},\ }\bibfield  {title} {\bibinfo {title} {Continuous-representation
  theory. ii. generalized relation between quantum and classical dynamics},\
  }\href@noop {} {\bibfield  {journal} {\bibinfo  {journal} {Journal of
  Mathematical Physics}\ }\textbf {\bibinfo {volume} {4}},\ \bibinfo {pages}
  {1058} (\bibinfo {year} {1963})}\BibitemShut {NoStop}%
\bibitem [{\citenamefont {Klauder}\ and\ \citenamefont
  {Skagerstam}(1985)}]{SM:klauder1985coherent}%
  \BibitemOpen
  \bibfield  {author} {\bibinfo {author} {\bibfnamefont {J.~R.}\ \bibnamefont
  {Klauder}}\ and\ \bibinfo {author} {\bibfnamefont {B.-S.}\ \bibnamefont
  {Skagerstam}},\ }\href@noop {} {\emph {\bibinfo {title} {Coherent states:
  applications in physics and mathematical physics}}}\ (\bibinfo  {publisher}
  {World scientific},\ \bibinfo {year} {1985})\BibitemShut {NoStop}%
\bibitem [{\citenamefont {Byrd}\ \emph {et~al.}(1995)\citenamefont {Byrd},
  \citenamefont {Lu}, \citenamefont {Nocedal},\ and\ \citenamefont
  {Zhu}}]{SM:byrd1995limited}%
  \BibitemOpen
  \bibfield  {author} {\bibinfo {author} {\bibfnamefont {R.~H.}\ \bibnamefont
  {Byrd}}, \bibinfo {author} {\bibfnamefont {P.}~\bibnamefont {Lu}}, \bibinfo
  {author} {\bibfnamefont {J.}~\bibnamefont {Nocedal}},\ and\ \bibinfo {author}
  {\bibfnamefont {C.}~\bibnamefont {Zhu}},\ }\bibfield  {title} {\bibinfo
  {title} {A limited memory algorithm for bound constrained optimization},\
  }\href@noop {} {\bibfield  {journal} {\bibinfo  {journal} {SIAM Journal on
  scientific computing}\ }\textbf {\bibinfo {volume} {16}},\ \bibinfo {pages}
  {1190} (\bibinfo {year} {1995})}\BibitemShut {NoStop}%
\bibitem [{\citenamefont {Fukui}\ \emph {et~al.}(2005)\citenamefont {Fukui},
  \citenamefont {Hatsugai},\ and\ \citenamefont {Suzuki}}]{SM:fukui2005chern}%
  \BibitemOpen
  \bibfield  {author} {\bibinfo {author} {\bibfnamefont {T.}~\bibnamefont
  {Fukui}}, \bibinfo {author} {\bibfnamefont {Y.}~\bibnamefont {Hatsugai}},\
  and\ \bibinfo {author} {\bibfnamefont {H.}~\bibnamefont {Suzuki}},\
  }\bibfield  {title} {\bibinfo {title} {Chern numbers in discretized brillouin
  zone: Efficient method of computing (spin) hall conductances},\ }\href@noop
  {} {\bibfield  {journal} {\bibinfo  {journal} {Journal of the Physical
  Society of Japan}\ }\textbf {\bibinfo {volume} {74}},\ \bibinfo {pages}
  {1674} (\bibinfo {year} {2005})}\BibitemShut {NoStop}%
\end{thebibliography}
\end{document}